\documentclass[11pt]{article}
\usepackage{amsmath, amssymb}
\usepackage{slashed}
\usepackage{verbatim}
\usepackage{latexsym}
\usepackage{euscript}
\usepackage{amsfonts}
\usepackage{verbatim}
\usepackage{cancel}
\usepackage{soul}
\usepackage[normalem]{ulem}
\usepackage[]{array}
\usepackage[]{mathrsfs}
\usepackage[utf8]{inputenc}
\usepackage[T1]{fontenc}
\usepackage{makeidx}
\makeindex
\usepackage{graphicx}
\usepackage{amsmath}
\usepackage{amssymb}
\usepackage{amsxtra}
\usepackage{color}
\def\be{\begin{eqnarray}}
\def\ee{\end{eqnarray}}
\def\0{\nonumber}

\def\tr{{\rm tr}}

\def\vx{\stackrel{\rightarrow}{x}}

\def\det{{\rm det}}

\usepackage{slashed}
\usepackage{verbatim}
\usepackage{latexsym}\usepackage{color}
\usepackage{euscript}
\usepackage[curve]{xypic}

\newcommand\ED{\EuScript{D}}

\newcommand\EV{\EuScript{V}}
\newcommand\EE{\EuScript{E}}

\def\sfe{{\sf e}}

\def\sfP{{\sf P}}

\def\sfG{{\sf G}}
\def\sfM{{\sf M}}

\def\sfU{{\sf U}}

\def\p{p\!\!/}
\normalfont\large
\def\parl{\!\buildrel \leftrightarrow\over\partial\!\!}
\begin{document}
\vskip 2cm
\begin{flushright}
{\it {In memory of Alexander Andreevich Andrianov}}
\end{flushright}
\vskip 2cm
\begin{center}

{\LARGE A fermion primer }

\vskip 1cm

{\large  Loriano Bonora$^{a}$\footnote{email:bonora@sissa.it}, Roberto Soldati$^{b}$\footnote{email:roberto.soldati@unibo.it}\\
\textit{${}^{a}$ International School for Advanced Studies (SISSA),\\Via
Bonomea 265, 34136 Trieste, Italy}\\ 
\textit{${}^{b}$ Universit\`a di Bologna, via Zamboni 33, 40126 Bologna, Italy}}
\end{center}
\vskip2cm
{
{\bf Abstract}.This is a review paper intended to illustrate a few critical issues concerning Dirac, Weyl and Majorana fermions and their differences. The first part consists in basic introductions to fermions, and more in detail to fermions in 4d, focusing in particular in what differentiate the three type of fermions: chirality, helicity, mass, field equations, properties under discrete symmetry transformations, Actions, Observables. On this basis we tackle a series of challenging and sometime controversial problems: the non-existence of Weyl fermion propagators, and the ways to circumvent it;  the regularizations for Weyl fermion amplitudes, in particular the appropriateness of using the PV regularization; the definition of a functional integral for Weyl fermion, the difference between Weyl and massless Majorana fermions
and the difference between the Dirac and Majorana mass terms. Most notably we come to the conclusion that the trick of replacing the non-existing Weyl fermion propagator with a massless Dirac one, although it may yield in some cases correct results, is flawed at the very origin by a logical 
loophole. We then show which is the correct way to proceed in this case. We consider also the topic of applying the Wick rotation at the classical Action level for Weyl fermions and conclude that this procedure leads to a nonequivalent theory. Finally, although anomalies are not the central focus here, we have deemed it useful to summarily review the relation, and its cohomological basis, that exists between the lack of a Weyl fermion propagator (non-invertibility of the Weyl-Dirac operator) and the appearance of dangerous anomalies in the theory.

\tableofcontents
\section{Introduction}

Fermions present surprising and unfamiliar aspects, with respect to standard macroscopic objects. When we take a turn of $360^o$ we see the same world as before turning. For a fermion this is not the case: in our anthropic language we say that it has to turn twice, i.e. $720^o$, in order to return to the initial situation. A fermion sees a world twice as large as our world. We say that fermions have half-integer spin, at variance with the other set of elementary particles, the bosons, which have integer spin. The consequence of this peculiarity is that the way we represent fermions is drastically different from the way we represent bosons. In order to represent fermions we use mathematical objects called spinors. The latter are called that way because they belong to representations of a spin group (see below), which has precisely to do with the revolving property alluded to above. Spin one-half spinors in 4d can be represented by two or four complex numbers, usually ordered in a column vector. As a consequence the action on them is represented by matrices, the Pauli or gamma matrices. The spinor wave functions or spinor fields are thus represented by columns of two or four functions. There are three distinct types of spinor wave-functions or fields: Weyl, Majorana and Dirac. The equations of motion are matrices of linear differential operator acting on them. The non-standard mathematics of spinors brings into the game some problems which do not seem to have been uniformly elaborated in the literature, or, at least, are resolved in ambiguous ways. This is a review  where we discuss such issues and the way we have resolved them. We utilize in part material taken or ri-elaborated  from our past research \cite{bonorasoldatizalel2020,bonorasoldati2019,LB}.

A first problem arises concerning the mass term for spinors, which can be vanishing or non-vanishing. Massive spinors are either Dirac or Majorana. They are bispinors with four components. In the Majorana case they can be cast in real form. Massless spinors can be of three types: Weyl spinors with two complex components or chiral bispinors; self-conjugate Majorana bispinors; Dirac bispinors. The Majorana mass term leads inevitably to a self-conjugate Majorana bispinor, for a Weyl spinor has perforce zero mass the two chiral components being decoupled (the mass terms does in fact couple them, as is well known). If a spinor has mass it cannot be Weyl, and viceversa. This is what we try to clarify in the simplest possible way, also to remove the debris generated by the illegal identification of massless Majorana and Weyl  fermions.

The second problem arises when one quantizes a theory containing Weyl fermions. Quantization requires the existence of the corresponding propagator. A propagator is the inverse of the kinetic operator, and  for Weyl fermions formally it does not exist because the kinetic operator contains a chiral projector. In perturbative field theory one resolves the puzzle by assigning to Weyl fermions the same propagator as for massless Dirac fermions (which does exist).  {As will be commented upon below this is an arbitrary way of proceeding and one does not have any warranty that the result so obtained makes any sense.  In fact experience shows that this trick may work as long as one does not meet critical issues such as anomalies (for Weyl fermions anomalies are the symptoms that a proper propagator does not exist). We will argue below that in such cases this procedure is based on a logical paradox. Nevertheless this may work in perturbative field theory because the fermion chiral nature is preserved by vertices which contain the chiral projector. An explanation will be provided for the cases in which this trick leads to correct results. In general it has to be assumed that the perturbative approach (Feynman diagrams) applied to Weyl fermion, in which the propagator employed is the Dirac one, is not reliable. For nonperturbative approaches (such as the heat kernel one) the same trick may lead astray because it changes the nature of the problem as will be explained below.}

Additional problems and ambiguities arise when one comes to the effective Action of a Weyl fermion theory. One can compute it perturbatively or non-perturbatively. In any case one faces the issue of regularization. {In the perturbative case one needs propagators and vertices, which, as we have just seen, becomes a formidable obstacle for Weyl fermions. The tentative remedy is to use the Dirac propagator instead of the missing Weyl one. Once this choice is made one faces the issue of regularization.}  We have several choices: dimensional \cite{DR1,DR2,DR3} Pauli-Villars \cite{PV,IZ} or cutoff regularizaton - below we consider these three in a simple enough example. 
Here we make clear a fundamental point: a PV regularized  propagator is not the inverse of any kinetic term of whatsoever classical local Action, independently of its being Dirac or Majorana (whose propagators coincide). A somewhat similar problem arises with the technique of Wick rotation: one may be tented to do the Wick rotation in the original classical Minkowski theory and formulate an equivalent problem in a Euclidean  spacetime. However we will show that the so obtained Euclidean theory, provided it exists, has very little to do with the original Minkowski one. In simple words, one cannot regularize the original local form of the Action. The regularization process is something separate, which has to do with the mathematical (distribution theoretical) properties of Feynman diagrams. In the case of Weyl fermions this process is particularly delicate, one must make sure that throughout it chirality is preserved. Careful not to be victim of a semantic trap: it is standard lore that in the regularization process a symmetry may be broken, but chirality is not a symmetry, it is the defining status of the fields in the theory. If `chirality is broken' it means that one, without realizing it,  has landed in a foreign theory. For this reason it is of upmost importance how we define the relevant Feynman diagrams or the relevant heat kernel operator. From this point of view PV regularization must be particularly under scrutiny. It needs masses and, as anticipated above, massive fermions can only be Dirac or Majorana, which carry both chiralities. PV can be therefore applied only as a mathematical algorithm to regularize chiral fermion integrals (see an example below),  where masses are simple regularizing parameters, without any reference to actual fields and particles. {In general a good amount of skepticism is in order when using a perturbative method based on Feynman diagrams for a theory involving Weyl fermions. Better methods are available.}

The paper is organized as follows. Section 2 is devoted to a general introduction of fermions as representations of a Clifford algebra and a spin group. This Section can be skipped by those who are not specifically interested in it without prejudice for the comprehension of the subsequent Sections.
Section 3 contains a very basic introduction to Weyl, Dirac and Majorana spinors in 4d\footnote{In this paper we limit ourselves to the traditional classification of fermions. We do not consider the enlargements of the spinor concept due to Lounesto \cite{lounesto}.}. Weyl fermions are the basis for the construction also of Dirac and Majorana ones. Classical solutions, symmetries and canonical quantization, with observables and causal Green's functions, are explicitly derived and exhibited. In Section 4 a summary is given of Dirac, Majorana and Weyl fermions in 4D with their properties in the language more appropriate for quantum field theory. Section 5 is devoted to explicit examples of regularizations (dimensional, Pauli-Villars and cutoff) for Weyl fermions. The subject of Section 6 are Dirac and Weyl propagators, together with the severe problems carried along by the latter, of which both false and true solutions are discussed. Two more critical issues are also discussed in this Section: the difference between Weyl and massless Majorana fermions, and the Dirac and Majorana mass terms. Section 7 is devoted to Euclidean fermion field theories and, in particular, to the impossibility to write down a real Action for Weyl fermions. Finally in Section 8  the Atiyah-Singer family's index theorem is presented and its relevance 
for the invertibility of the Dirac-Weyl operator (=existence of propagator) is highlighted.

\vskip 1cm

{\bf Notation.} The metric $g_{\mu\nu}$ has mostly - signature. Gamma
matrices satisfy $\{\gamma^\mu,\gamma^\nu\}= 2 g^{\mu\nu}$ and
\be
\gamma_{\mu}^\dagger= \gamma_0\gamma_\mu\gamma_0\0
\ee
The generators of the Lorentz group are $\Sigma_{\mu\nu} =\frac 14
[\gamma_\mu,\gamma_\nu]$.
The charge conjugation operator $C$ is defined to satisfy
\be
\gamma_{\mu}^T = -C^{-1} \gamma_\mu C, \quad\quad CC^*=-1, \quad\quad
CC^\dagger=1\label{C}
\ee 
The chiral matrix $\gamma_5=i\gamma^0\gamma^1\gamma^2\gamma^3$ has the
properties
\be
\gamma_5^\dagger=\gamma_5, \quad\quad (\gamma_5)^2=1, \quad\quad C^{-1} \gamma_5
C= \gamma_5^T\0
\ee

In the sequel, when more convenient, we shall use a more explicit representation of the gamma matrices, see beginning of section 3.

\section{Clifford algebras and spinors}
\label{s:Clifford}

Spinor fields in quantum field theories are founded on the mathematical concept of Clifford algebra and its representations. Spinors are in fact representations both of a Clifford algebra and of a spin group, which is in turn a subgroup of the Clifford algebra. This Section is meant to be a short introduction to Clifford algebras and to its relevance for spin and spinor fields (more complete accounts can be found, for instance, in \cite{Lawson,Vaz}). The uninterested reader can skip it without prejudice. The only essential thing to be kept in mind for the sequel is the table at the end.

In any vector space $V$ with a quadratic form $q(v)$, defined $ \forall v\in V$,  one can introduce the tensor algebra ${\EuScript T} (V)=\sum_{r=0}^\infty \otimes ^r V$. The ideal ${\EuScript I}_q(V)$ is formed by all elements of the type $ v\otimes v - q(v) 1$. The quotient  of ${\EuScript T} (V)$ by the latter defines the Clifford algebra
\be
C\ell(V,q)= {\EuScript T}(V)/{\EuScript I}_q(V)\0
\ee
$C\ell(V,q)$ can also be defined as the algebra generated by $V$ (and 1), subject to the relation
\be
u\cdot v + v\cdot u= 2 g(u,v)\label{qvw}
\ee
for $u,v\in V$. In turn $g(u,v)$ is given by $2g(u,v)= q(u+v)-q(u)-q(v)$, and the dot in \eqref{qvw} is the product 
\be
u\!\cdot\!v = u\wedge v +g(u,v)  \label{cliff}
\ee

Let $\{\sfe_i\}, i=1,\ldots n$ of $V$ be a basis of $V$ with $g(\sfe_i,\sfe_j)= \eta_{ij}$, where $\eta_{ij}$ is the diagonal matrix with $r$ entries 1 and $s$ entries -1,  and let the quadratic form $q$ be defined by means of this basis. Then the relation \eqref{qvw} becomes
\be
\sfe_i\cdot \sfe_j + \sfe_j\cdot \sfe_i = 2 \eta_{ij}\label{eiejetaij}
\ee 
Eq.\eqref{qvw} has the form of  a $\gamma$-matrix algebra of field theory, which is not accidental
because the latter is precisely a representation of the relevant Clifford algebra. When we wish to refer specifically the form \eqref{eiejetaij}, we denote Clifford algebras by  $C\ell_{r,s}$. The latter have been classified i terms of (algebraic) fields and of real and complex  $2\times 2$ matrix rings,  ${\EuScript M}(2, \mathbb R)$ and ${\EuScript M}(2, \mathbb C)$, respectively.

For instance
\be
C\ell_{3,0}= {\EuScript M}(2, \mathbb C),\quad\quad C\ell_{3,1}= {\EuScript M}(2, \mathbb R)\otimes {\EuScript M}(2, \mathbb R), \quad\quad {\rm etc.}\0
\ee
The Clifford algebras considered so far are real. Complex Clifford algebra are obtained by complexifying real ones. They are denoted $C\ell_{\mathbb C} (n)$. Their classification is simple
\be
C\ell_{\mathbb C} (2k)= {\EuScript M}(2^k,{\mathbb C}), \quad\quad C\ell_{\mathbb C}(2k+1)=  {\EuScript M}(2^k,{\mathbb C})\oplus  {\EuScript M}(2^k,{\mathbb C})\0
\ee

There is an automorphism of $C\ell(V,q)$ induced by $u\to \tilde u= -u$ for $u\in V$, which we use to decompose it into even and odd subspaces
\be
C\ell(V,q)= C\ell^0(V,q)\oplus C\ell^1(V,q)\label{CLsplit}
\ee
It gives $C\ell(V,q))$ a ${\mathbb Z}_2$-graded algebra structure: $C\ell^i(V,q)\cdot C\ell^j(V,q)\subseteq C\ell^{i+j}(V,q)$, $i,j=0,1$.

In a Clifford algebra one can recognize various groups. The largest one is the group of invertible elements
\be
C\ell_{r,s}^\times = \{a \in C\ell_{r,s}|\,\exists\, a^{-1}\}\label{invertibles}
\ee
An important subgroup is the Clifford-Lipschitz group
\be
{\cal C}_{r,s} = \{ a\in C\ell_{r,s}^\times|\,a\,v\,a^{-1}\in V,\, \, \forall v \in V= {\mathbb R}^{r,s}\}\label{Gammapq}
\ee
whose definition hinges on the adjoint action in $V$, and can be extended to the full Clifford algebra 
\be
\sigma_a (x) = a\, x\, a^{-1},\quad\quad \forall a \in{\cal C}_{r,s}, \, \forall x\in  C\ell_{r,s}\label{adjointaction}
\ee
$\sigma_a$ is an automorphism of  ${\cal C}_{r,s}$. A similar but more useful operation is the twisted adjoint 
\be
\widetilde \sigma_a(x) = \tilde a \, x \, a^{-1}\label{twistedadjoint}
\ee
where $\tilde a=-a$ as above. 

With the above tools we can now define more familiar groups. First the group
\be
O(V,q) = \{ t \in GL(V): t^\ast q=q\}\0
\ee 
where $t^\ast q$ is the pullback of the quadratic form $q$ by the linear map $t:V\to V$. Therefore 
$O(V,q) $ is the group of those linear invertible maps in $GL(V)$ that  leave  the quadratic form $q$ unchanged. llowing this we define the subgroup
\be 
SO(V,q)=\{ t\in O(V,q): {\det} (t)=1\}\0
\ee
One can prove the correspondences
\be
\widetilde \sigma ({\cal C}_{r,s} )= O(r,s), \quad\quad \widetilde \sigma ({\cal C}^0_{r,s} )= SO(r,s)
\label{sigmaOSO}
\ee
where ${\cal C}^0_{r,s}= {\cal C}_{r,s}\cap C\ell^0_{r,s} $ is the even Clifford subalgebra.

Now let us come to the spin groups. For any element $a\in C\ell(V,q)$one can define a norm $N(a)$, such that $N(u)= q(u)$ for any $u\in V$. Then iif $V$ is endowed with a quadratic form $q$,  the spin group of $V$  
\be
Spin(V,q) = \{a\in {\cal C}^0_{r,s}| N(a)=\pm 1\}\label{spingroupdef}
\ee
The restricted spin group $Spin^+(V,q)$  is identified by  $N(a)=1$. 

One can prove that the following sequence is exact
\be
0\longrightarrow {\mathbb Z}_2 \longrightarrow Spin(V,q) \stackrel{\widetilde \sigma}{ \longrightarrow} SO(V,q)  \longrightarrow 0\0
\ee
where the twisted adjoint map $\widetilde\sigma$, for any $u,v\in V$, can also be represented as
\be
\widetilde\sigma_u(v)= v -2 \frac{q(u,v)}{q(u)} u\label{Ad}
\ee
As follows from the above exact sequence, $Spin(V,q)$ is a double covering of $SO(V,q)$. 

Let us come now to representations. A $\mathbb K$ ($\mathbb K$=$\mathbb C$ or $\mathbb R$) representation of $C\ell (V,q) $ is a  $\mathbb K$-algebra homomorphism
\be
C\ell(V,q) \longrightarrow  Hom_{\mathbb K}(W,W)\0
\ee
of $C\ell(V,q)$ into the linear transformations of a finite dimensional vector space $W$. $W$ is then called a $\mathbb K$-{\it module} of $C\ell(V,q)$. It is evident that this defines also representations of $Spin(V,q)$ and $SO(V,q)$ as groups contained in $C\ell(V,q)$. 

We are now ready to define spinors. In the literature there are various definitions.  We choose the following:

Let us consider a vector space $V={\mathbb R}^{r,s}$ with canonical quadratic form $q$  and the associated Clifford algebra $C\ell(V,q)$, 
\begin{itemize}
\item {\it A spinor is any vector belonging to an irreducible representation space of the associated spin group $Spin(V,q)$.}
\end{itemize}

In the literature one can find a classification of all spin representations. Here we limit ourselves to the classification of complex representations. They are given by
\begin{itemize}
\item  $n=2k, \quad\quad {\mathbb C}^{2^{k-1}}\oplus  {\mathbb C}^{2^{k-1}}$
\item  $n=2k+1, \quad\quad {\mathbb C}^{2^{k}}$
\end{itemize}

For $n=2k$ The decomposition into two irreducible representations is related  to the volume element. The latter , in the complex case, is given by
\be
\omega_{\mathbb C} =i^k \sfe_1 \ldots \sfe_{2k},\label{volumeelement}
\ee
where $\sfe_i$ is an orthonormal basis of $V={\mathbb R}^{r,s}$ (see above).
$\omega_{\mathbb C}$ is characterized by the property $\omega_{\mathbb C}^2=1$. These two irreducible representations are identified  by the eigenvalues $\pm 1$ of $\omega_{\mathbb C}$.  The reelevant projectors are $\pi_\pm =\frac 12 (1\pm \omega_{\mathbb C})$.

It is evident that the element $\omega_{\mathbb C}$ corresponds to the chirality operator in quantum field theory.

\subsection{Spinor representations in even dimension}

The properties of spinors in even dimensional spaces are summarized in the following table \ref{tab:spinorsinevend}, \cite{LB}:

\begin{table}[h]
\begin{tabular}{||l|l|l|l|l|l||}
\hline\hline
metric sign.&Weyl, ${\mathbb C}$&Conjugacy&Dirac, ${\mathbb C}$&Majorana-Weyl, ${\mathbb R}$& Majorana
, ${\mathbb R}$\\\hline\hline
\quad\quad(2,0)&\quad 1L\quad \quad 1R&Mutual&\quad\quad   2&\quad-\quad \quad\quad\quad- &\quad 2\\ \hline
\quad\quad(1,1)&\quad 1L\quad\quad   1R&Self&\quad\quad    2&\quad 1L\quad\quad\quad  1R&\quad 2\\ \hline
\quad\quad(4,0)&\quad 2L\quad\quad 2R&Self& \quad\quad 4&\quad-\quad \quad\quad\quad- &\quad -\\\hline
\quad\quad(3,1)&\quad 2L\quad\quad 2R&Mutual&\quad\quad 4&\quad-\quad \quad\quad\quad- &\quad 4\\ \hline
\quad\quad(6,0)&\quad 4L\quad \quad4R& Mutual& \quad\quad 8&\quad-\quad \quad\quad\quad- &\quad 8\\\hline
\quad\quad(5,1)&\quad 4L\quad\quad 4R&Self&\quad\quad 8&\quad-\quad \quad\quad\quad- &\quad -\\ \hline
\quad\quad(8,0)&\quad 8L\quad \quad8R& Self&\quad\quad 16&\quad8L\quad\quad\quad  8R &\quad 16\\\hline
\quad\quad(7,1)&\quad 8L\quad\quad 8R&Mutual&\quad\quad 16&\quad-\quad \quad\quad\quad- &\quad -\\ \hline
\!\!\quad\quad(10,0)&\!\!\quad 16L~\quad16R& Mutual&\quad\quad 32&\quad-\quad \quad\quad\quad- &\quad 32\\\hline
\quad\quad(9,1)&\!\!\quad 16L~\quad16R&Self&\quad\quad 32&\!\!\quad 16L\!\!\quad\quad\quad 16R&\quad 32\\ \hline\hline
\end{tabular}
\caption{Spinors in even dimension}
\label{tab:spinorsinevend}
\end{table}
Captions for this table: the labels $L,R$ denote left-handed or right-handed spinors; the numbers in the columns with $\mathbb C$ or $\mathbb R$ tag represent the number of complex or real components according to whether they are in column $\mathbb C$ or $\mathbb R$, respectively. For instance in the (3,1) case Weyl fermions can be left or right with two complex components each, Dirac fermions have four complex components and Majorana four real (with no definite handedness).  `Self' means self-conjugate spinor and `Mutual' means that the $L$ and $R$ representations are conjugate to each other.

After this general and abstract introduction we focus in the sequel on 4d spinors and fields.

\section{Spinors in 4 D}

In this Section we shall analyze in depth the two basic forms of spinors: namely, the Weyl 2-component spinors, their chiral bispinor representations
and the corresponding dynamical properties, as well as the charge self-conjugated Majorana bispinors. 
As we shall see, the longstanding well-known Dirac bispinors
will be obtained as complex combinations of Majorana bispinors and in terms of a local Bogoliubov transformation. For the sake of clarity and completeness, we would like to specify that in the sequel we shall employ the following (more detailed) conventions and notations.

The Pauli spin matrices are the three Hermitean $2\times2$ matrices
$$
\sigma_1=\left\lgroup
\begin{array}{cc}
0 & 1\\
1 & 0
\end{array}
\right\rgroup\quad
\sigma_2=\left\lgroup
\begin{array}{cc}
0 & - i\\
i & 0
\end{array}
\right\rgroup\quad
\sigma_3=\left\lgroup
\begin{array}{cc}
1 & 0\\
0 & -1
\end{array}
\right\rgroup
$$
which satisfy
$$
\sigma_j\,\sigma_k=\delta_{\,jk}\ +\ i\,\varepsilon_{\,jk\ell}\,\sigma_\ell
\qquad\quad(\,j,k,\ell=1,2,3\,)
$$
where $ \varepsilon_{\,123}=1=-\,\varepsilon^{\,123} $ so that
$$
[\,\sigma_j\,,\,\sigma_k\,]\ =\ 2 i\varepsilon_{jk\ell}\,\sigma_\ell\qquad
\{\sigma_i\,,\,\sigma_j\}\ =\ 2\delta_{\,ij}
$$
The Dirac matrices in the Weyl, or spinorial, or even chiral representation are given by
$$
\gamma^0 =
\left\lgroup\begin{array}{cc}
0 &  {\bf 1}\\
{\bf 1} & 0\\
\end{array}\right\rgroup\qquad\quad
\gamma^{\,k} =
\left\lgroup\begin{array}{cc}
0 & \sigma_k\\
-\,\sigma_k  & 0\\
\end{array}\right\rgroup\qquad\quad
\gamma_{5} = \left\lgroup\begin{array}{cc}
- \mathbf{1} &  0\\
0 & \mathbf{1}\\
\end{array}\right\rgroup
$$
with the Hermitean conjugation property
$$
\gamma^{\,\mu\,\dagger}=\gamma_{0}\,\gamma^{\,\mu}\,\gamma_{0}\qquad\quad
\{\gamma^{\,\mu}\,,\,\gamma^{\,\nu}\}=2\, g^{\,\mu\nu}\,{\mathbb I}
$$
so that
\[
\beta=\gamma^0 =\gamma_{0}=
\left\lgroup\begin{array}{cc}
0 &  {\bf 1}\\
{\bf 1} & 0\\
\end{array}\right\rgroup\qquad\quad
\alpha^{\,k} = \gamma_{0}\gamma^{k} =
\left\lgroup\begin{array}{cc}
-\,\sigma_k & 0\\
0 & \sigma_k\\
\end{array}\right\rgroup\qquad\quad(\,k=1,2,3\,)
\]
We denote by $ \psi_{L} $ the 2-component spinor belonging to the irreducible bidimensional left representation $ D(\frac12,0) $
of the Lorentz group, while $ \psi_{R} $ belongs to its nonequivalent right counterpart $ D(0,\frac12)\,. $ The same holds true for the spin-states
$ u_{L}(\mathbf{p}) $ and $ u_{R}(\mathbf{p}) $ \textit{et cetera}. The four-component spinors, or bispinors, $ \psi(x) $ do belong to the reducible tetradimensional
representation $ D(\frac12,0)\bigoplus D(0,\frac12) $
\[
\psi(x)= \left\lgroup
\begin{array}{c}
\psi_L(x) \\ \psi_R(x)
\end{array}
\right\rgroup
\]
We denote the rank-two chiral projectors with $ \mathbb{P}_{\pm}\equiv\frac12(1\pm\gamma_{5})\,, $
in such a manner that for any four-component bispinor $ \psi(x) $
\[
\psi_{-}(x)=\mathbb{P}_{-}\,\psi(x)=
\left\lgroup
\begin{array}{c}
\psi_L(x) \\ 0
\end{array}
\right\rgroup
\]
\[
\psi_{+}(x)=\mathbb{P}_{+}\,\psi(x)=
\left\lgroup
\begin{array}{c}
0 \\ \psi_{R}(x)
\end{array}
\right\rgroup
\]
and the chiral spin-states $ u_{\,\pm}(\mathbf{p}) $ \textit{et cetera} are defined in the very same way.

\medskip\noindent
\subsection{Weyl Spinor Fields}

Let us first analyze the massless and chiral Weyl spinor fields and their specific properties at the classical and quantum level.

Consider a massless Dirac bispinor, i.e., any complex and Grassmann valued bispinor in the Minkowski space, 
which is a solution of the differential equation 
$$ 
\partial\!\!\!/\,\psi(x)=0 \qquad\quad
\psi(x)= \left\lgroup
\begin{array}{c}
\psi_L(x) \\ \psi_R(x)
\end{array}
\right\rgroup
$$

If we want to derive the most general solution of the above equation, we have to abandon to a considerable extent the conventional approach and formalism familiar for the celebrated Dirac equation
$$
i\partial\!\!\!/\,\psi(x)=M\psi(x)
$$
We start by building as follows a very convenient set of spin-states in the chiral representation  for the gamma matrices.

Consider the matrix $ \alpha^{\,\nu}\equiv \gamma^{0}\gamma^{\,\nu}=(\,\mathbb{I}, \vec{\alpha}\,) $ with matrix elements
\[
\alpha^{\,\nu}=\left\lgroup
\begin{array}{cc}
\sigma^{\,\nu} & 0 
\\
0 & \tilde{\sigma}^{\,\nu} 
\end{array}\right\rgroup
\]
where we have set $ \sigma^{\,\nu}=(\mathbb{I}, -\,\sigma_{k}) $ with $ k=1,2,3\,, $ while
$ \tilde{\sigma}^{\,\nu}=(\mathbb{I}, \sigma_{k})\,. $ Then the massless Dirac equation can also
be cast in the form
\begin{equation}
\alpha\cdot\partial\,\psi(x)=(\,\partial_{0} + \alpha^{k} \partial_{k}\,)\,\psi(x)=0\qquad
\Longleftrightarrow\qquad
\left\lbrace 
\begin{array}{c}
\sigma\cdot\partial\,\psi_{L}(x)=0
\\
\bar{\sigma}\cdot\partial\,\psi_{R}(x)=0
\end{array}
\right. 
\label{Dirac_Weyl_eqs} 
\end{equation}
in such a manner that it can be rewritten \emph{\`{a} la} Schr\"{o}dinger 
\[
i\hslash\,\frac{\partial\psi}{\partial t}=H_0\,\psi
\]
where the massless Dirac Hamiltonian operator reads, in physical units, 
$$
H_0=c\,\alpha^{k}\,(-\,i\hslash\nabla_{k})=c\,\alpha^{k}\hat{p}^{\,k}
$$
Solving the massless Dirac equation is nothing but finding the general solutions of the pair of decoupled Weyl equations (\ref{Dirac_Weyl_eqs}).
{
Notice that the Weyl field equations do involve the pair of first order rank-two matrix differential operators acting upon
left- and right-handed Weyl spinors respectively: namely,
\begin{equation}
\sigma\cdot\partial=\sigma^{\mu}\partial_{\mu}=\partial_{0}-\sigma_{k}\partial_{k} \equiv \partial_{L}
\qquad\quad
\tilde\sigma\cdot\partial=\tilde\sigma^{\mu}\partial_{\mu}=\partial_{0}+\sigma_{k}\partial_{k} \equiv \partial_{R}
\label{Weyldiffop} 
\end{equation}
which satisfy 
\begin{eqnarray}
\partial_{L}\partial_{R}=\partial_{R}\partial_{L}=\square
\label{D'Alembert} 
\\
\partial_{L}^{2}=\square-2(\sigma_{k}\partial_{k}\partial_{0} - \vartriangle)
\qquad\quad
\partial_{R}^{2}=\square+2(\sigma_{\ell}\partial_{\ell}\partial_{0} + \vartriangle)
\end{eqnarray}
}

\subsubsection{Spin-States}
{ 
We can search the plane waves solutions of eq.~(\ref{Dirac_Weyl_eqs}) in the form
\begin{equation}
\psi(x)=\Gamma(p)\,\exp\lbrace -\,ip_{\mu} x^{\,\mu}\rbrace=\Gamma(p)\,e^{-\,ip\cdot x}
\qquad\quad p_{\mu}=\tilde{p}^{\,\mu}=(p_{0}, -\,\mathbf{p})
\end{equation}
where $ \Gamma(p) $ are the so called spin-states associated to the plane wave solutions of the massless Dirac-Weyl equation (\ref{Dirac_Weyl_eqs}).
Now, if we set $ H_{0}(\mathbf{p})\equiv\alpha^{k}p^{k}=\vec{\alpha}\cdot\vec{p}=H^{\dagger}_{0}(\mathbf{p})\,, $ from the commutation relation $ [\,\gamma_{5}, H_{0}\,]=0 $
we see that we can always select a basis in the 4d complex linear space of the bispinor spin-states according to their chirality 
$ \chi=\mp\, 1 $ and energy (or frequency) $ p_{0}=\pm\, \vert\,\mathbf{p}\,\vert $ pair of eigenvalues.

\medskip
Thus, in order to realize a basis,  we have to search among states of opposite chirality and frequency.
Then, in so doing,  we will be able to have at hand a complete and  orthogonal quartet of spin-states for the 4d massless bispinor space.
The construction goes as follows.
Explicit evaluation of the relevant spin-matrices yields for $ p_0 = \vert\,\vec{p}\,\vert = \wp $
\[
\sigma\cdot p =\sigma_{\mu}\,p^{\,\mu} = \wp\,\mathbb{I}+\sigma_{k}\, p^{\,k} = \wp + \vec{\sigma}\cdot\vec{p} =
\left\lgroup
\begin{array}{cc}
\wp + p_z & p_x - ip_y
\\
p_x + ip_y & \wp - p_z
\end{array}
\right\rgroup
\]
\[
\tilde\sigma\cdot p = {\sigma}_{\mu}\,\tilde p^{\,\mu}= \wp\,\mathbb{I} - \sigma_{k}\, p^{\,k} = \wp - \vec{\sigma}\cdot\vec{p} =
\left\lgroup
\begin{array}{cc}
\wp - p_z & -\,p_x + ip_y
\\
-\,p_x - ip_y & \wp + p_z
\end{array}
\right\rgroup
\]
\begin{eqnarray*}
\gamma^{\,\nu} p_{\nu} &\equiv& p\!\!/ =
\left\lgroup
\begin{array}{cc}
0 & \tilde{\sigma}\cdot p
\\
\sigma\cdot p & 0
\end{array}
\right\rgroup
\\
&=&
\left\lgroup
\begin{array}{cccc}
0 & 0 & \wp - p_z & -\,p_x + ip_y  
\\
0 & 0 & -\,p_x - ip_y & \wp + p_z 
\\
\wp + p_z & p_x - ip_y & 0 & 0 
\\
p_x + ip_y & \wp - p_z & 0 & 0 
\end{array}
\right\rgroup
\\ \\
\gamma^{\,\nu} \tilde{p}_\nu &\equiv & \tilde{p}\!\!/ =
\left\lgroup
\begin{array}{cc}
0 & {\sigma}\cdot p
\\
\tilde\sigma\cdot p & 0
\end{array}
\right\rgroup
\\
&=& \left\lgroup
\begin{array}{cccc}
0 & 0 & \wp + p_z & p_x - ip_y 
\\
0 & 0 & p_x + ip_y & \wp - p_z 
\\
\wp - p_z &  -\,p_x + ip_y & 0 & 0 
\\
-\,p_x - ip_y & \wp + p_z & 0 & 0 
\end{array}
\right\rgroup
\end{eqnarray*}
so that we find
\begin{eqnarray}
\sigma\cdot p\;\tilde{\sigma}\cdot p=\tilde{\sigma}\cdot p\;\sigma\cdot p=p^{2}=\wp^{2}-\vec{p}^{\,2}
= p\!\!/^{\,2} = \tilde{p}\!\!/^{\,2} = 0
\end{eqnarray}
\[
(\,\sigma\cdot p\,)^{2}=2\wp\,\sigma\cdot p
\qquad\quad
(\,\tilde\sigma\cdot p\,)^{2}=2\wp\,\tilde\sigma\cdot p
\]
Notice that $ \tilde{p}^{\,\nu}=(\,\wp,-\,\vec{p}\,)= p_{\nu}\, $ is the dual light-cone tetra-vector which satisfies
\[
\tilde{p}^{\,2}=p^{\,2}=0\qquad\quad p\cdot\tilde{p}=2\vec{p}^{\;2} = 2\wp^{2}
\]
Moreover it is useful to define the on-shell $ (\,p_{0}=\wp\,) $ Weyl-Dirac 2d projectors
\begin{eqnarray}
{\Pi}(\mathbf{p})\equiv \frac{ \tilde{p}\!\!\!/\,p\!\!\!/ }{4\wp^{2}}  
=\frac{1}{2\wp}
\left\lgroup
\begin{array}{cc}
\sigma\cdot p & 0
\\
0 & \tilde\sigma\cdot p
\end{array}
\right\rgroup
= \frac12\left( 1 - \frac{\vec{\alpha}\cdot\vec{p}}{\wp}\right) 
=\frac{\alpha\cdot p}{2\wp}
\\
\widetilde\Pi(\mathbf{p})\equiv \frac{p\!\!\!/\,\tilde{p}\!\!\!/}{4\wp^{2}} 
=\frac{1}{2\wp}
\left\lgroup
\begin{array}{cc}
\tilde\sigma\cdot p & 0
\\
0 & \sigma\cdot p
\end{array}
\right\rgroup
= \frac12\left( 1 + \frac{\vec{\alpha}\cdot\vec{p}}{\wp}\right) 
=\frac{\alpha\cdot \tilde p}{2\wp}
\end{eqnarray}
which fulfill by construction
\begin{eqnarray}
\Pi=\Pi^{\dagger}\qquad\quad \Pi^{\,2}=\Pi\qquad\quad \mathrm{Tr}\,\Pi=2
\\
\widetilde{\Pi}=\widetilde{\Pi}^{\dagger}\qquad\quad \widetilde{\Pi}^{\,2}=\widetilde\Pi\qquad\quad \mathrm{Tr}\,\widetilde{\Pi}=2
\end{eqnarray}
\begin{equation}
\Pi\,\widetilde{\Pi}=0=\widetilde{\Pi}\,\Pi\qquad\quad [\,\gamma_{5},\,\Pi\,]=[\,\widetilde{\Pi},\,\gamma_{5}\,]=0
\end{equation}
\begin{equation}
\gamma_{0}\,\Pi\,\gamma_{0}=\widetilde{\Pi}\qquad\quad \gamma_{0}\,\widetilde{\Pi}\,\gamma_{0}=\Pi
\end{equation}
It means that $ \Pi $ and $ \widetilde{\Pi} $ are projectors onto 2d orthogonal chirally invariant 
though parity exchanging spaces. The momentum space Weyl-Dirac equation (\ref{Dirac_Weyl_eqs}) for the positive energy spin-states 
evidently reads $ \Pi(\mathbf{p})\,\Gamma(\wp,\mathbf{p})=0\,, $ the general solution of which being
$ \Gamma(\wp,\mathbf{p})=\widetilde{\Pi}(\mathbf{p})\,\xi\,, $ where $ \xi $ is any arbitrary constant bispinor.
Hence, without loss of generality, we can set, for example, with $ p^{\,\nu}=(\wp,\mathbf{p}) = (\wp,\vec{p}\,)$
\begin{eqnarray}
u_{-}( \mathbf{p} ) &=&
\frac{1}{\sqrt{\wp - p_z}} \,\alpha^{\nu}\tilde{p}_{\nu} \,\xi
= \left\lgroup
\begin{array}{c}
u_L( \mathbf{p} ) \\  0 \\ 0
\end{array}
\right\rgroup
\nonumber \\
&=& \frac{1}{\sqrt{\wp - p_z}} \left\lgroup
\begin{array}{c}
\wp - p_z \\ -\,p_x - ip_y \\ 0 \\ 0
\end{array}
\right\rgroup
\\
\xi &=& \left\lgroup
\begin{array}{c}
1 \\ 0 \\ 0 \\ 0
\end{array}
\right\rgroup
\qquad\quad 
u_L( \mathbf{p} ) = \frac{\sigma\cdot\tilde p }{\sqrt{\wp - p_z}} \,
\left\vert
\begin{array}{c}
1 \\ 0
\end{array}
\right\vert  
\nonumber
\end{eqnarray}
The above spin-state is a positive frequency solution of equation (\ref{Dirac_Weyl_eqs}) and exhibits negative chirality
\begin{eqnarray}
&& H_{0}(\mathbf{p})\,u_{-}( \mathbf{p} ) = \wp\,u_{-}( \mathbf{p} )  \qquad\quad \gamma_5\, u_{-}( \mathbf{p} ) = -\,u_{-}( \mathbf{p} ) 
\\
&& \Pi(\mathbf{p})\,u_{-}( \mathbf{p} ) = 0\qquad\quad 
\widetilde{\Pi}(\mathbf{p})\,u_{-}( \mathbf{p} )=u_{-}( \mathbf{p} )
\end{eqnarray}
Moreover, the 2-component left-handed Weyl spinor $ u_L( \mathbf{p} )  $ does fulfill
\begin{equation}
u^{\dagger}_L( \mathbf{p} )\, u_L( \mathbf{p} ) = 2\wp
\qquad\quad
u_L( \mathbf{p} ) \otimes u^{\dagger}_L( \mathbf{p} ) = 2\wp\,\widetilde{P}_{L}
\qquad\quad(\,p_0=\wp\,)
\end{equation}
where the rank-two square matrix
\[
\widetilde{P}_{L}\equiv \frac{ \tilde{\sigma}\cdot p }{ 2\wp } =\frac{{\sigma}\cdot \tilde p }{ 2\wp } 
\]
does satisfy for $ p_0=\wp $
\[
\sigma\cdot p\;\widetilde{P}_{L}=0
\qquad\quad
\widetilde{P}_{L}^{\,\dagger}=\widetilde{P}_{L}
\qquad\quad
\widetilde{P}_{L}^{\,2}=\widetilde{P}_{L}
\qquad\quad
\mathrm{Tr}\,\widetilde{P}_{L}=1
\]
so that it corresponds to the \textsf{one dimensional projector} on positive energy and left-handed spin-states.
{ It follows therefrom that we have a pair of \textbf{one-dimensional projectors} upon positive energy left- and right-handed spin states
respectively: namely,
\begin{equation}
\widetilde{P}_{L}\equiv \frac{ \tilde{\sigma}\cdot p }{ 2\wp } 
\qquad\quad
P_{R}\equiv \frac{ \sigma\cdot p }{ 2\wp } 
\label{1dimprojectors} 
\end{equation}
}

\medskip
In order to set up a right-handed positive energy spin-state, we can involve the charge conjugated  of the above constructed spin-state: namely, 
\begin{equation}
u_{+}(\mathbf{p}) \equiv i\gamma^{\,2} u_{-}^{\ast}(\mathbf{p}) = u_{-}^{\,c}(\mathbf{p})
\label{u+definition} 
\end{equation}
As a matter of fact, it is easy to check that the chiral bispinor $ u_{+}(\mathbf{p}) $ satisfies the distinguished features
\begin{equation}
\gamma_{5}\,u_{+}(\mathbf{p}) =u_{+}(\mathbf{p}) 
\qquad\quad
H_{0}(\mathbf{p})\,u_{+}(\mathbf{p}) = \wp\,u_{+}(\mathbf{p}) 
\label{u+property} 
\end{equation}

\medskip
\begin{footnotesize}
\texttt{Proof}. We immediately find
\[
\gamma_{5}\,u_{+}(\mathbf{p}) = i\,\gamma_{5}\gamma^{\,2} u_{-}^{\ast}(\mathbf{p})
= -\,i\gamma^{\,2} [\,\gamma_{5} u_{-}(\mathbf{p})\,]^{\ast} =  i\,\gamma^{\,2} u_{-}^{\ast}(\mathbf{p}) = u_{+}(\mathbf{p})
\]
Furthermore we can write
\begin{eqnarray*}
H_{0}(\mathbf{p})\,u_{+}(\mathbf{p})=H_{0}(\mathbf{p})\, i\gamma^{\,2} u_{-}^{\ast}(\mathbf{p})
= \left[ \,H^{\ast}_{0}(\mathbf{p})\, i\gamma^{\,2} u_{-}(\mathbf{p})\,\right]^{\ast} 
\\
H^{\ast}_{0}(\mathbf{p})\,\gamma^{\,2} = ( \alpha^{1} p_{x} + \alpha^{3} p_{z} - \alpha^{2} p_{y} )\,\gamma^{\,2} = \gamma^{\,2} H_{0}(\mathbf{p})
\\
H_{0}(\mathbf{p})\,u_{+}(\mathbf{p}) = \left[ \, i\gamma^{\,2} \wp\,u_{-}(\mathbf{p})\,\right]^{\ast} =  i\gamma^{\,2} \wp\,u^{\ast}_{-}(\mathbf{p})
= \wp\,u_{+}(\mathbf{p})
\end{eqnarray*}
that proves the above statements (\ref{u+property}).
\end{footnotesize}

\medskip
On the other side, it is immediate to realize that a right-handed spin-state with negative energy is provided by
\begin{equation}
v_{+}(\mathbf{p}) \equiv \gamma_{0}\,u_{-}(\mathbf{p})
\label{v+definition} 
\end{equation}
which obviously fulfills
\begin{equation}
\gamma_{5}\,v_{+}(\mathbf{p}) =v_{+}(\mathbf{p}) 
\qquad\quad
H_{0}(\mathbf{p})\,v_{+}(\mathbf{p}) = -\,\wp\,v_{+}(\mathbf{p}) 
\label{v+property} 
\end{equation}
its charge conjugated being left-handed, i.e.,
\begin{equation}
v_{-}(\mathbf{p}) \equiv i\gamma^{\,2} v_{+}^{\ast}(\mathbf{p}) = v_{+}^{\,c}(\mathbf{p})
\label{v-definition} 
\end{equation}
and satisfying
\begin{equation}
\gamma_{5}\,v_{-}(\mathbf{p}) = -\,v_{-}(\mathbf{p}) 
\qquad\quad
H_{0}(\mathbf{p})\,v_{-}(\mathbf{p}) = -\,\wp\,v_{-}(\mathbf{p}) 
\label{v-property} 
\end{equation}
The ultimate meaning of the above construction is that, once we have fixed the initial left-handed spin-state of positive energy,
the rest of the spin-state orthogonal basis can be fully determined by the charge conjugation and parity discrete symmetry operations.
Notice that we can always write
\begin{equation}
u_{\mp}(-\,\mathbf{p})=v_{\mp}(\mathbf{p})
\qquad\quad
v_{\mp}(-\,\mathbf{p})=u_{\mp}(\mathbf{p})
\label{momentumchange} 
\end{equation}
because under the momentum inversion $ \mathbf{p}\,\longleftrightarrow\,-\,\mathbf{p} $ the chirality remains manifestly untouched while
we have $ H_{0}(-\,\mathbf{p})=-\,H_{0}(\mathbf{p}) $ and e.g.
$$ 
H_{0}(\mathbf{q})\,u_{\mp}(-\,\mathbf{q}) = H_{0}(-\,\mathbf{p})\,u_{\mp}(\mathbf{p}) 
= -\,H_{0}(\mathbf{p})\,u_{\mp}(\mathbf{p}) = -\,\wp\,u_{\mp}(\mathbf{p}) =  -\,\wp\,u_{\mp}(-\,\mathbf{q})
$$
in such a manner that the spin-states with opposite momentum share the very same chirality but change  the sign of energy,
which endorses (\ref{momentumchange}). As a consequence we can always write the identity
\begin{equation}
\gamma_{0}\,u_{\pm}(-\,\mathbf{p})=u_{\mp}(\mathbf{p})
\qquad\quad
\gamma_{0}\,v_{\mp}(-\,\mathbf{p})=v_{\pm}(\mathbf{p})
\end{equation}
\begin{equation}
\bar{u}_{\pm}(-\,\mathbf{p}) = u^{\dagger}_{\mp}(\mathbf{p})
\qquad\quad
\bar{v}_{\mp}(-\,\mathbf{p})=v^{\dagger}_{\pm}(\mathbf{p})
\end{equation}

\medskip
It turns out that the above defined quartet of massless and chiral bispinor spin-states does fulfill
orthogonality and closure relations. As a matter of fact, spin-states of opposite chirality are
always and necessarily orthogonal, \textit{viz.,}
\begin{eqnarray}
\left.
\begin{array}{c}
u_{\,\pm}^{\,\dagger}( \mathbf{p} )\,u_{\,\mp} ( \mathbf{q}\,)\equiv 0 \equiv
v_{\pm} ^{\,\dagger}( \mathbf{p} )\,v_{\mp} ( \mathbf{q}\,)
\\ \\
u_{\pm}^{\,\dagger}( \mathbf{p} )\,v_{\mp} ( \mathbf{q}\,) \equiv 0
\end{array}
\right\rbrace
\end{eqnarray}
 Moreover we have
\begin{eqnarray}
u_{\pm} ^{\,\dagger}( \mathbf{p} )\, u_{\pm} ( \mathbf{p} ) =
v_{\pm}^{\,\dagger}( \mathbf{p} )\,v_{\pm} ( \mathbf{p} ) = 2\wp
\end{eqnarray}
that exhausts the orthogonality relations. 

Concerning the closure or completeness relations, on the one hand we clearly get
\begin{eqnarray}
(2\wp)^{-1} [\,u_{\pm} ( \mathbf{p} ) \otimes u_{\pm}^{\,\dagger} ( \mathbf{p} ) + 
v_{\pm}(\mathbf{p} ) \otimes v_{\pm}^{\,\dagger}(\mathbf{p} ) \,] = \mathbb{P}_{\pm}
\label{chiralityclosure} 
\end{eqnarray}
which represent the \textbf{chirality closure or completeness relations} for the massless and chiral bispinor spin-states basis quartet.
On the other hand, in order to set up  the \textbf{energy or frequency closure or completeness relations} we notice that
\begin{equation}
\bar{u}_{\pm}(\mathbf{p})\,u_{\pm}(\mathbf{q})\equiv 0 \equiv
\bar{v}_{\pm}(\mathbf{p})\,v_{\pm}(\mathbf{q})
\qquad\quad
\bar{u}_{\pm}(\mathbf{p})\,v_{\pm}(\mathbf{q})\equiv 0
\end{equation}
\begin{equation}
\bar{u}_{\pm}(\mathbf{p})\,u_{\mp}(-\,\mathbf{p})=u_{\pm}^{\dagger}(\mathbf{p})\,\gamma_{0}\,u_{\mp}(-\,\mathbf{p})
=u_{\pm}^{\dagger}(\mathbf{p})\,\,u_{\pm}(\mathbf{p}) =2\wp
\end{equation}
\begin{equation}
\bar{v}_{\pm}(\mathbf{p})\,v_{\mp}(-\,\mathbf{p})=v_{\pm}^{\dagger}(\mathbf{p})\,\gamma_{0}\,v_{\mp}(-\,\mathbf{p})
=v_{\pm}^{\dagger}(\mathbf{p})\,\,v_{\pm}(\mathbf{p}) =2\wp
\end{equation}
\begin{equation}
\bar{u}_{\pm}(\mathbf{p})\,v_{\mp}(-\,\mathbf{p})=u_{\pm}^{\dagger}(\mathbf{p})\,\gamma_{0}\,v_{\mp}(-\,\mathbf{p})
=u_{\pm}^{\dagger}(\mathbf{p})\,\,v_{\pm}(\mathbf{p}) = 0
\end{equation}
It follows that we can firmly set
\begin{eqnarray}
\EE_{>}(\mathbf{p}) &=& (2\wp)^{-1} [\,u_{-}(\mathbf{p})\otimes u_{-}^{\dagger}(\mathbf{p}) + u_{+}(\mathbf{p})\otimes u_{+}^{\dagger}(\mathbf{p})\,]
\\
&=& (2\wp)^{-1} [\,u_{-}(\mathbf{p})\otimes \bar u_{+}(-\,\mathbf{p}) + u_{+}(\mathbf{p})\otimes \bar u_{-}(-\,\mathbf{p})\,]
= \widetilde\Pi(\mathbf{p})
\label{E>closure} 
\\ \\
\EE_{<}(\mathbf{p}) &=& (2\wp)^{-1} [\,v_{-}(\mathbf{p})\otimes v_{-}^{\dagger}(\mathbf{p}) + v_{+}(\mathbf{p})\otimes v_{+}^{\dagger}(\mathbf{p})\,]
\\
&=& (2\wp)^{-1} [\,v_{-}(\mathbf{p})\otimes \bar v_{+}(-\,\mathbf{p}) + v_{+}(\mathbf{p})\otimes \bar v_{-}(-\,\mathbf{p})\,]
= \Pi(\mathbf{p})
\label{E<closure} 
\end{eqnarray}

\bigskip

The corresponding plane-wave functions, which are normal-mode solutions of the massless Dirac equation, read
\begin{eqnarray}
\left\lbrace 
\begin{array}{c}
u_{\pm ,\,\mathbf{p}}(x)=[\,(2\pi)^{3} 2\wp\,]^{-\frac12} u_{\pm} ( \mathbf{p} ) \,e^{\,-\,i\wp t + i\mathbf{p}\cdot \mathbf{x}}
\\ \\
v_{\pm,\,\mathbf{p}}(x)=[\,(2\pi)^{3} 2\wp\,]^{-\frac12}\, v_{\pm}(-\,\mathbf{p} ) \,
e^{\,i\wp t - i\mathbf{p}\cdot \mathbf{x}}
\label{Weylspinorwavefunctions} 
\end{array}
\right. 
\end{eqnarray}
and fulfill in turn orthonormality and closure relation with respect to the usual Poincar\'{e} invariant inner product:
for instance
\begin{eqnarray*}
\int \mathrm{d}\mathbf{x}\,u^{\,\dagger}_{\imath,\,\mathbf{p}}(x)\,
v_{\jmath,\,\mathbf{p}  ^{\prime} }(x)=0
\\
\int \mathrm{d}\mathbf{x}\,u^{\,\dagger}_{\imath,\,\mathbf{p}}(x)\,
u_{\jmath,\,\mathbf{p}  ^{\prime} }(x) = \delta_{\,\imath\jmath}\,\delta( \mathbf{p} - \mathbf{p}^{\prime} )
=\int \mathrm{d}\mathbf{x}\,v^{\,\dagger}_{\imath,\,\mathbf{p}}(x)\,
v_{\jmath,\,\mathbf{p}  ^{\prime} }(x)
\\
(\,\forall\,\imath,\jmath=+,-\,\vee\,\mathbf{p},\mathbf{p}^{\,\prime}\in \mathbb{R}^{3}\,)
\end{eqnarray*}
\textit{et cetera}. 

We recall that, according to the general principles of quantum field theory, the plane wave functions $ v_{\pm,\,\mathbf{p}}(x) $
of negative energy or frequency are associated to the anti-particles of the corresponding spin $ \frac12 $ charged quantum field.

\medskip
In the Standard Model of the fundamental interactions the neutrino and anti-neutrino particles are supposed 
to be massless\footnote{We recall that the neutrino and/or anti-neutrino  mass has not yet been directly measured: nonetheless, the
present experimental limit is severely stringent $ < $ 0.45 eV, i.e. a million times lighter than the lightest known particles.} 
and charge conjugated one to each other, the charge being identified
with the so called lepton number. Moreover, the \emph{statu quo} of the present phenomenological frame \cite{Goldhaber} does require
that neutrino particles are always left-handed and with a negative helicity, while anti-neutrinos always appear to right-handed and with  a positive helicity.
\emph{Sic stantibus rebus} it turns out to be natural to
associate $ u_{-} ( \mathbf{p} )\,,u_{-,\,\mathbf{p}}(x) $ to the left-handed spin-state and normal mode, respectively,
of the \textsf{Standard Model neutrino Weyl field},  while 
$ v_{+} (\mathbf{p} )\,,v_{+,\,\mathbf{p}}(x) $ will refer to the corresponding negative frequency
spin-state and normal modes, which will be related to the anti-neutrino particles of opposite chirality and helicity, as far as their masses can
eventually be neglected. \textsf{It is worthwhile to stress once more that this mandatory strong correlation between chirality and helicity does actually
occur only for the Weyl spinor field and never for any other kind of relativistic spin $ \frac12 $ wave fields.}

The remaining pair of chiral spin-states and normal-modes $ u_{+}, v_{-} $ would correspond
to hypothetical right-handed neutrinos and related left-handed anti-neutrino spin-states and normal modes, 
\textbf{that have never been observed so far},
although their existence does  represent, if any, one of the most intriguing and exciting actual possibilities to go beyond the Standard Model.}

\subsubsection{Symmetries}

The classical Lagrange density for the massless Dirac field can be written in different forms: namely,
\begin{eqnarray}
\mathcal{L}_{0} &=& \psi^{\dagger}(x)\,i\alpha\cdot\partial\,\psi(x)
\doteq\textstyle\frac12\,\psi^{\dagger}(x)\,i\alpha\cdot\!\buildrel \leftrightarrow\over \partial\!\psi(x)
\label{massless_Dirac_lagrangian} \\
&\doteq& \textstyle\frac12\,\psi_{L}^{\dagger}(x)\,i\sigma\cdot\!\buildrel \leftrightarrow\over \partial\!\psi_{L}(x)
+ \textstyle\frac12\,\psi_{R}^{\dagger}(x)\,i\bar\sigma\cdot\!\buildrel \leftrightarrow\over \partial\!\psi_{R}(x)
\label{Weyl_lagrangian} 
\end{eqnarray}
up to irrelevant tetra-divergences, where the spinor fields are understood as Grassmann valued functions on the Minkowski space,
whereas
\[
\psi(x)=\left\lgroup
\begin{array}{c}
\psi_L(x) \\ \psi_R(x)
\end{array}
\right\rgroup
\qquad\quad
\alpha^{\nu}=\left\lgroup
\begin{array}{cc}
\sigma^{\,\nu} & 0
\\
0 & \bar{\sigma}^{\,\nu}
\end{array}
\right\rgroup
\]
The Euler-Lagrange field equations are
\[
\alpha\cdot\partial\,\psi(x)=0\quad\Leftrightarrow\quad\sigma\cdot\partial\,\psi_L(x)=\bar{\sigma}\cdot\partial\,\psi_R(x)=0
\]
the general plane-wave solutions of which have been obtained and discussed in the previous Section.
The canonical energy-momentum tensor reads
\begin{eqnarray}
T_{\mu\nu} =\psi^{\dagger}(x)\,\alpha_{\mu}\,i\partial_{\nu}\psi(x)
=\psi^{\dagger}_L(x)\,\sigma_{\mu}\,i\partial_{\nu}\psi_L (x)
+ \psi^{\dagger}_R (x)\,\bar{\sigma}_\mu\,i\partial_{\nu}\psi_R (x)
\end{eqnarray}
which is not symmetric and drives to the conserved energy-momentum vector
\begin{eqnarray}
P_0 &=& \int\mathrm{d}\mathbf{x}\,
\psi^{\dagger}(x)\,i\partial_{0}\psi(x)
=\int\mathrm{d}\mathbf{x}\,
\psi^{\dagger}(x) H_{0} \psi(x)
\\
&=&\int\mathrm{d}\mathbf{x}\,\left[ \,
\psi_L^{\dagger}(x) H_{L} \psi_L(x) +
\psi_R^{\dagger}(x) H_{R} \psi_R(x) \,\right] 
\\
\mathbf{P} &=& \int\mathrm{d}\mathbf{x}\,
\psi^{\dagger}(x)(-\,i\hslash\nabla)\psi(x)
\end{eqnarray}
Notice that the energy-momentum tensor is trace-less
$ g^{\,\mu\nu} T_{\mu\nu}=0\,, $ owing to the validity of the field equations.
The canonical total angular momentum density tensor can always be written in the form
\begin{eqnarray}
M^{\,\lambda\mu\nu}=x^{\,\mu} T^{\lambda\nu} - x^{\,\nu} T^{\lambda\mu} + S^{\,\lambda\mu\nu}
\\
S^{\,\lambda\mu\nu} =
\textstyle\frac12\,\psi^{\dagger}(x)\,\gamma^{0}\lbrace\gamma^{\lambda},\,\sigma^{\,\mu\nu}\rbrace\, \psi(x)
\end{eqnarray}
so that from Noether theorem related to the Lorentz symmetry we get
\[
T_{\nu\mu} - T_{\mu\nu}=\partial^{\,\lambda}S_{\lambda\mu\nu}
\]
which notoriously entails the helicity conservation in any direction along which the energy-momentum tensor appears to be symmetric. 
In order to determine the helicity assignments we can proceed as follows.
Since the generators of the rotation group for both the
irreducible Weyl spinor representations
are $S_{\,jk}=\frac12\,\varepsilon_{\,jk\ell}\,\sigma_\ell$
it follows that the corresponding Pauli-Lubanski  operator reads
\[
W_{0}=\mathbf P\cdot\mathbf S =
\vec{p}\cdot\textstyle\frac12\,\hslash\,\vec{\sigma}
\]
\[
W_{j}=\textstyle\frac12\,\wp\sigma_j-\textstyle\frac12\,i\varepsilon_{jk\ell}\,p_k\,\sigma_\ell
\]
which entails
\begin{eqnarray*}
g^{\,\mu\nu}\,W_\mu W_\nu=\textstyle\frac14\,p_j\,p_k\,\sigma_j\sigma_k
-\textstyle\frac14\left(\,\wp\sigma_j-i\varepsilon_{jk\ell}\,p_k\,\sigma_\ell\right)
\left(\,\wp\sigma_j-i\varepsilon_{jrs}\,p_r\,\sigma_s\right)
\\
=\,-\,\textstyle\frac12\,\wp^2 +
\frac14\left(\,\delta_{\,kr}\delta_{\,\ell s}-\delta_{\,ks}\delta_{\,\ell r}\right)p_k\,p_r\,\sigma_\ell\,\sigma_s
=0
\end{eqnarray*}
where the $2\times2$ identity matrix has been understood. 
Then, thanks to the light-like nature
of the Pauli-Lubanski operator, we can identify the \textsf{helicity operator} to be
\[
\mathrm{h}\equiv\frac{W_0}{\wp}=\frac{|{\bf W}|}{\wp}=
{\textstyle\frac12}\,\hslash\,\hat{n}\cdot\vec{\sigma}
\qquad\quad \hat{n}=\vec{p}/\wp
\]
together with the Weyl Hamiltonian operators
\[
H_0=\left\lgroup 
\begin{array}{cc}
H_L & 0 \\ 0 & H_R
\end{array}
\right\rgroup
\qquad\quad
H_R = c\,\hslash\, \vec{p}\cdot\vec{\sigma}= -\,H_L
\]
in such a manner that we find $H_L = -\,\hslash c \wp\mathrm{h}\,,\,H_R=\hslash c\wp\mathrm h$ 
and consequently
\begin{eqnarray*}
&& H_L\,u_L({\bf p})=\hslash c\wp\,u_L({\bf p})\qquad\quad
\mathrm{h} u_L({\bf p})= -\,\textstyle\frac12\,u_L({\bf p})
\\
&& H_L\,v_L({\bf p})=-\,\hslash c\wp\,v_L({\bf p})\qquad\quad
\mathrm{h} v_L({\bf p})=\textstyle\frac12\,v_L({\bf p})
\\ \\
&& H_R \,u_R({\bf p})=\hslash c\wp\,u_R({\bf p})\qquad\quad
\mathrm{h} u_R({\bf p}) = \textstyle\frac12\,u_R({\bf p})
\\
&& H_R\,v_R({\bf p})= -\,\hslash c\wp\,v_R({\bf p})\qquad\quad
\mathrm{h} v_R({\bf p})=-\,\textstyle\frac12\,v_R({\bf p})
\end{eqnarray*}
which means that  :

{
\begin{itemize}
\item 
neutrino spin-states carry negative helicity, while the
anti-neutrino spin-states carry positive helicity, 
i.e. $ \gamma_{5}u_{-}(\mathbf{p})=-\,\gamma_{5} v_{+}(\mathbf{p})= -\,1\,.$
\item The situation is reversed for the remaining pair of the spin-states. Thus, 
positive helicity is exhibited by positive frequency particles - the would-be right neutrinos - while negative helicity by negative frequency anti-particles,
i.e. $ \gamma_{5} u_{+}(\mathbf{p})=-\,\gamma_{5} v_{-}(\mathbf{p})=+\,1\,.$
\end{itemize}
}

The Weyl Action integrals 
\[
S_L=\int \mathrm{d}^4x\,\textstyle\frac12\,\psi_{L}^{\dagger}(x)\,i\sigma\cdot\!\buildrel \leftrightarrow\over \partial\!\psi_{L}(x)
\qquad\quad
S_R=\int \mathrm{d}^4x\,\textstyle\frac12\,\psi_{R}^{\dagger}(x)\,i\bar\sigma\cdot\!\buildrel \leftrightarrow\over \partial\!\psi_{R}(x)
\]
and the massless Dirac Action integral, 
which is the sum of the left- and right-handed Weyl Action integrals, have a common property, i.e. they
are both invariant under scale or dilatation symmetry.
To be more specific a scale transformation acts upon any inertial coordinate system and any spinor field in a Minkowski spacetime as follows
\begin{eqnarray}
x^{\lambda}\qquad\longmapsto\qquad x^{\,\prime\,\lambda}=e^{-\,\varrho} x^{\lambda}
\qquad\quad(\,\varrho\in\mathbb{R}\,)
\\
\Psi(x)\qquad\longmapsto\qquad \Psi^{\,\prime}(x)=e^{\,\frac32\varrho}\, \Psi(e^{\,\varrho}x)
\qquad\quad(\,\Psi=\psi,\psi_L,\psi_R\,)
\end{eqnarray}
In order to obtain the infinitesimal operators of the scale transformations we consider an infinitesimal dilatation/contraction, \textit{viz.,}
\[
 x^{\,\prime\,\mu}=( 1 - \delta\varrho + \cdots )\, x^{\,\mu}
\qquad\quad(\,\vert\,\delta\varrho\,\vert \ll 1\,)
\]
\begin{eqnarray*}
\Psi^{\,\prime}(x)=\left( 1+\textstyle\frac32\delta\varrho + \cdots \right) \Psi(x + x \delta\varrho + \cdots )
\\
= \left( 1+\textstyle\frac32\delta\varrho \right) \left[ \,\Psi(x) + x\cdot\partial\,\Psi(x)\,\delta\varrho\,\right] + \cdots
\\
= \Psi(x) + \textstyle\frac32\delta\varrho \,\Psi(x) + x\cdot\partial\,\Psi(x)\,\delta\varrho + \cdots
\end{eqnarray*}
and thereby
\[
\delta\Psi(x)=\delta\varrho\,\left[ \,\textstyle\frac32 + x\cdot\partial\,\right] \,\Psi(x)
=\textstyle\frac32\,\delta\varrho\,\Psi(x) - \mathrm{d}\Psi(x)
\]
because $ \delta x= -\,x\delta\varrho\,. $
Then we can write the infinitesimal total variation of the spinor fields in the form
\begin{equation}
\Delta\Psi(x)=(\,\delta + \mathrm{d}\,)\Psi(x)=\textstyle\frac32\,\delta\varrho\,\Psi(x) 
\end{equation}
in accordance with the well-known general rule that connects local ($ \delta $ ) and total ($ \Delta $)
variations of any classical relativistic wave field. Hence, we are allowed to identify 
\begin{equation}
\mathrm{D}_{\Psi}\equiv \textstyle\frac32 + x\cdot\partial
\end{equation}
with the generator of the local scale transformations on the classical  spinor fields.
Now, since $ \Psi^{\,\prime}(x^{\,\prime})=e^{\,\frac32\varrho}\, \Psi(x)\,, $ it follows that the Lagrangian
(\ref{massless_Dirac_lagrangian}) does satisfy 
\[
\mathcal{L}_{0}^{\,\prime}(x^{\,\prime}) = e^{\,4\varrho} \mathcal{L}_{0}(x)
\]

It follows that the Action integral for the massless Dirac spinor field is invariant, \textit{viz.,}
\begin{equation}
S=\int\mathrm{d}^{4}x^{\,\prime}\,\mathcal{L}_{0}^{\,\prime}(x^{\,\prime}) 
= \int\mathrm{d}^{4}x\,\mathcal{L}_{0}(x)
\end{equation}
The four-current for dilatation is derived via Noether theorem and can be written in various
equivalent forms: namely,
\begin{eqnarray}
\mathcal{D}^{\,\mu}(x) &=& \frac{\delta \mathcal{L}_0 }{ \delta\partial_{\mu} \psi(x) }\,\mathrm{D}_{\psi} \,\psi(x) - \mathcal{L}_0 (x) \,x^{\,\mu}
\nonumber\\
&=& \frac{\delta \mathcal{L}_0  }{ \delta\partial_{\mu} \psi(x) }\,\textstyle\frac32\,\psi(x) + x_{\nu} \,T^{\mu\nu}(x)
\nonumber\\
&=& \textstyle\frac32\,\psi^{\dagger}(x)i\alpha^{\,\mu}\psi(x) + x_{\nu} \,T^{\mu\nu}(x)
\end{eqnarray}
so that
\begin{equation}
\partial\cdot\mathcal{D}=g^{\,\mu\nu} T_{\mu\nu}=0
\end{equation}

Finally, the massless Dirac Action integral is invariant as well under the
U(1) groups of the ordinary and chiral phase changes,\textit{viz.,}

\begin{eqnarray}
\psi(x)\qquad\longmapsto\qquad
\left\lbrace 
\begin{array}{c}
\psi^{\,\prime}(x)=e^{-\,i\theta} \psi(x)
\\ \\
\psi^{\,\prime}(x)=e^{-\,i\gamma_{5}\theta} \psi(x)
\end{array}
\right. 
\qquad\quad(\,0\le\theta<2\pi\,)
\end{eqnarray}
or equivalently
\begin{eqnarray}
\psi_L(x)\qquad\longmapsto\qquad
\left\lbrace 
\begin{array}{c}
\psi_L^{\,\prime}(x)=e^{-\,i\theta} \psi_L(x)
\\ \\
\psi_L^{\,\prime}(x)=e^{\,i\theta} \psi_L(x)
\end{array}
\right. 
\qquad\quad(\,0\le\theta<2\pi\,)
\\ \nonumber\\
\psi_R(x)\qquad\longmapsto\qquad\psi_R^{\,\prime}(x)=e^{-\,i\theta} \psi_R(x)
\qquad\quad(\,0\le\theta<2\pi\,)
\end{eqnarray}
in such a manner that Noether theorem drives to the internal currents
\begin{equation}
J^{\mu}(x)=\psi^{\dagger}(x)\alpha^{\,\mu}\psi(x)\qquad\quad
J_{5}^{\mu}(x)=\psi^{\dagger}(x)\alpha^{\,\mu}\gamma_{5}\psi(x)
\end{equation}
which satisfy the continuity equation so that the internal charges
\begin{eqnarray}
\mathcal{Q}=\int\mathrm{d}\mathbf{x}\,\left[ \,
\psi_L^{\dagger}(x)\,\psi_L(x) + \psi_R^{\dagger}(x)\,\psi_R(x) \,\right] 
\\
\mathcal{Q}_{5}=\int\mathrm{d}\mathbf{x}\,\left[ \,
-\,\psi_L^{\dagger}(x)\,\psi_L(x) + \psi_R^{\dagger}(x)\,\psi_R(x) \,\right] 
\end{eqnarray}
are conserved in time $ \dot{\mathcal{Q}}=\dot{\mathcal{Q}}_5=0\,. $
To sum up, so far for a massless Dirac field we have seen that it is characterized by invariance under  
a 13-parameter Lie group of space-time and internal symmetries, which implies thirteen conserved charges. But it can be shown, that scale invariance may imply invariance under a larger group with 15-parameter, the \textsf{conformal group}, including in particular the ten dimensional Poincar\'e group. More precisely scale invariance implies conformal invariance if and only if there exists in the theory a \textsf{symmetric traceless conserved energy-momentum tensor} $ \Theta_{\mu\nu}\,, $
(the improved energy-momentum tensor) such that $ \partial\cdot\mathcal{D} = g^{\,\mu\nu} \Theta_{\mu\nu} =0 \,.$ It can be verified that this is the case for all power-counting renormalizable massless models involving fields of spin $ \le 1\,. $ It follows that  the massless Dirac Action integral is symmetric under a 17-parameter Lie group involving a space-time 15-parameter group plus two $U(1)$ symmetries.

As for discrete symmetries, the Action integral $ S=\int\mathrm{d}^{4}x\mathcal{L}_0(x) $
is invariant under $ \mathcal{C}, \mathcal{P} $ and $ \mathcal{T} $ transformations separately.
Charge conjugation $ \mathcal{C} $ acts on the classical Dirac spinors as follows
\[
\psi(x)\qquad\longmapsto\qquad \psi^{\,c}(x)=\gamma^{\,2}\psi^{\,\ast}(x)
\]
while parity $ \mathcal{P} $ transforms yields $ \psi^{\,\prime}(t,-\,\mathbf{x})=\gamma^0\psi(x)\,, $
whereas time reversal $ \mathcal{T} $ is provided  by $ \psi^{\,\prime}(-\,t,\mathbf{x})=\Theta\,\psi(x)\,,\ \Theta=\gamma^{\,3}\gamma^{1}\,. $
It is straightforward to verify by direct inspection that the Action $S$ is invariant. Notice that the above mentioned discrete transforms are
not even well defined for chiral spinors, for they mix the chiral components and necessarily drive out of the specific chiral spinor spaces
of the two irreducible nonequivalent Weyl representations of the Lorentz group $ \mathrm{D}(\frac12,0) $ and $ \mathrm{D}(0,\frac12)\,. $

\subsubsection{Quantum Theory}

{ The setting up of a canonical quantum theory for the massless Dirac-Weyl spinor field is far from being straightforward and trivial.
First of all it is not at all unique: there are several options available in the market, the best buy to be selected and picked up through the attainment
to the general principles of quantum field theory as well as by the correspondence with the actual phenomenology.
In order to be appropriate and applicable to the concrete high energy and particle Physics context, the operation of parity and charge conjugation
must be actually well defined and truly  implementable: this basic requirement clearly rules out \emph{a priori} the possibility to formulate the quantum theory
in terms of the 2-component Weyl spinor field $ \psi_L $ and $ \psi_R\,, $ as the afore mentioned operations are not supported by the 2d Weyl spinors.
Hence, the only possible consistent and physically meaningful formulation is in terms of the chiral bispinors $ \psi_{\mp}\,. $
}

The massless Dirac quantum field is the  operator valued tempered distribution
\begin{eqnarray}
\psi(x)=\psi_{-}(x) + \psi_{+}(x) 
\label{masslessDiracfield} 
\\
\psi^{(-)}(x)=\int\mathrm{d}\mathbf{p}\,[\,(2\pi)^{3}\,2\wp\,]^{-\frac12}
\left[ \,c_{-,\,\mathbf{p}}\, u_{-}(\mathbf{p}) + c_{+,\,\mathbf{p}}\, u_{+}(\mathbf{p}) \,\right] \,e^{\,i\mathbf{p}\cdot\mathbf{x} - i\wp t    }
\\
\psi^{(+)}(x)=\int\mathrm{d}\mathbf{p}\,[\,(2\pi)^{3}\,2\wp\,]^{-\frac12}
\left[ \,d^{\,\dagger}_{-,\,\mathbf{p}}\, v_{-}( \mathbf{p}) 
+ d^{\,\dagger}_{+,\,\mathbf{p}}\, v_{+}( \mathbf{p}) \,\right] \,e^{\,i\wp t - i\mathbf{p}\cdot\mathbf{x}    }
\\
\psi_{\mp}(x) = \mathbb{P}_{\mp}\,\psi(x) =
\int\mathrm{d}\mathbf{p}\,\left[ \, c_{\mp,\,\mathbf{p}}\, u_{\mp,\,\mathbf{p}}(x)
+ d^{\,\dagger}_{\mp,\,\mathbf{p}}\, v_{\mp,\,\mathbf{p}}(x) \,\right]
\label{WeylQuantumFields} 
\\
\psi^{\,\dagger}_{\mp}(y) =   \psi^{\,\dagger} (y)\, \mathbb{P}_{\mp}   =
\int\mathrm{d}\mathbf{p}^\prime\,\left[ \,c^{\,\dagger}_{\mp,\,\mathbf{p}^\prime }\, u^{\top\,\ast}_{\mp,\,\mathbf{p}^\prime } (y)
+ d_{\mp,\,\mathbf{p}^\prime }\,v^{\top\,\ast}_{\mp,\,\mathbf{p}^\prime }(y) \,\right] 
\end{eqnarray}
which corresponds to the most general solution of the Weyl-Dirac equations (\ref{Dirac_Weyl_eqs}),
in which the creation and destruction operators satisfy the canonical anti-commutation relations
\begin{equation}
\lbrace c_{\mp,\,\mathbf{p}}\,,c^{\,\dagger}_{\mp,\,\mathbf{p}^{\prime} } \rbrace
=\lbrace d_{\mp,\,\mathbf{p}}\,,d^{\,\dagger}_{\mp,\,\mathbf{p}^{\prime} } \rbrace
= \delta( \mathbf{p} - \mathbf{p}^{\prime} )
\end{equation}
all the other anti-commutators being null. It follows that, for example,
\begin{eqnarray*}
&& \lbrace \psi_{-}(x)\,,\psi_{-}^{\,\dagger}(y) \rbrace_{x_0=y_0}
= \int\mathrm{d}\mathbf{p}\int\mathrm{d}\mathbf{p}^\prime
\\
&\times& \!\!\left[  \lbrace c_{-,\,\mathbf{p}}\,,c^{\,\dagger}_{-,\,\mathbf{p}^\prime} \rbrace\,
u_{-,\,\mathbf{p}}(x)\,u^{\,\dagger}_{-,\,\mathbf{p}^\prime }(y) 
+ \lbrace d^{\,\dagger}_{-,\,\mathbf{p}}\,, d_{-,\,\mathbf{p}^\prime } \rbrace\,
v_{-,\,\mathbf{p}}(x)\,v^{\,\dagger}_{-,\,\mathbf{p}^\prime }(y) \right] _{x_0=y_0}
\\
&=& \int\mathrm{d}\mathbf{p}\,
\left[ \,u_{-,\,\mathbf{p}}(x)\,u^{\,\dagger}_{-,\,\mathbf{p} }(y) 
+ v_{-,\,\mathbf{p}}(x)\,v^{\,\dagger}_{-,\,\mathbf{p} }(y) \,\right] _{x_0=y_0}
\\
&=& \int
\frac{  \mathrm{d}\mathbf{p}  }{  (2\pi)^{3} 2\wp  } \,\left[ \,
u_{-} ( \mathbf{p} ) \, u_{-}^{\,\dagger} ( \mathbf{p} ) 
\,e^{\, i\mathbf{p}\cdot ( \mathbf{x} - \mathbf{y} ) }
+ v_{-} ( -\,\mathbf{p} ) \, v_{-}^{\,\dagger} ( -\,\mathbf{p} ) 
\,e^{\,-\, i\mathbf{p}\cdot ( \mathbf{x} - \mathbf{y} ) } \,\right] 
\\
&=& \int
\frac{  \mathrm{d}\mathbf{p}  }{  (2\pi)^{3} 2\wp  } \,\left[ \,
u_{-} ( \mathbf{p} ) \, u_{-}^{\,\dagger} ( \mathbf{p} ) 
+ v_{-} ( \mathbf{p} ) \, v_{-}^{\,\dagger} ( \mathbf{p} ) \,\right] 
\,e^{\, i\mathbf{p}\cdot ( \mathbf{x} - \mathbf{y} ) }
\\
&=& \int
\frac{  \mathrm{d}\mathbf{p}  }{  (2\pi)^{3} 2\wp  } \,2\wp\,\mathbb{P}_-
\,e^{\, i\mathbf{p}\cdot ( \mathbf{x} - \mathbf{y} ) }
= \mathbb{P}_-\,\delta( \mathbf{x} - \mathbf{y} )
\end{eqnarray*}
and quite analogously
\[
\lbrace \psi_{+}(x)\,,\psi_{+}^{\,\dagger}(y) \rbrace_{x_0=y_0}=\mathbb{P}_+\,\delta( \mathbf{x} - \mathbf{y} )
\]
while all the remaining anti-commutators do vanish at arbitrary times, e.g.
$ \lbrace \psi_{-}(x)\,,\psi_{+}^{\,\dagger}(y) \rbrace=0 $ \textit{et cetera.} 
{ It follows that the canonical equal time anticommutation relations for the \textsf{massless Dirac quantum field} (\ref{masslessDiracfield})
will be provided by
\[
\lbrace \psi(x)\,,\psi^{\,\dagger}(y) \rbrace_{x_0=y_0}\,=\,\delta( \mathbf{x} - \mathbf{y} )
\qquad\quad 
\lbrace \psi(x)\,,\overline{\psi}(y) \rbrace_{x_0=y_0}\,=\,\gamma_{0}\,\delta( \mathbf{x} - \mathbf{y} )
\]
while the general canonical anticommutation relations take the standard usual form
\begin{eqnarray}
\lbrace \psi(x)\,,\psi(y) \rbrace =0=\lbrace \overline{\psi}(x)\,,\overline{\psi}(y)\rbrace
\end{eqnarray}
\begin{equation}
\lbrace \psi(x)\,,\overline{\psi}(y) \rbrace\equiv S_{0}(x-y)=
\gamma^{\,\rho} \partial D_{0}(x-y)/\partial x^{\,\rho}
\end{equation}
where $ D_{0}(x-y) $ is the well known massless Pauli-Jordan distribution, which fulfill the D'Alembert wave equation and endorses the
microcausality property
\begin{equation}
\gamma_{0} S_{0}(x-y)=\partial_{0} D_{0}(x-y)\qquad\quad ( x-y )^{2}<0
\end{equation}

Now the main point. A consistent and reliable canonical quantum theory for a free non interacting field must give rise to the causal Green's function,
or Feynman propagator, in the absence of which the perturbative expansion and the renormalization procedure cannot be implemented,  so that any
meaningful physical interpretation and predictions are prevented. The causal Green's functions is the time-ordered product of field operators and 
provides the specific inversion of the kinetic operator which guarantees causality: namely, first a particle of the quantum field is created out of the
vacuum and only later it is destroyed. Moreover, the causal Green's function always admits the smooth  transition to the Euclidean formulation,
which is essential in order to evaluate the Feynman integrals in perturbation theory, in particular the divergent and regularized ones.
As we shall see below, it turns out that only the canonical quantum theory of the massless Dirac spinor field can provide this featuring \emph{bonus}:
the Weyl chiral bispinor quantum fields (\ref{WeylQuantumFields}) or any other combination of those ones but (\ref{masslessDiracfield}) can do the job.
Let's analyze this crucial issue in more detail.}

\subsubsection{Causal Green's functions}

With the vacuum expectation value of the time ordered product of  a pair of e.g. right-handed quantum Weyl fields  let us form the rank-two square matrix
\begin{equation}
S_{c}(x-y)\equiv \langle 0\vert T\,\psi_R(x)\,\psi_R^{\dagger}(y)\vert0\rangle
\end{equation}
Inserting the normal-mode expansion one gets
\begin{footnotesize}
\begin{eqnarray*}
S_{c}(x-y) &=&
\theta(x_0-y_0)\langle\psi_R(x)\,\psi_R^{\dagger}(y)\rangle_0
- \theta(y_0-x_0)\langle\psi_R^{\dagger}(y)\,\psi_R(x)\rangle_0
\\
&=&\theta(x_0-y_0)\int\mathrm{d}\mathbf{p}\,[\,(2\pi)^3\, 2\wp \,]^{-1}
u_R( \mathbf{p} )\otimes u^{\dagger}_R( \mathbf{p} )
\\
&\times& \exp \left\lbrace -\,i\wp(x_0-y_0) + i\mathbf{p}\cdot(\mathbf{x}-\mathbf{y})\right\rbrace 
\\
&-&\theta(y_0-x_0)\int\mathrm{d}\mathbf{p}\,[\,(2\pi)^3\, 2\wp \,]^{-1}
v_R( -\mathbf{p} )\otimes v^{\dagger}_R( -\mathbf{p} )
\\
&\times& \exp \left\lbrace \,i\wp(x_0-y_0) - i\mathbf{p}\cdot(\mathbf{x}-\mathbf{y})\right\rbrace 
\\
&=&\theta(x_0-y_0)\; i\partial\cdot\sigma \int\mathrm{d}\mathbf{p}\,[\,(2\pi)^3\, 2\wp \,]^{- 1} \,
\exp \left\lbrace -\,i\wp(x_0-y_0) + i\mathbf{p}\cdot(\mathbf{x}-\mathbf{y})\right\rbrace 
\\
&+& \theta(y_0-x_0)\; i\partial\cdot\sigma \int\mathrm{d}\mathbf{p}\,[\,(2\pi)^3\, 2\wp \,]^{- 1}
\exp \left\lbrace \,i\wp(x_0-y_0) - i\mathbf{p}\cdot(\mathbf{x}-\mathbf{y})\right\rbrace 
\end{eqnarray*}
Now, let us recall the definition of massless Wightman distributions \cite{Bogol}
\begin{eqnarray*}
i D_0^{(\pm)}(x-y) &\equiv& \left. \pm\int Dp\,\exp \left\lbrace \pm\, ip\cdot x\right\rbrace  \right\rfloor_{\,p_0=\wp}
\\
&=& \pm\int \frac{  \mathrm{d}\mathbf{p} }{ (2\pi)^3\, 2\wp  } \,
\exp \left\lbrace \pm\,i\wp(x_0-y_0) \mp\, i\mathbf{p}\cdot(\mathbf{x}-\mathbf{y})\right\rbrace 
\end{eqnarray*}
From the above we get 
\begin{eqnarray*}
S_{c}(x-y) &=& \theta(x_0-y_0)\;i\partial\cdot\sigma\,(-\,i)D_0^{(-)}(x-y) 
+ \theta(y_0-x_0)\;i\partial\cdot\sigma\,iD_0^{(+)}(x-y) 
\\
&=& i\partial\cdot\sigma [\,\theta(y_0-x_0)\,iD_0^{(+)}(x-y) - \theta(x_0-y_0)\,iD_0^{(-)}(x-y)\,]
\\
&-& \delta(x_0-y_0)[\,D_0^{(-)}(x-y) + D_0^{(+)}(x-y)\,] 
\\
&=& i\partial_{x}\cdot\sigma\,D^{\,c}_0(x-y)
\end{eqnarray*}
Here $ D_0(\xi) = D_0^{(-)}(\xi) + D_0^{(+)}(\xi) $ is the massless Pauli-Jordan distribution \cite{Bogol}, which satisfies
$ D_0(0,\vec{\xi}\;)=0 $. On the other hand
\[
D^{\,c}_0(x-y)\equiv \theta(y_0-x_0)\,iD_0^{(+)}(x-y) - \theta(x_0-y_0)\,iD_0^{(-)}(x-y)
\]
is the Feynman propagator, or causal Green's function, of a massless scalar field, which has Fourier transform representation
\[
D^{\,c}_0(x-y)=\frac{i}{(2\pi)^4}\int\mathrm{d}^4p\;\frac{ e^{-ip\cdot ( x - y )} }{p^{\,2}+i\varepsilon}
\]
Therefore $ iD^{\,c}_0(\xi) $ is the unique fully causal inverse of the d'Alembert differential operator.
\end{footnotesize}
From the previous definitions and identities one can check that { [\,in this paragraph we use the notation $ \tilde{\sigma}_{\mu}=\bar{\sigma}_{\mu} $\,] }
\begin{equation}
\bar{\sigma}\cdot \partial_{x}\,S_{c}(x-y) = \delta^{\, (4)}(x-y)
\end{equation}
This means that the right-handed differential operator $ \bar{\sigma}\cdot\partial\,.$ is inverted by the rank-two matrix-valued causal Green's function $ S_{c} $ while fully preserving causality.  Now repeating step-by-step the above derivation, with obvious changes, it is not difficult to prove that for the left-handed two-rank square matrix
\begin{equation}
\bar S_{c}(x-y)\equiv \langle 0\vert T\,\psi_L(x)\,\psi_L^{\dagger}(y)\vert0\rangle
\end{equation}
one gets
\begin{eqnarray}
\bar S_{c}(x-y) = i\partial_{x}\cdot\bar\sigma\,D^{\,c}_0(x-y)
\qquad\quad
\sigma\cdot\partial_x\,\bar S_{c}(x-y) =\,\delta^{\, (4)}(x-y)
\end{eqnarray}
{
Now the crucial remark is in order. According to the notations of Eq.(\ref{Weyldiffop}) one can rewrite the above obtained relations in the quite suggestive form
\begin{equation}
S_{c}(x-y)  =  i\partial_{L}\,D^{\,c}_0(x-y)\qquad\quad \partial_{R} S_{c}(x-y)= \delta^{\, (4)}(x-y)
\end{equation}
\begin{equation}
\bar S_{c}(x-y) = i\partial_{R}\,D^{\,c}_0(x-y)
\qquad\quad
\partial_L\,\bar S_{c}(x-y) =\,\delta^{\, (4)}(x-y)
\end{equation}
Consider now the classical Action integral for a left-handed Weyl spinor, \textit{viz.,}
\begin{equation}
S_L=\int \mathrm{d}^4x\,\psi_{L}^{\dagger}(x)\,i\sigma\cdot \partial\,\psi_{L}(x)
=\int \mathrm{d}^4x\,\psi_{L}^{\dagger}(x)\,i\partial_{L}\psi_{L}(x)
\end{equation}
It turns out that the inversion of the causal Green's function $ \langle 0\vert T\,\psi_L(x)\,\psi_L^{\dagger}(y)\vert0\rangle $ for a left-handed Weyl spinor 
is not related to the inverse of the kinetic operator $ i\partial_{L} $ appearing in the corresponding classical Action integral but to the kinetic operator
$ i\partial_{R} $ appearing in the classical Action integral of a right-handed Weyl spinor and \textit{viceversa}.
\textsf{This means that, in the process of inversion of a kinetic operator or of a causal Green's function for a Weyl spinor, it is impossible to restrict
ourselves only to one of the two inequivalent two-dimensionional irreducible representations of the Lorentz group}: \textbf{both of them must be necessarily 
involved}. This quite apparent and evident fact can also be rephrased in terms of the general transformation laws under the the inhomogeneous Lorentz group.
}
To this aim let us see next the Poincar\'e properties of the previous objects. 
Let a change of inertial reference frame be given by $ x^{\,\prime\mu}=\Lambda^{\mu}_{\phantom{\mu}\nu}(x+\mathrm{a})^{\nu} $. We have
\[
D_{0}^{\,c\,\prime}(x'-y')=D_{0}^{\,c}(x-y)
\]
\[
\Lambda_{L}^{-1}\,\bar{\sigma}^{\,\mu}\,\Lambda_{R}\,\partial^{\,\prime}_{\mu} 
=\Lambda_{R}^{\dagger}\,\bar{\sigma}^{\,\mu}\,\Lambda_{R}\,\partial^{\,\prime}_{\mu} 
=\Lambda^{\mu}_{\phantom{\mu}\kappa}\,\bar{\sigma}^{\,\kappa}\Lambda_{\mu}^{\phantom{\mu}\rho}\,\partial_{\rho} 
=\bar{\sigma}^{\,\mu}\,\partial_{\mu}
\]
and, as a consequence,
$$
\bar{\sigma}^{\,\mu}\,\partial^{\,\prime}_{\mu} = \Lambda_{L}\,\bar{\sigma}^{\,\mu}\,\partial_{\mu}\,\Lambda_{R}^{-1}
$$
Thus a transformation law for the Weyl causal Green's functions is
\begin{eqnarray}
\bar S_{c}^{\,\prime}(x'-y')= \Lambda_{L}\,\bar S_c(x-y)\,\Lambda_{R}^{-1}
\\
S_{c}^{\,\prime}(x'-y')= \Lambda_{R}\,S_c(x-y)\,\Lambda_{L}^{-1}
\end{eqnarray}

{
We see that these transformations involve the $ SL(2,\mathbb{C}) $ matrix representatives of both IRREPs $ D(0,\frac12) $ and $ D(\frac12,0)\,. $
\textbf{This is a delicate and crucial issue}. To better grasp its implications let us rewrite the causal Green's function using its Fourier integral representation, i.e.,
\begin{eqnarray}
&& \bar S_{\,c}(x-y) =\frac{i}{(2\pi)^4}\int\mathrm{d}^4p\;\frac{ p_0 - \vec{\sigma}\cdot\vec{p}  }{ p_0^2 - \vec{p}^{\;2} + i\varepsilon }\;e^{-ip\cdot (x - y)}
\nonumber\\
&=& \frac{i}{(2\pi)^4}\int\mathrm{d}^4p\;e^{-ip\cdot (x - y)}\;\frac{ p_0 - \vec{\sigma}\cdot\vec{p}  }{ (\,p_0 - \wp + i\varepsilon)(\,p_0 + \wp - i\varepsilon) }
\nonumber\\
&=& \int\frac{  \mathrm{d}^4p }{ (2\pi)^4  }     \;e^{-ip\cdot (x - y)}
\left( \frac{i}{ p_0 - \wp + i\varepsilon } - \frac{i}{ p_0 + \wp - i\varepsilon } \right) 
\;\frac{ p_0 - \vec{\sigma}\cdot\vec{p}  }{ 2\wp }
\end{eqnarray}
From this off-shell expression one sees that the rank-two square matrix
\begin{equation}
\mathbb{W} = \frac{ p_0 - \vec{\sigma}\cdot\vec{p}  }{ 2\wp }
\end{equation}
\textbf{is not anymore a projector on left-handed two-component spin states} because
\begin{equation}
\mathbb{W}^{\,2}=\frac{  p_0^{\,2} - 2p_0\vec{\sigma}\cdot\vec{p} + \vec{p}^{\;2}   }{ 4\vec{p}^{\;2} } \not= \mathbb{W}
\qquad\quad
\mathrm{Tr}\,\mathbb{W} = \frac{  p_0 }{ \wp } \in\mathbb{R}
\end{equation}
which, on the contrary, was true for the corresponding on-shell matrix with $ p_0=\wp\,. $ This leads to an unavoidable conclusion: \textsf{in the process
of inverting Weyl differential operators, to obtain the Weyl causal Green's functions, the notion itself of chirality cannot be preserved and is definitely lost}, due to the nature of the time ordered products of the Weyl quantum fields.

\textsf{The lesson that we learn from the above derivations and calculations is that the two-component formalism is not faithful and trustable at the quantum level,
just because the physical notions of chirality and parity are not at all defined, as they necessarily require the four-component formalism.}
}

\medskip

\medskip
In order to obtain a trustable fully causal inverted kinetic differential operator of a massless spinor we do not have other choices but turn back to Dirac bispinors.
For consider the $4\times 4$ matrix
\begin{footnotesize}
\begin{eqnarray*}
S^{\,c}_0(x-y) &=& \langle0\vert T \psi(x)\,\bar{\psi}(y)\vert0\rangle
=\left\lgroup
\begin{array}{cc}
0 & \bar{S}_c(x-y)
\\
S_c(x-y) & 0
\end{array}
\right\rgroup
\\
&=&\left\lgroup
\begin{array}{cc}
0 & i\partial\cdot\bar{\sigma}
\\
i\partial\cdot \sigma & 0
\end{array}
\right\rgroup
D_{0}^{\,c}(x-y)
\end{eqnarray*}
\end{footnotesize}
and apply $i\slashed \partial$ 
\begin{eqnarray}
i\partial\!\!\!/S^{\,c}_0(x-y) =
(-1)\left\lgroup
\begin{array}{cc}
\square & 0
\\
0 & \square
\end{array}
\right\rgroup
D_{0}^{\,c}(x-y)
=i\delta^{\, (4)}(x-y)
\end{eqnarray}
As expected, this proves that that the massless Schwinger propagator $ S_0(x-y) $ is the sole inverse of the massless Dirac differential operator
$ \partial\!\!\!/ $, with the property of fulfilling causality forward and backward in time. Its Fourier representation is
\begin{equation}
S^{\,c}_0 (x-y) = \frac{i}{(2\pi)^4}\int \mathrm{d}^4p\;\frac{ p\!\!\!/ }{p^{\,2} + i\varepsilon}\;
e^{-\,ip\cdot ( x - y )}
\end{equation}

\subsubsection{Observables}

{ As we have seen above, the consistent quantum theory of the massless Weyl-Dirac free field necessarily involves four types of quanta,
although the phenomenological constraints seem to indicate that only two of them are physical. The 1-particle state associated to a real left-handed neutrino 
of momentum $ \mathbf{p} $ and a negative helicity is given by
$ c^{\,\dagger}_{-,\,\mathbf{p}}\,\vert 0\rangle\,, $ while a real right-handed anti-neutrino of momentum $ \mathbf{q} $ and positive helicity is provided by
$ d^{\,\dagger}_{+,\,\mathbf{q}}\,\vert 0\rangle\,. $ The plane wave function that describes an incoming neutrino in a scattering experiment or decay process is therefore
\begin{equation}
\langle0\vert\,\psi(x)\,c^{\,\dagger}_{-,\,\mathbf{p}}\,\vert 0\rangle = u_{-,\,\mathbf{p}}(x)
\end{equation}
Quite analogously, an incoming right-handed anti-neutrino of momentum $ \mathbf{q} $ and positive helicity will be described by the plane wave function
\begin{equation}
\langle0\vert\,\overline{\psi}(x)\,d^{\,\dagger}_{+,\,\mathbf{q}}\,\vert 0\rangle = \bar v_{+,\,\mathbf{q}}(x)
\end{equation}
The outgoing states for the very same physical particles will be clearly described by the plane wave functions
\begin{equation}
\langle0\vert\,c_{-,\,\mathbf{p}}\,\overline{\psi}(x)\,\vert0\rangle = \bar{u}_{-,\,\mathbf{p}}(x)
\qquad\quad
\langle0\vert\,d^{\,\dagger}_{+,\,\mathbf{q}}\,\psi(x)\,\vert0\rangle = {v}_{+,\,\mathbf{q}}(x)
\end{equation}
All the remaining 1-particle state are not physical, as far as we know from experiments, in such a manner that the Standard Model gauge interactions 
will forbid the transitions from physical massless left-handed neutrinos and right-handed anti-neutrinos to unphysical Weyl particles, as it does.
}

Let us come now to the evaluation of the Observables - or better to say invariant and conserved charges - in the massless spinor quantum theory.
For example, since  we have
\begin{eqnarray*}
\psi^{\,\dagger}_{\pm}(y) i\partial_{\mu} \psi_{\pm}(x) =\
:\psi^{\,\dagger}_{\pm}(y) i\partial_{\mu} \psi_{\pm}(x) :\
- \int\mathrm{d}\mathbf{p}\,p_{\mu}\,v^{\top\,\ast}_{\pm,\,\mathbf{p} }(y)\,v_{\pm,\,\mathbf{p}}(x) 
\end{eqnarray*}
with $ p_0=\wp \,,$ one can easily obtain
\begin{eqnarray}
P^{\,\mu} = \sum_{\iota = -,+} \int\mathrm{d}\mathbf{p}\,p^{\,\mu}
\left[ \, c^{\,\dagger}_{\iota,\,\mathbf{p} }\, c_{ \iota,\,\mathbf{p} }
+ d^{\,\dagger}_{\iota,\,\mathbf{p} }\, d_{ \iota,\,\mathbf{p} } \,\right] _{p_0=\wp}
= P^{\,\mu}_L + P^{\,\mu}_R
\end{eqnarray}
In order to set up the spin operator, consider for instance a left Weyl field $ \psi_{-}(t,z) $ travelling along
the $ Oz- $axis so that
\[
T_{-}^{12}(t,z) - T_{-}^{21}(t,z)=0=\partial_{\nu} S_{-}^{\,\nu21}(t,z)
\]
\[
\Sigma_{-}=\int_{-\infty}^{\infty}\mathrm{d}z :S_{-}^{\,021}(t,z):
\qquad\quad \dot{\Sigma}_{-}=0
\]
\[
\Sigma_{-}=-\,\frac12\int_{-\infty}^{\infty}\mathrm{d}z :\psi_{-}^{\,\dagger}(0,z)\,\Sigma_{3}\,\psi_{-}(0,z):\
=-\,\frac12\int_{-\infty}^{\infty}\mathrm{d}z :\psi_{L}^{\dagger}(0,z)\,\sigma_{3}\,\psi_{L}(0,z):
\]
Now, since for a 1d motion we have for $ \wp = p_z\,\mathrm{sgn}(p_z) $
\begin{eqnarray*}
&& \psi_L(t,z) = \int_{-\infty}^{\infty}\mathrm{d}p_z\,[\,4\pi\wp \,]^{-\frac12}
\nonumber\\
&\times& \left[ \,c_{-,\,p_z}\, u_L( p_z )\,e^{-\,i\wp t + izp_z}
+ d^{\,\dagger}_{-,\,p_z}\,v_L(-\,p_z ) \,e^{\,i\wp t - izp_z} \,\right] 
\\ \\
&& u_L( p_z ) = \frac{ \wp -\,\sigma_{3} \, p_z }{ \sqrt{ \wp - p_z} }\,
\left\vert
\begin{array}{c}
1 \\ 0
\end{array}
\right\vert 
= \sqrt{ \wp - p_z}\,\left\vert
\begin{array}{c}
1 \\ 0
\end{array}
\right\vert 
\qquad\quad
v_L(-\,p_z ) = \frac{ \wp - \sigma_{3} \, p_z }{ \sqrt{\wp + p_z} }\,
\left\vert
\begin{array}{c}
0 \\ 1
\end{array}
\right\vert
= \sqrt{\wp + p_z}\,\left\vert
\begin{array}{c}
0 \\ 1
\end{array}
\right\vert
\end{eqnarray*}
and thereby
\[
\sigma_3\,u_L( p_z )  = u_L( p_z ), \qquad\quad
\sigma_3\,v_L(-\,p_z )  = -\,v_L(-\,p_z ),
\]
\begin{eqnarray*}
&& \sigma_3\,\psi_L(t,z) = \int_{-\infty}^{\infty}\mathrm{d}p_z\,[\,4\pi\wp \,]^{-\frac12}
\nonumber\\
&\times& \left[ \,c_{-,\,p_z}\, u_L( p_z )\,e^{-\,i\wp t + izp_z}
- d^{\,\dagger}_{-,\,p_z}\,v_L(-\,p_z ) \,e^{\,i\wp t - izp_z} \,\right] 
\end{eqnarray*}
This entails that the general form of the spin operator for a one dimensional motion along e.g. the $ Oz- $axis
becomes
\begin{eqnarray}
\Sigma_{z} &=& -\,\frac12\int_{-\infty}^{\infty}\mathrm{d} p_z\, 
\left[ \,  c^{\,\dagger}_{-,\,p_z }\, c_{ -,\,p_z }
- d^{\,\dagger}_{+,\,p_z }\, d_{ +,\,p_z } 
\right. 
\nonumber\\
&-& \left.  c^{\,\dagger}_{+,\,p_z }\, c_{ +,\,p_z }
+ d^{\,\dagger}_{-,\,p_z }\, d_{ -,\,p_z } 
\,\right] 
\end{eqnarray}
which endorses the fact that neutrino particles are of negative helicity, while the antineutrinos exhibit
positive helicity, owing to the opposite verse of the wave vector, the situation being reversed for the
hypothetical right-handed neutrinos of positive chirality. 

Turning to the internal symmetries and charges we find
\begin{eqnarray}
\mathcal{Q} &=&
\int\mathrm{d}\mathbf{x} :\psi_{-}^{\,\dagger}(x)\,\psi_{-}(x) 
+ \psi_{+}^{\,\dagger}(x)\,\psi_{+}(x):\  \equiv \mathcal{Q}_{-} + \mathcal{Q}_{+}
\\
\mathcal{Q}_{\mp} &=& \int\mathrm{d}\mathbf{x}\int\mathrm{d}\mathbf{p}^{\prime} : \left[ \,
c^{\,\dagger}_{\mp,\,\mathbf{p}^\prime }\, u^{\top\,\ast}_{\mp,\,\mathbf{p}^\prime } (x)
+ d_{\mp,\,\mathbf{p}^\prime }\,v^{\top\,\ast}_{\mp,\,\mathbf{p}^\prime }(x) \,\right] 
\nonumber \\
&\times& \int\mathrm{d}\mathbf{p}\,\left[ \, c_{\mp,\,\mathbf{p}}\, u_{\mp,\,\mathbf{p}}(x)
+ d^{\,\dagger}_{\mp,\,\mathbf{p}}\, v_{\mp,\,\mathbf{p}}(x) \,\right] :
\nonumber \\
&=& \int\mathrm{d}\mathbf{p}\,\left[ \,c^{\,\dagger}_{\mp,\,\mathbf{p} }\,c_{\mp,\,\mathbf{p}}
- d^{\,\dagger}_{\mp,\,\mathbf{p} }\,d_{\mp,\,\mathbf{p}} \,\right] 
\\
\mathcal{Q}_5 &=& \mathcal{Q}_{+} - \mathcal{Q}_{-}
\end{eqnarray}
Finally, the dilatation charge actually reads
\begin{eqnarray}
\mathcal{D}&=&\int\mathrm{d}\mathbf{x} : \textstyle\frac32 i \psi^{\dagger}(x)\psi(x)+x_{\nu}T^{0\nu}(x) :
\nonumber \\
&=& x_0P_0 + i \left\lbrace {\textstyle\frac32}\,\mathcal{Q} 
-  \int\mathrm{d}\mathbf{x}\, x^{\,k} :  \psi^{\dagger}(x) \nabla_{k} \psi(x) : \right\rbrace 
\end{eqnarray}
We find
\begin{eqnarray*}
\partial_0\mathcal{D} &=& P_0 + \int\mathrm{d}\mathbf{x}\, x^{\,k} :  \left[ \,i\partial_0\psi(x)\,\right] ^{\dagger}\nabla_{k} \psi(x) : 
- \int\mathrm{d}\mathbf{x}\, x^{\,k} :  \psi^{\dagger}(x) \nabla_{k} i\partial_0\psi(x) : 
\\
&=& P_0 + \int\mathrm{d}\mathbf{x}\, x^{\,k} :  \psi^{\dagger}(x) H_0 \nabla_{k} \psi(x) : 
- \int\mathrm{d}\mathbf{x}\, x^{\,k} :  \psi^{\dagger}(x) \nabla_{k} H_0 \psi(x) : 
\\
&=& P_0
\end{eqnarray*}
Since we have the well-known relation
\[
\partial_{0}{\mathcal{D}} = i \,[\,\mathcal{D}, P_0\,] = P_0
\]
it follows that we can write, as expected,
$$
\dot{\mathcal{D}}= \frac{ \mathrm{d} \mathcal{D}   }{ \mathrm{d} t  }= \partial_0\mathcal{D} + i \,[\,P_0,\mathcal{D}\,] = 0
$$
so that the dilatation charge is globally time independent.

From the previous analysis, we learn that the 1-particle states $ c_{-}^{\,\dagger}(\mathbf{p})\vert 0 \rangle $
and $ d^{\,\dagger}_{+}(\mathbf{p})\vert 0 \rangle $
belonging to the Fock space of the massless Dirac spinor field represent the physical particles to which the following quantum number are assigned:

\textit{i}) opposite chirality $ \chi=\mp\,1 $; 

\textit{ii}) the same energy-momentum $ p^{\,\nu}=(\wp, \mathbf{p})\,;$ 

\textit{iii}) opposite internal charge, called \textsf{lepton number}, $ \ell=\pm\,1\,; $

\textit{iv}) opposite helicity $ \mathrm{h}=\mp\,1\,. $
}

\medskip
\subsubsection{Discrete Symmetries}

Let us now consider the properties of massless Dirac spinor fields under discrete symmetry transformations. Let us start from charge conjugation. This is represented by a unitary transformation 
of the quantum theory that {\sf simultaneous exchanges chirality and particle-antiparticle (creation and annihilation) operators}, up to a phase factor, i.e. 

\begin{eqnarray}
\mathcal{C}\,c_{\mp,\,\mathbf{p}}\,\mathcal{C}^{-1} = e^{\,i\theta } d_{\pm,\,\mathbf{p}}
\qquad\quad
\mathcal{C}\,d_{\mp,\,\mathbf{p}}\,\mathcal{C}^{-1} = e^{-\, i\theta } c_{\pm,\,\mathbf{p}}
\\
\mathcal{C}^{\,2} = \mathbb{I} \qquad\quad \mathcal{C}=\mathcal{C}^{\,\dagger} = \mathcal{C}^{-1}\qquad\quad0\le \theta<  2\pi
\end{eqnarray}
that yields
\[
\mathcal{C}\,\psi_{\mp}(x)\,\mathcal{C} = e^{\, i\theta}
\int\mathrm{d}\mathbf{p}\,\left[ \, d_{\pm,\,\mathbf{p}}\, u_{\mp,\,\mathbf{p}}(x)
+ c^{\,\dagger}_{\pm,\,\mathbf{p}}\, v_{\mp,\,\mathbf{p}}(x) \,\right]
\]
The  charge conjugation relations for the plane wave Weyl spinor functions read
\begin{equation}
u_{\mp,\,\mathbf{p}}(x) = i \gamma^{\,2}  v^{\ast}_{\pm,\,\mathbf{p}}(x) 
\qquad\quad
v_{\mp,\,\mathbf{p}}(x) = i \gamma^{\,2}  u^{\ast}_{\pm,\,\mathbf{p}}(x) 
\end{equation}

\medskip
\begin{footnotesize}
\texttt{Proof}. Consider for instance
\[
v^{\ast}_{\pm,\,\mathbf{p}}(x) = [\,(2\pi)^{3} 2\wp\,]^{-1} v^{\ast}_{\pm}(-\,\mathbf{p})\,e^{-\,i\wp t + i\vec{p}\cdot\vec{x}}
\]
We readily find $ \gamma_{5}\gamma^{\,2}  v^{\ast}_{\pm}(-\,\mathbf{p}) = \mp\,\gamma^{\,2}  v^{\ast}_{\pm}(-\,\mathbf{p})\,. $
Moreover we can write
\[
H_{0}(\mathbf{p})\, i \gamma^{\,2} v^{\ast}_{\pm}(-\,\mathbf{p})  =
\left[ \,H^{\ast}_{0}(\mathbf{p})\, i \gamma^{\,2}  v_{\pm}(-\,\mathbf{p}) \,\right] ^{\ast}
=\left[ \, i \gamma^{\,2} H_{0}(\mathbf{p})\,  v_{\pm}(-\,\mathbf{p})  \,\right] ^{\ast}
=\wp\, i \gamma^{\,2} v^{\ast}_{\pm}(-\,\mathbf{p})  
\]
It follows that the spin-state $ i \gamma^{\,2}  v_{\pm}(-\,\mathbf{p}) $ is a basis spinor with chirality $ \chi=\mp 1 $
and positive energy, in such a manner that we can safely identify $ u_{\mp,\,\mathbf{p}}(x) = i \gamma^{\,2}  v^{\ast}_{\pm,\,\mathbf{p}}(x)  $
as above claimed.
\end{footnotesize}

\medskip
Hence we can write
\begin{eqnarray*}
\mathcal{C}\,\psi_{\mp}(x)\,\mathcal{C} &\equiv& \psi_{\mp}^{\,c}(x) = e^{\, i\theta }
\int\mathrm{d}\mathbf{p}\,\left[ \, d_{\pm,\,\mathbf{p}}\, u_{\mp,\,\mathbf{p}}(x)
+ c^{\,\dagger}_{\pm,\,\mathbf{p}}\, v_{\mp,\,\mathbf{p}}(x) \,\right]
\\
&=& i \gamma^{\,2} \,e^{\, i\theta  }
\int\mathrm{d}\mathbf{p}\,\left[ \, c^{\,\dagger}_{\pm,\,\mathbf{p}}\,  u^{\ast}_{\pm,\,\mathbf{p}}(x) 
 + d_{\pm,\,\mathbf{p}}\, v^{\ast}_{\pm,\,\mathbf{p}}(x) \,\right]
 \\
 &=& \gamma^{\,2} \left( \psi_{\pm}^{\,\dagger}(x)\right) ^{\top}
 \qquad\quad\mathrm{for}\ \theta = -\,\pi/2
\end{eqnarray*}
Now, since we have $ \psi(x) =\psi_{-}(x) + \psi_{+}(x) $, we come to the result
\begin{equation}
\psi^{\,c}(x) = \mathcal{C}\,\psi(x)\,\mathcal{C} = \gamma^{\,2} \left( \psi^{\,\dagger}(x)\right) ^{\top}
\end{equation}
which is nothing but the quantum generalization of the charge conjugation rule for classical spinor fields,
i.e. $ \psi^{\,c}(x) = \gamma^{\,2} \psi^{\ast}(x)\,, $ as expected.

{ Thus the charge conjugated of a particle of e.g. negative chirality, negative helicity and
positive lepton number is an anti-particle of opposite chirality, positive helicity and
negative lepton number. It turns out that the charge conjugation transformation necessarily
involves both chiralities, in order to be implemented. Hence the $ \mathcal{C} $ transform is well defined
only for 4-component bispinors and not for 2-component Weyl spinors belonging to the two 
irreducible 2d representations.}

\medskip
The parity transformation is a discrete space-time symmetry of the Action integral $ S=\int\mathrm{d}^{4}x\,\mathcal{L}_0(x) $
that can be implemented at the quantum field theory level as
\begin{eqnarray}
\mathcal{P}\, c_{\mp,\,\mathbf{p}} \,\mathcal{P}^{\dagger} = e^{\,i\eta_{\mp} } \,c_{\pm, -\,\mathbf{p}}
\qquad\quad
\mathcal{P}\, d_{\mp,\,\mathbf{p}} \,\mathcal{P}^{\dagger} = e^{\,i\theta_{\mp} } \,d_{\pm, -\,\mathbf{p}}
\\
\qquad
0\le \eta_{\mp} < 2\pi\qquad\quad 0\le \theta_{\mp} < 2\pi
\nonumber
\end{eqnarray}
After insertion of the normal mode expansion of the massless Dirac quantum field, then we find with
$ 	x^{\,\prime\mu} = \tilde{x}^{\,\mu}=(x_0,-\,\vec{x} \,) = x_\mu $
\begin{eqnarray*}
\psi^{\prime}(x^{\prime}) &=& \mathcal{P} \,\psi(\tilde{x}) \,\mathcal{P}^{\dagger}
\\
&=& \int\mathrm{d}\mathbf{p}\,\left[ \,
\mathcal{P}\, c_{\mp,\,\mathbf{p}} \,\mathcal{P}^{\dagger}\,u_{\mp,\,\mathbf{p}}( \tilde{x} ) 
+ \mathcal{P}\, d^{\,\dagger}_{\mp,\,\mathbf{p}} \,\mathcal{P}^{\dagger}\,v_{\mp,\,\mathbf{p}}( \tilde{x} ) 
\,\right] 
\\
&=& \int\mathrm{d}\mathbf{p}\,\left[ \,
e^{\,i\eta_{\mp} } \,c_{\pm, -\,\mathbf{p}} \,u_{\mp,\,\mathbf{p}}( \tilde{x} ) 
+ e^{ -\,i\theta_{\mp} } \,d^{\,\dagger}_{\pm, -\,\mathbf{p}} \,v_{\mp,\,\mathbf{p}}( \tilde{x} ) \,\right] 
\\
&=& \int\mathrm{d}\mathbf{p}\,\left[ \,
e^{\,i\eta_{\mp} } \,c_{\pm, \,\mathbf{p}} \,u_{\mp, -\,\mathbf{p}}( \tilde{x} ) 
+ e^{ -\,i\theta_{\mp} } \,d^{\,\dagger}_{\pm,\,\mathbf{p}} \,v_{\mp,-\,\mathbf{p}}( \tilde{x} ) \,\right] 
\end{eqnarray*}
Now we have
\begin{eqnarray*}
u_{\mp , -\,\mathbf{p}}( \tilde{x} ) =
[\,(2\pi)^{3} 2\wp\,]^{-\frac12} \,u_{\mp} ( -\,\mathbf{p} ) \,e^{\,-\,i\wp t + i\mathbf{p}\cdot \mathbf{x}}
= -\,\gamma^0 \,u_{\pm , \,\mathbf{p}}(x) 
\\ \\
v_{\mp, -\,\mathbf{p}}(\tilde{x})=[\,(2\pi)^{3} 2\wp\,]^{-\frac12}\, v_{\mp}( -\,\mathbf{p} ) \,
e^{\,i\wp t - i\mathbf{p}\cdot \mathbf{x}}
= \gamma^0 \,v_{\pm , \,\mathbf{p}}(x) 
\end{eqnarray*}
so that, after choosing
\[
\eta_{\mp}  = \mp\,k\pi
\qquad\quad
\theta_{\mp}  = \mp\,2k\pi\qquad\quad(\,k\in\mathbb{Z}\,)
\]
we definitely obtain the usual transformation law
\begin{eqnarray}
\psi_{\mp}^{\,\prime}(x^{\prime}) &=& \mathcal{P} \,\psi_{\mp} ( \tilde{x} ) \,\mathcal{P}
= \gamma^{0}\,\psi_{\pm}(x)
\\
\psi^{\,\prime}(x^{\prime})&=& \mathcal{P} \,\psi ( \tilde{x} ) \,\mathcal{P} = \gamma^0\,\psi(x)
\\
&&\mathcal{P}^{\,2} = \mathbb{I} \qquad \mathcal{P}=\mathcal{P}^{\,\dagger} = \mathcal{P}^{\,-1}
\end{eqnarray}
Once again, it is worthwhile to remark that even the parity transform is not a well defined operation for a two-component
Weyl fermion, because it necessarily involves both chiralities, i.e. both irreducible 2d representations of the proper Lorentz group.

\medskip
In quantum field theory the time reversal operator $ \mathcal{T} $ corresponds to an antilinear and antiunitary transformation
\cite{Merz,PS} which acts as follows: namely,
\begin{eqnarray}
[\,P_0,\,\mathcal{T}\,]=0\qquad\quad\lbrace\vec{P},\,\mathcal{T}\,\rbrace=0
\qquad\quad\lbrace M^{\rho\sigma},\,\mathcal{T}\,\rbrace=0
\end{eqnarray}
where, as usual, $ (\,P^{\,\mu},\,M^{\rho\sigma}\,) $ are the ten generators of the Poincar\'{e} group.
It follows therefrom that the antilinear and antiunitary time reversal operation does reverse the signs of all
particle momenta and spin angular momenta, the same being true for antiparticles. 
In this regard we need to explain what is meant by spin-flip \cite{PS}. 
{
Consider for example the spin state
\begin{eqnarray}
u_{-}( \mathbf{p} ) &=& \left\lgroup
\begin{array}{c}
u_L( \mathbf{p} ) \\  0 \\ 0
\end{array}
\right\rgroup
=\frac{1}{\sqrt{\wp - p_z}} \left\lgroup
\begin{array}{c}
\wp - p_z \\ -\,p_x - ip_y \\ 0 \\ 0
\end{array}
\right\rgroup
\\
u_L( \mathbf{p} ) &=&  \frac{ {\sigma}\cdot\tilde{p} }{\sqrt{\wp - p_z}} \, 
\left\vert
\begin{array}{c}
1 \\ 0
\end{array}
\right\vert  
\nonumber
\end{eqnarray}
This positive energy, left-handed spin-state of negative chirality } does involve the spin-up constant
spinor $ \left\vert
\begin{array}{c}
1 \\ 0
\end{array}
\right\vert  $ and  to our purpose it will be better labelled as $ u_{-}( \mathbf{p},\uparrow ) \,. $
Its corresponding spin reversed state $ u_{-}( \mathbf{p},\downarrow ) $ has the very same structure, but for the constant spinor that
will be turned to the spin-down $ \left\vert
\begin{array}{c}
0 \\ 1
\end{array}
\right\vert \,. $
Thus we can denote the pair of normalized spin flip states according to
\begin{eqnarray}
u_{-}( \mathbf{p},\uparrow )  =
\frac{1}{\sqrt{\wp - p_z}} \left\lgroup
\begin{array}{c}
\wp - p_z \\ -\,p_x - ip_y \\ 0 \\ 0
\end{array}
\right\rgroup
\label{spinup} 
\\
u^{\ast}_{-}( -\mathbf{p},\downarrow )  =
\frac{1}{\sqrt{\wp - p_z}} \left\lgroup
\begin{array}{c}
p_x + ip_y \\ \wp - p_z \\ 0 \\ 0
\end{array}
\right\rgroup
\label{spindown} 
\end{eqnarray}
and quite analogous relationships for the remaining spin states.
Then we have
\begin{eqnarray}
&&\mathcal{T}\,c_{\mp,\,\mathbf{p}}\,\mathcal{T}^{\,-1} = e^{ \,i\eta_{\mp} } \,c_{\mp, -\,\mathbf{p}}
\qquad\quad
\mathcal{T}\,d_{\mp,\,\mathbf{p}}\,\mathcal{T}^{\,-1} = e^{ \,i\theta_{\mp} } \,d_{\mp, -\,\mathbf{p}}
\\ \nonumber\\
&&\psi_{\mp}^{\,\prime}(x^{\prime})=\mathcal{T}\,\psi_{\mp}(-\,x_0,\vec{x}\,)\,\mathcal{T}^{\,-1}
\nonumber \\
&=& \int\mathrm{d}\mathbf{p}\,\left[ \,
e^{\,i\eta_{\mp} } \,c_{\mp, -\,\mathbf{p}} \,u^{\,\ast}_{\mp,\,\mathbf{p}}( -\,\tilde{x} ) 
+ e^{ -\,i\theta_{\mp} } \,d^{\,\dagger}_{\mp, -\,\mathbf{p}} \,v^{\,\ast}_{\mp,\,\mathbf{p}}( -\,\tilde{x} ) \,\right] 
\\
&=& \int\mathrm{d}\mathbf{p}\,\left[ \,
e^{\,i\eta_{\mp} } \,c_{\mp, \,\mathbf{p}} \,u^{\,\ast}_{\mp, -\,\mathbf{p}}( -\,\tilde{x} ) 
+ e^{ -\,i\theta_{\mp} } \,d^{\,\dagger}_{\mp, \,\mathbf{p}} \,v^{\,\ast}_{\mp, -\,\mathbf{p}}( -\,\tilde{x} ) \,\right] 
\end{eqnarray}
the complex conjugation originates from the antilinear and antiunitary nature of the $ \mathcal{T}- $operation, while we have understood the spin-flip operation in the spinor plane wave functions. Acting on $\psi^{(-)}$, for instance, we have
\begin{eqnarray*}
&& \mathcal{T}\,\psi^{(-)}(-\,t,\mathbf{x}\,)\,\mathcal{T}^{\,-1} \ =\
\int\mathrm{d}\mathbf{p}\,[\,(2\pi)^{3}\,2\wp\,]^{-\frac12}
\\
&\times&\left[ \,e^{\,i\eta_{-} } \,c_{-, \,\mathbf{p}}\,u^{\,\ast}_{-}( -\,\mathbf{p} \downarrow) 
+ e^{\,i\eta_{+} } \,c_{+, \,\mathbf{p}}\,u^{\,\ast}_{+}( -\,\mathbf{p} \downarrow) \,\right] 
\,e^{-\,i\wp t + i\mathbf{p}\cdot\mathbf{x} }
\end{eqnarray*}
We can see that the right-hand side becomes the local representation of a quantum field at the instant $ t $,
to be denoted $ \Theta\,\psi^{(-)}(t,\mathbf{x}\,)\,, $ if we can find a $ 4\times4 $ matrix for which
\begin{equation}
e^{\,i\eta_{\mp} } \,u^{\,\ast}_{\mp}( -\,\mathbf{p} \downarrow) =\Theta\,u_{\mp}( \mathbf{p} \uparrow)
\end{equation}
The explicit forms (\ref{spinup}-\ref{spindown}) tell us that the above relations hold true for $ \eta_{\mp}=0\,. $
For orthogonality and closure relations for the spin-states we need the unitarity property
$ \Theta\,\Theta^{\dagger}=\mathbb{I} = \Theta^{\dagger}\Theta\,. $ In addition from the spin-state equations we get
\begin{eqnarray*}
\vec{\alpha}\cdot\vec{p}\;u_{\mp}(\vec{p}\,)=\wp\,u_{\mp}(\vec{p}\,)
\\
-\,\vec{\alpha}^{\,\ast}\cdot\vec{p}\; u^{\,\ast}_{\mp}( -\,\vec{p}\,)=\wp\,u^{\,\ast}_{\mp}( -\,\vec{p}\,)
\\
-\,\vec{\alpha}^{\,\ast}\cdot\vec{p}\; \Theta\,u_{\mp}(\vec{p}\,) = \wp\,\Theta\,u_{\mp}(\vec{p}\,)
\\
-\,\Theta^{\dagger}\,\vec{\alpha}^{\,\ast}\Theta = \vec{\alpha}
\end{eqnarray*}
A unitary solution of the last equation is unique except for an irrelevant phase factor. In particular in the Weyl representation of the Clifford algebra, we have a solution endowed with the important property $ \Theta^{\ast}\Theta+\mathbb{I}=0\,, $ which does not depend on the specific representation of the Dirac matrices.. It is the real matrix 
\begin{equation}
\Theta = -\,\gamma^{1}\gamma^{3} = -\,i\,\Sigma_{2}
\end{equation}

Concluding we have
\begin{equation}
\psi_{\pm}^{\,\prime}(x^{\prime})=\mathcal{T}\,\psi_{\pm}(-\,t,\mathbf{x}\,)\,\mathcal{T}^{\,-1} = \Theta\,\psi_{\pm}(t,\mathbf{x})
\qquad\quad(\,x^{\prime}=-\,\tilde{x}\,)
\end{equation}

which shows that the time reversal is a chirality preserving transformation.
Moreover, time reversal is well defined even for two component Weyl spinors, since the spin-flip operation does not mix left and right components
of a massless Dirac bispinor. This means that also the $ \mathcal{CP} $ and $ \mathcal{CPT} $ transformations are well defined for two-component
Weyl spinors. In summary, the quantum theory which arises from the classical Action integral
\[
S_0\equiv \int\mathrm{d}^4x\, \mathcal{L}_0(x) = \int\mathrm{d}^4x\, \psi^{\dagger}(x)\,i\alpha\cdot\partial\,\psi(x)
\]
is invariant under the discrete $ \mathcal{C,P,T} $ transformation separately, while the Weyl Action integrals 
\[
S_L=\int\mathrm{d}^4x\, \textstyle\frac12\,\psi_{L}^{\dagger}(x)\,i\sigma\cdot\!\buildrel \leftrightarrow\over \partial\!\psi_{L}(x)
\qquad\quad
S_R=\int\mathrm{d}^4x\, \textstyle\frac12\,\psi_{R}^{\dagger}(x)\,i\bar\sigma\cdot\!\buildrel \leftrightarrow\over \partial\!\psi_{R}(x)
\]

and the corresponding quantum theories are invariant under $ \mathcal{T} $  and $ \mathcal{CP} $ transforms. They are  well defined and good symmetries for left and right-handed Weyl spinors.

\subsection{Majorana Spinor Field}

The second part of the section will be devoted to Majorana spinor fields and relevant quantum theories. We can start by writing down a relativistic
invariant field equation for a massive 2-component  spinor
field. To this end let us start from  a left Weyl spinor $ \psi_L\in D(\frac12,0) $ that transforms with the matrix $ \Lambda_{L}\,. $ belonging to $ SL(2,\mathbb{C}) $ 
Let us call such a 2-component spinor field $\chi_a(x)\ (a=1,2)$. 

\subsubsection{Definitions and Lagrangians}

The starting point is a Weyl spinor wave field, considered as a classical anti-commuting field, 
i.e. a {\sf Gra\ss mann valued}  left Weyl spinor field over the
Minkowski spacetime. We assume it satisfies
$$
\{\,\chi_a(x)\,,\,\chi_b(y)\,\}=0\qquad\quad\left(\,x,y\in{\mathcal  M}\ \vert\ a,b=1,2\,\right)
$$
and we assume also the complex conjugation rule
\begin{equation}
(\chi_1\chi_2)^\ast=\chi_2^\ast\chi_1^\ast=-\,\chi_1^\ast\chi_2^\ast
\label{ccrule} 
\end{equation}
The idea is to imitate the Hermitean conjugation of quantum fields.
Now {\sf a  Majorana classical spinor field is a self-conjugated bispinor}. It can be constructed, for instance, from the above left-handed spinor $ \chi_a(x)\ (a=1,2) $ as follows:

\[
\chi_{M}(x)=\left\lgroup
\begin{array}{c}
\chi(x)
\\
-\sigma_{2}\chi^{\ast}(x)
\end{array}
\right\rgroup
=\chi_{M}^{\,c}(x)
\]
the charge conjugation rule for any classical bispinors $ \psi $ being defined by the general relationship
$$ 
\psi^{\,c}(x) = e^{\,i\theta}\,\gamma^{\,2}\,\psi^{\,\ast}(x)
\qquad\quad(\,0\le\theta<2\pi\,)
$$ 
which is a discrete internal - i.e. space-time point independent - symmetry transformation. Here below we shall suitably choose $ \theta=0\,. $
The Majorana bispinor has a right-handed lower Weyl spinor component $ -\sigma_{2}\chi^{\ast}\in D(0,\frac12)\,, $
albeit functional dependent, due to the charge self-conjugation constraint,
in such a manner that $ \chi_{M} $ possesses both chiralities and polarizations, at variance with its left-handed Weyl  building spinor $ \chi(x)\,. $
There is another kind of self-conjugated Majorana bispinor, which can be set-up out of a right-handed Weyl building spinor
$ \varphi\in D(0,\frac12) $
\[
\varphi_{M}(x)=\left\lgroup
\begin{array}{c}
\sigma_{2}\varphi^{\ast}(x)
\\
\varphi(x)
\end{array}
\right\rgroup
=\varphi_{M}^{\,c}(x)
\]

From the Majorana self-conjugated bispinors, one can readily construct the most general Poincar\'e invariant and power counting
renormalizable Lagrangian. For instance, by starting from the bispinor $ \chi_{M}(x) $ we have
$$
{\mathcal L}_M =
\textstyle\frac14\,{\chi}^{\dagger}_{M}(x)\,\alpha^{\,\mu}\,i\parl_\mu\chi^{\,c}_{M}(x) 
- \frac12\,m\overline{\chi}_{M}(x)\chi^{\,c}_{M}(x) 
$$
From the notation it is clear that the upper and lower components of a Majorana bispinor are related to each other by the self-conjugation constraint, therefore 
\sf{they can never be treated as functionally independent, even formally}.

As a consequence of its construction \textsf{the Majorana mass term} can be written in the two equivalent forms

\begin{eqnarray*}
\mathcal{L}_{M}^{\,m} &=& \textstyle\frac12\,m\left[ \,
\chi^{\top}(x) \sigma_{2}\, \chi(x) + \chi^{\dagger}(x) \sigma_{2}\, \chi^{\ast}(x)\,\right] 
\\
&=& -\,im\left[ \,\chi_{1}(x)\chi_{2}(x) + \chi^{\ast}_{1}(x)\chi^{\ast}_{2}(x)\,\right] 
\\
&=& im\left[ \,\chi^{\ast}_{2}(x)\chi^{\ast}_{1}(x) + \chi_{2}(x)\chi_{1}(x)\,\right] 
= \left( \mathcal{L}_{M}^{\,m} \right) ^{\ast}
\end{eqnarray*}
whence it is clear that the corresponding integral $ \int\mathrm{d}^{4}x\,\mathcal{L}_{M}^{\,m} (x) $
is not invariant under the internal $ U(1) $ phase transformation
\[
\chi(x)\quad\mapsto\quad \chi^{\,\prime}(x)=\chi(x)\,e^{\,i\theta}
\qquad\quad(\,0\le\theta<2\pi\,)
\]
Concerning the kinetic term, from the relations
\[
\chi^{\dagger}_{M} = \left\lgroup \chi^{\dagger}\quad -\,\chi^{\top}\,\sigma_{2} \right\rgroup
\qquad\quad\chi^{\dagger}=(\,\chi^{\ast})^{\top}
\]
together with
\[
\alpha^{\nu}=\gamma_{0}\gamma^{\,\nu}=\left\lgroup
\begin{array}{cc}
\sigma^{\,\nu} & 0
\\
0 & \bar{\sigma}^{\,\nu}
\end{array}
\right\rgroup
\qquad\quad
\sigma_{\mu}=\bar{\sigma}^{\,\mu}=(1,\vec{\sigma}\,)
\]
\[
\alpha^{\,\nu}\gamma^{\,2}=\left\lgroup
\begin{array}{cc}
0 &\sigma^{\,\nu}\sigma_{2}
\\
-\,\bar{\sigma}^{\,\nu}\sigma_{2} & 0
\end{array}
\right\rgroup
\]
one can obtain the Majorana kinetic term in the 2-component formalism
\begin{eqnarray}
&& \textstyle\frac14\,\chi_{M}^{\dagger}(x)\,\alpha^{\,\mu}\gamma^{\,2}\,i\parl_\mu\chi^{\,\ast}_{M}(x)\label{MajoWeyl}
\\
&=&\textstyle\frac14\,\chi^{\dagger}(x)\,\sigma^{\,\mu}\,i\parl_\mu\chi(x) +
\frac14\,\chi^{\top}(x)\,\sigma_{2}\,\bar{\sigma}^{\,\mu}\,\sigma_{2}\,i\parl_\mu\chi^{\ast}(x) \0
\\
&=& \textstyle\frac14\,\chi^{\dagger}(x)\,\sigma^{\,\mu}\,i\parl_\mu\chi(x) +
\frac14\,\chi^{\top}(x) ( \bar{\sigma}^{\,\mu} )^{\ast}\, i\parl_\mu\chi^{\ast}(x)\0 
\\
&=& \textstyle\frac12\,\chi^{\dagger}(x)\,\sigma^{\,\mu}\,i\partial_\mu\chi(x) +
\frac12\,\chi^{\top}(x) ( \bar{\sigma}^{\,\mu} )^{\ast}\, i\partial_\mu\chi^{\ast}(x) \0
\\
&=& \textstyle\chi^{\dagger}(x)\,\sigma^{\,\mu}\,i\partial_\mu\chi(x) 
- \frac12\,i\partial_\mu\left[ \,\chi^{\dagger}(x)\,\sigma^{\,\mu}\,\chi(x)\,\right] \0
\end{eqnarray}
which coincides with the Lagrangian for a left-handed Weyl spinor, as expected, where the above complex conjugation rule
(\ref{ccrule}) for Grassmann valued field has been used.

\medskip
It turns out that even the Majorana  Action integral
\begin{eqnarray}
S_{M} &=& \textstyle\frac14\int\mathrm{d}^{4}x\,\Big[\, \chi_{M}^{\dagger}(x)\,\alpha^{\,\mu}\gamma^{\,2}\,i\parl_\mu\chi^{\,\ast}_{M}(x)
- 2m \chi_{M}^{\dagger}(x)\,\alpha^{2} \chi_{M}^{\ast}(x)\Big]
\\
&=& \int\mathrm{d}^{4}x\,\left\lbrace \chi^{\dagger}(x)\,\sigma^{\,\mu}\,i\partial_\mu\chi(x) 
+ \textstyle\frac12\,m\left[ \,\chi^{\dagger}(x)\,\sigma_{2}\, \chi^{\ast}(x)
+ \chi^{\top}(x)\,\sigma_{2}\, \chi(x)\,\right] \right\rbrace 
\label{MajoranaAction} 
\end{eqnarray}
\textsf{is not invariant under the overall phase transformation} $ \chi^{\,\prime}_{M}(x)=e^{\,i\theta}\chi_M(x) $
of the Majorana bispinor, just like the previously discussed mass term. Hence, as it will be further endorsed after the transition
to the Majorana representation of the Dirac matrices, \textsf{there is no invariant scalar charge or lepton number for a Majorana \textbf{massive} spinor}.
According to the general Noether theorem for a Lie group of global internal symmetry transformations
\[
\partial_{\mu} J^{\mu}_{a}=\delta\mathcal{L}/\delta\omega_{a}\qquad\quad(a=1,2,\ldots,s)
\]
where $ \omega_{a} $ are the parameters of the internal symmetry Lie group of dimension $ s $, we find
\begin{equation}
J^{\mu}_{M}=(\delta\mathcal{L}_{M}/\delta\partial_{\mu}\chi)\,i\chi= - \chi^{\dagger}\sigma^{\mu}\chi
\qquad\quad
\delta\mathcal{L}_{M}/\delta\theta=im\left( \chi^{\top}\sigma_{2}\chi - \chi^{\dagger}\sigma_{2}\chi^{\ast}\right) 
\end{equation}
so that
\begin{equation}
\partial_{\mu} J^{\mu}_{M} = \partial_{\mu}(\chi^{\dagger}\sigma^{\mu}\chi)=im( \chi^{\dagger}\sigma_{2}\chi^{\ast} - \chi^{\top}\sigma_{2}\chi) 
\end{equation}

\medskip
\begin{footnotesize}
We recall that the canonical dimensions of any kind of spinor $ \psi $ in physical units are 
$ [ \psi ] = \sqrt{\mathrm{eV}}\,\mathrm{cm}^{-1}\,. $
It follows that the Noether charge is the \textsf{lepton number}, a pure number $ \ell\,, $ which can be identified as follows: namely,
\begin{equation}
\ell(t)= -\,\frac{1}{\hslash c}\int \mathrm{d}\mathbf{x}\,\chi^{\dagger}(t,\mathbf{x}) \chi(t,\mathbf{x})
\end{equation}
which satisfies the evolution equation
\begin{equation}
\frac{\mathrm{d\ell}}{\mathrm{d}t} = \frac{im}{\hslash^{2}}\int \mathrm{d}\mathbf{x}\,
\left[\,\chi^{\top}(x)\sigma_{2}\chi(x) - \chi^{\dagger}(x)\sigma_{2}\chi^{\ast}(x)\,\right]
\label{leptonnumberviolation} 
\end{equation}
that leads to the \textsf{lepton number conservation only in the massless case.}
\end{footnotesize}

Actually, as it will be discussed and clarified in the sequel, \textsf{there is a relic continuous U(1) symmetry only for Majorana massless spinors}, which drives to
the existence of a conserved pseudo-scalar charge, the meaning of which will be better focused further on.

\medskip
Another key observation is now in order. Starting with a left-handed Weyl 2-component spinor $ \chi(x)\in D(\frac12,\,0) $ one can also 
build up a \textsf{chiral left-handed bispinor} 
\[
\nu_{L}(x)=P_{L}\chi_{M}(x)=\left\lgroup
\begin{array}{c}
\chi(x)
\\
0
\end{array}
\right\rgroup\qquad\quad P_{R}\,\nu_{L}=\textstyle\frac12 (1+\gamma_{5})\nu_{L}=0
\]
together with its right-handed counterpart
\[
\nu^{\,c}_{L}(x)=\gamma^{\,2} \nu_{L}^{\,\ast}(x) 
=\left\lgroup
\begin{array}{c}
0
\\
-\sigma_{2}\chi^{\ast}(x)
\end{array}
\right\rgroup\equiv \nu_{R}(x)
\qquad\quad P_{L}\,\nu_{R}=\textstyle\frac12 (1-\gamma_{5})\nu_{R}=0
\]
which \textsf{still involves only the very same left-handed Weyl 2-component spinor} $ \chi(x)\in D(\frac12,\,0)\,. $
With the help of those two chiral left- and right-handed bispinors, one can suitably describe a \textsf{massive neutrino}, once we employ the related Lagrangian density
\begin{eqnarray}
{\mathcal L}_{L} &=&
\textstyle\frac12\,\overline{\nu}_{L}(x)\,\gamma^{\,\mu}\,i\parl_\mu\nu_{L}(x) 
- \frac12\,m\overline{\nu}_{L}(x)\nu_{R}(x) + \mathrm{c.c.}\\
&=&\textstyle\frac12\,\chi^{\dagger}(x)\,\sigma^{\,\mu}\,i\partial_\mu\chi(x) 
+ \textstyle\frac12\,m\chi^{\dagger}(x)\,\sigma_{2}\,\chi^{\ast}(x) + \mathrm{c.c.}
\end{eqnarray}
However, it turns out that the above Lagrangian does actually coincides with the Majorana Lagrangian in (\ref{MajoranaAction}) in such a manner
that the field equations for a chiral left-handed massive bispinor $\nu_{L}(x)=P_{L}\chi_{M}(x)$ are nothing but the very same as 
the massive Majorana spinor $\chi(x)\,,$ the general solutions of which will be fully discussed in the sequel, after the transition to the Majorana representation.

\medskip
The full Majorana Lagrangian in the 2-component formalism reads
\be
\widehat{\mathcal{L}}_{M}=\textstyle\frac12\,\chi^{\dagger}(x)\,\sigma^{\,\mu}\,i\partial_\mu\chi(x) 
+ \textstyle\frac12\,m\chi^{\top}(x)\,\sigma_{2}\,\chi(x) + \mathrm{c.c.}\label{Majoranaaction}
\ee
so that the Euler-Lagrange field equation becomes
\begin{equation}
i\,\sigma^{\,\mu}\,\partial_{\,\mu}\chi(x) + m\,\sigma_2\,\chi^{\ast}(x)=0
\label{MWE1} 
\end{equation}
which is known as the Majorana field equation.
Multiplying to the left by $ \sigma_{2} $ and taking complex conjugation yields
\[
i\sigma_2\,\partial_0 \chi^\ast(x) - i\sigma_2\,\sigma_k^\ast\,\partial_{k}\chi^{\ast}(x) + m\chi(x)=0
\]
Remembering that we have $ \sigma_2\,\vec{\sigma}\,\sigma_2=-\,\vec{\sigma} $ and that $ \bar{\sigma}^{\,\mu}=(1,\vec{\sigma}\,) $
we come to the equivalent form of the Majorana wave equation, \textit{viz.,}
\begin{equation}
i\bar{\sigma}^{\,\mu}\,\sigma_{2}\,\partial_{\mu}\chi^{\ast}(x) + m\chi(x)=0
\label{MWE2} 
\end{equation}
Now, if we act from the left with the operator $ i\bar{\sigma}^{\,\nu}\partial_{\nu} $ to equation (\ref{MWE1} ) and use equation
(\ref{MWE2} ) we find
\[
\bar{\sigma}^{\,\nu}\,\sigma^{\,\mu}\,\partial_{\nu}\partial_{\mu}\chi(x) + m^{2}\chi(x)
=(\,\square + m^2\,)\chi(x)=0
\]
which means that the left-handed spinor $ \chi\in D(\frac12,0)\,, $ i.e. the building block
of the self-conjugated Majorana bispinor, is actually  solution of the Klein-Gordon wave equation. 
Moreover, by understanding the Majorana bispinor to be defined by the self-conjugation constraint
$ \chi_{M}(x)=\chi_{M}^{\,c}(x) \,, $ it is easy to check that the pair of coupled wave equations
(\ref{MWE1} ) and (\ref{MWE2} ) is equivalent to the single bispinor wave equation
\begin{equation}
(\,\alpha^{\nu}i\partial_{\nu} - \beta m\,)\chi_{M}(x)=0
\label{MWE3} 
\end{equation}
where use has been made of the Dirac notation $ \beta=\gamma_0\,, $
while the Majorana Lagrangian can be recast in a further 4-component form
\begin{equation}
\mathcal{L}_{M} = \textstyle\frac14\,\chi_{M}^{\dagger}(x)\,\alpha^{\,\mu}\,i\parl_\mu\chi^{\,c}_{M}(x) 
- \frac12\,m\,\chi_{M}^{\dagger}(x)\,\beta\,\chi^{\,c}_{M}(x) 
\end{equation}
It is immediate to verify that the bispinor form (\ref{MWE3} ) of the field equations does coincide with the pair of 
functionally dependent and equivalent forms (\ref{MWE1} ) and (\ref{MWE2} ) of the Majorana spinor wave equation.
Notice that the Majorana self-conjugated bispinor transforms under the Poincar\'e group according to
\begin{equation}
\chi_{M}^{\,\prime}(x')=\Lambda_{\frac12}\,\chi_{M}^{\,c}(x)
\qquad\quad
\Lambda_{\frac12}=\left\lgroup
\begin{array}{cc}
\Lambda_{L} & 0
\\
0 & \Lambda_{R}
\end{array}
\right\rgroup
\end{equation}
with $ x'=\Lambda(x+\mathrm{a})\,. $ Hence it follows that the generators of the Lorentz group acting upon any kind of bispinor are provided 
by the usual expressions
\[
S^{\,0k}=\frac{i\hslash}{2} \gamma_{0}\gamma^{\,k}=\frac{\hslash}{2}
\left\lgroup
\begin{array}{cc}
-i\sigma_{k} & 0
\\
0 & i\sigma_{k}
\end{array}
\right\rgroup
\]
\[
S^{\,\jmath k}=\frac{i\hslash}{4} [\,\gamma^{\,\jmath}\,,\gamma^{\,k}\,]=
\textstyle\frac{1}{2}\hslash\varepsilon^{\jmath k\ell}  
\left\lgroup
\begin{array}{cc}
\sigma_{\ell} & 0
\\
0 & \sigma_{\ell}
\end{array}
\right\rgroup = \frac{1}{2}\hslash\varepsilon^{\jmath k\ell}\,  \Sigma^{\ell}
\]
in such a manner that we obtain the infinitesimal variation
\[
\Delta\chi_{M}(x)= -{\textstyle\frac12} i\sigma_{\mu\nu} \chi_{M}(x)\,\delta\omega^{\mu\nu}
\qquad\quad
\sigma_{\mu\nu}=\frac{i}{4}[\,\gamma_{\mu} , \gamma_{\nu}\,]
\]

\subsubsection{The Majorana real representation}

Let us now search the general complex solutions of the Majorana linear wave equation (\ref{MWE3}) for a self-conjugated bispinor $ \chi_{M}(x)\,. $
Once obtained, we can readily get the general complex solution for the massive chiral left-handed neutrino $\nu_{L}(x)$  after a projection on the upper component
of $ \chi_{M}(x)\,. $ To implement this task, it appears utmost convenient to employ the so called \textsf{Majorana representation} \cite{Majorana}.
It turns out that, by definition, the Majorana bispinor $ \chi_{M}(x)=\chi_{M}^{\,c}(x) $ 
must fulfill the self-conjugation constraint, which linearly relates the lower 
spinor component to the complex conjugate of the upper spinor component.
Then, \textsf{a representation must exist which makes the Majorana bispinor 
real}, with the previously introduced two independent complex functions
$\chi_a\in{\mathbb C}\ (a=1,2)$ replaced by the
four real functions
$\psi_{M,\alpha}\in{\mathbb R}\ (\alpha=1,2,3,4)\,.$
To obtain this real representation, we notice that for any complex bispinor $ \psi $ we find
\[
\psi^{c}=\gamma^{2}\psi^{\ast}\qquad\quad
\psi^{\ast}=-\gamma^{2} \psi^{c}
\]
in such a manner that for the self-conjugated complex bispinor $ \chi_{M} $ we can write
\[
\chi_{M}^{\ast}(x) = -\,\gamma^{2}\chi_{M}(x)
\]
Then a linear transformation to a real bispinor field $\psi_{M}=\psi_{M}^{\,\ast}$
can be found by noting that
$$
\chi_M = S\,\psi_{M}
\qquad\quad
\chi_M^\ast = S^{\,\ast}\,\psi_{M}=S^{\,\ast}\,S^{-1}\,\chi_M 
\qquad\quad
\gamma^{2}+S^{\,\ast} S^{-1}=0
$$
The explicit form of the unitary transformation matrix can be readily found to be
\begin{equation}
S=\frac{\sqrt2}{2}\,\left\lgroup
\begin{array}{cc}
1 & \sigma_{2}
\\
-\,\sigma_{2} & 1
\end{array}
\right\rgroup\
=\ \frac{\sqrt2}{2} (1+\gamma^{\,2})
\qquad\quad
S^{\,\dagger}=S^{-1}=\ \frac{\sqrt2}{2} (1-\gamma^{\,2})
\end{equation}
or even more explicitly
$$
S=\frac{\sqrt2}{2}\,\left\lgroup
\begin{array}{cccc}
1 & 0 & 0 & -i
\\
0 & 1 & i& 0
\\
0 & i& 1 & 0
\\
-i & 0 & 0 & 1
\end{array}
\right\rgroup
$$
From the above relation $ \psi_{M}=S^{-1}\chi_{M}=\psi_{M}^{\,\ast} $ one can immediately obtain
the correspondence rule between the complex and real forms of the self-conjugated Majorana bispinor: namely,
\begin{eqnarray}
\begin{array}{cc}
\psi_{M1}=\sqrt{2}\,\Re\mathrm{e}\,\chi_{1} & \psi_{M2}=\sqrt{2}\,\Re\mathrm{e}\,\chi_{2}
\\
\psi_{M3}= \sqrt{2}\,\Im\mathrm{m}\,\chi_{2} & \psi_{M4}=-\,\sqrt{2}\,\Im\mathrm{m}\,\chi_{1}
\end{array}
\end{eqnarray}
\begin{equation}
\begin{array}{c}
\chi_{1} = (\,\psi_{M1} - i\psi_{M4}\,)/\sqrt{2}
\\
\chi_{2} = (\,\psi_{M2} + i\psi_{M3}\,)/\sqrt{2}
\end{array}
\end{equation}
Thus we can suitably make use of the so called
{\sf Majorana representation for the gamma matrices}, which
is purely imaginary, and provided by a similarity transformation acting on the 
$ \gamma- $matrices in the Weyl representation, {\em viz.,}
$$
\gamma^{\,\mu}_M \equiv\ S^{\,\dagger}\,\gamma^{\,\mu}\,S
$$
that yields
\[
\gamma_{M}^{\,0}=
\left\lgroup
\begin{array}{cc}
-\,\sigma_{2} & 0\\
0 & \sigma_{2}
\end{array}
\right\rgroup
\qquad\quad
\gamma^{\,5}_M =
\left\lgroup
\begin{array}{cc}
0 & -\,\sigma_2
\\
-\,\sigma_2 & 0
\end{array}
\right\rgroup
\]
\[
\gamma^{\,1}_M=
\left\lgroup
\begin{array}{cc}
-\,i\sigma_{3} & 0
\\
0 & -\,i\sigma_{3}
\end{array}
\right\rgroup 
\qquad\quad
\gamma^{\,2}_M=
\left\lgroup
\begin{array}{cc}
0 & \sigma_{2}
\\
-\,\sigma_{2} & 0
\end{array}
\right\rgroup
\qquad\quad
\gamma^{\,3}_M =
\left\lgroup
\begin{array}{cc}
i\sigma_{1} & 0
\\
0 & i\sigma_{1}
\end{array}
\right\rgroup
\]
which satisfy by direct inspection the Clifford algebra
$$
\{\gamma^{\,\mu}_M\,,\,\gamma^{\,\nu}_M\}=2 g^{\,\mu\nu}
\qquad
\{\gamma^{\,\nu}_M\,,\,\gamma_M^{\,5}\}=0
$$
$$
\gamma^{\,0}_M=\beta^{\,\dagger}_M
\qquad
\gamma^{\,k}_M=-\,\gamma^{\,k\,\dagger}_M
\qquad
\gamma_{\,M}^{\,5}=\gamma_M^{\,5\,\dagger}
$$
together with 
$$
\gamma^{\,\nu}_M\ =\ -\;\gamma^{\,\nu\,\ast}_M
\qquad
\gamma_{M}^{\,5}\ =\ -\;\gamma_{M}^{\,5\,\ast}
$$
The result is that, at the place of a complex self-conjugated bispinor,
which has been constructed out of a left-handed Weyl spinor $ \chi_{M}(x)\,, $ one can safely and more suitably employ a real Majorana bispinor: namely,
\[
\chi_M(x)=\chi_M^{\,c}(x)\qquad\leftrightarrow\quad
\psi_M(x)=S^{\,\dagger} \chi_M(x)=\psi_M^\ast(x)
\]
a quite analogous construction being obviously there, 
had we started from a right-handed Weyl spinor $ \varphi\in D(0\,,\frac12)\,. $
Then, the Majorana Lagrangian and the ensuing Majorana
wave equation take the \textbf{manifestly real forms}
\begin{equation}
{\mathcal L}_M={\textstyle\frac14}\;\psi^{\top}_{\,M}(x)\,\alpha_M^{\,\nu}\,\parl_\nu\,\psi_M(x)
+ {\textstyle\frac12}\,im\;\psi^{\top}_{ M}(x)\,\beta_{M}\,\psi_M(x)
\label{MajLag} 
\end{equation}
\begin{equation}
( i\partial\!\!\!/_M-m)\psi_M(x)=0
\qquad\quad 
\psi_M(x)=\psi^{\,\ast}_M(x)
\label{MWE} 
\end{equation}
$$
\alpha_M^{\,\nu}=\gamma_M^{\,0}\,\gamma_M^{\,\nu}\qquad
\alpha_M^{\,0}={\mathbb I}\qquad\beta_{M}\equiv\gamma_M^{\,0}
$$
with the real symmetric matrices
\[
\alpha^{\,1}_{M} = \left\lgroup
\begin{array}{cc}
-\sigma_{1} & 0
\\
0 & \sigma_{1}
\end{array}
\right\rgroup
\qquad\quad
\alpha^{\,2}_{M} = \left\lgroup
\begin{array}{cc}
0  & -1
\\
-1 & 0
\end{array}
\right\rgroup
\qquad\quad
\alpha^{\,3}_{M} = \left\lgroup
\begin{array}{cc}
-\sigma_{3} & 0
\\
0 & \sigma_{3}
\end{array}
\right\rgroup
\]
which manifestly fulfill $ \vec{\alpha}_{M} = \vec{\alpha}_{M}^{\,\ast} = \vec{\alpha}_{M}^{\,\top}\,. $

\medskip

One conclusion can be drawn from the manifestly real form of the Majorana Lagrangian:
\textsf{the only residual internal symmetry of the Majorana's Action integral is the discrete
${\mathbb Z}_2$ symmetry, i.e. $\psi_M(x)\,\longmapsto -\,\psi_M(x)\,,$ - except in the massless case.}

\medskip
\begin{footnotesize}
\texttt{Singular massless limit.} Actually, \textbf{but only in the massless case, }  another internal symmetry  of the
Action integral manifests itself.  To see it let us consider the U(1) internal symmetry group

\[
\psi_{M}(x)\quad\mapsto\quad \psi^{\,\prime}_{M}(x)=\exp \left\lbrace \pm\,i\theta\,\gamma_{M}^{\,5}\right\rbrace\psi_{M}(x)
\qquad\quad(\,0\le\theta<2\pi\,)
\]
the imaginary unit being convenient to keep the reality of the transformed Majorana bispinor. From the Noether theorem we get the
corresponding real tetra-current, which satisfies the continuity equation
\[
\jmath^{\,\mu}_{5}(x)=\textstyle\frac12\,\psi_M^{\,\top}(x)\,\alpha_M^{\,\mu}\, i\gamma_M^{\,5}\,\psi_M(x)
\qquad\quad \partial\cdot\jmath_{5}(x)=0
\]
as well as the ensuing conserved pseudo-scalar charge
\[
\pm\,Q_5 = \pm\,\frac12\int \mathrm{d}\mathbf{x}\, \psi_{M}^{\,\top}(t,\mathbf{x})\,i\gamma_{M}^{\,5}\psi_M(t,\mathbf{x})
\qquad\quad\dot{Q}_{5}=0
\]
the overall $ \pm $ sign is conventional and can be deopped. To gain a  better understanding  of this charge let us return to the complex form $ \chi_M(x) $ of the self-conjugated bispinor. Then  one can can compare the above U(1) internal symmetry of the 
massless case with the ordinary phase transformation for the independent component spinor $ \chi(x)\,. $ They are the same. As a matter of fact, the phase transformation
$ \chi^{\,\prime}(x)=e^{-\,i\theta}\chi(x)  $ corresponds to the chiral transformation $ \chi_{M}^{\,\prime}(x)=e^{\,i\theta\gamma_{5}}\chi_{M}(x) $
on the complex form of the Majorana bispinor. Notice that the quantity

\[
\psi_{M}^{\,\top}\,i\gamma_{M}^{\,5}\psi_M=
\left\lgroup
\psi_{M\uparrow} \quad \psi_{M\downarrow}
\right\rgroup 
\left\lgroup
\begin{array}{cc}
0 & -\,i\sigma_{2}
\\
 -\,i\sigma_{2} & 0
\end{array}
\right\rgroup 
\left\lgroup
\begin{array}{c}
\psi_{M\uparrow} 
\\
\psi_{M\downarrow}
\end{array}
\right\rgroup 
\]
does vanish for ordinary real functions $ \psi_{M,\,\alpha}\in\mathbb{R}\ (\,\alpha=1,2,3,4\,) $ 
while in the case of anti-commuting Grassmann-valued functions we get
\[
\pm\,Q_5 = \pm\int \mathrm{d}\mathbf{x}\, \left[ \,
\psi_{M,1}(t,\mathbf{x})\psi_{M,4}(t,\mathbf{x}) - \psi_{M,2}(t,\mathbf{x})\psi_{M,3}(t,\mathbf{x}) \,\right] 
\]
the integrated quantity being nothing but than $ -\,i\chi^{\,\dagger}(x)\chi(x) $ as expected.
\end{footnotesize}

\medskip\noindent
A very important remark is now in order. The Majorana Lagrangian (\ref{MajLag}) and wave equation (\ref{MWE}) have been derived starting from
an original two-component left-handed spinor $ \chi(x) $ or, equivalently, a self-conjugated bispinor $ \chi_{M}(x)=\chi_{M}^{\,c}(x)\,. $
However, a closely related and still real Lagrange density, just leading to the very same field equation, can be immediately written even for any
arbitraty complex bispinor $ \Psi_{M}(x)\,: $ namely,
\begin{equation}
{\mathcal L}_M={\textstyle\frac12}\;\Psi_{M}^{\,\dagger}(x)\,\alpha_M^{\,\nu}\,i\parl_\nu\,\Psi_{M}(x)
- \,m\;\Psi_{M}^{\,\dagger}(x)\,\beta_{M}\,\Psi_M(x)
\label{PsiMajLag} 
\end{equation}
\begin{equation}
( i\partial\!\!\!/_M-m)\Psi_M(x)=0
\label{PsiMWE} 
\end{equation}
At variance with (\ref{MajLag}) the Lagrange density (\ref{PsiMajLag}) does enjoy the invariance under the global U(1) symmetry
transformations
$ \Psi_{M}^{\,\prime}(x)=e^{-i\theta} \Psi_{M}(x)\ (\,0\le\theta<2\pi\,) $ driving eventually to the charge or number conservation.
\textsf{It follows therefrom that the real Majorana wave equation will admit in general both real as well as complex solutions,
with quite different physical properties and behaviors.}

\subsubsection{Solution of the Majorana Wave Equation} 

One can rewrite the Majorana wave equation (\ref{MWE}) in the Schr\"{o}dinger form and in physical units
\[
i\hslash\frac{\partial\psi_{M}}{\partial t}=mc^{2}\beta_{M}\psi_{M}-\hslash c\,\alpha_{M}^{k} i\nabla_{k}\psi_{M}
= H_{M}\,\psi_M(x)
\]
whence we can read the Majorana one-particle Hamiltonian operator
\begin{equation}
H_M=\alpha_M^{\,k}\,\hat p^{\,k} c + mc^{2}\,\beta_{M}
\qquad\quad(\,\hat p^{\,k}=\,-\,i\hslash\nabla_k\,)
\end{equation}
To solve the Majorana linear and real wave equation  (\ref{MWE}) we turn to the Fourier representation
$$
\psi_M(x) = \int \frac{\mathrm{d}^4p}{(2\pi)^{3/2}}\ 
\widetilde\psi_M(p)\,\exp\{-\,ip\cdot x\}
= \int \frac{\mathrm{d}^4p}{(2\pi)^{3/2}}\ 
\widetilde\psi_M(-\,p)\,\exp\{\,ip\cdot x\}
$$ 
so that from (\ref{MWE}) we get
\begin{equation}
\left\lbrace 
\begin{array}{c}
(\,\p_M-m\,)\,\widetilde\psi_M(p)=0
\\
(\,\p_M + m\,)\,\widetilde\psi_M(-\,p)=0
\end{array}
\right. 
\qquad\quad
\p_M\equiv p_{\nu} \gamma_M^{\,\nu} 
\label{spin_states_eq} 
\end{equation}
which implies
\begin{eqnarray*}
&&\widetilde\psi_{M,\,\alpha}(p)=(\,\p_M+m\,)_{\alpha\beta}\,
\widetilde\phi_{\beta}(p)
\\
&&\left(\, p^2-m^2\,\right) \widetilde\phi_{\alpha}(p)=0
\qquad\qquad\quad (\alpha=1,2,3,4)
\\
&& \widetilde\phi_{\alpha}(p)=\delta\left(\, p^2-m^2\,\right)\,f_\alpha(p)
\end{eqnarray*}
and in turn
\begin{eqnarray*}
\widetilde\psi_{M,\,\alpha}(-\,p) &=&
(\,m - \p_{M}\,)_{\alpha\beta}\,
\widetilde\phi_{\beta}(-\,p) = (\,m - \p_{M}\,)_{\alpha\beta}\,f_{\beta}(-\,p)\,\delta\left(\, p^2-m^2\,\right)
\end{eqnarray*}
The momentum space Majorana equations (\ref{spin_states_eq}) are a pair of  algebraic equation
\[
\left\lbrace 
\begin{array}{c}
(\,p_{0}\beta_{M} - p^{\,k}\gamma^{\,k}_{M} - mc\,)\,\widetilde\psi_M(p) = 0
\\
(\,p_{0}\beta_{M} - p^{\,k}\gamma^{\,k}_{M} + mc\,)\,\widetilde\psi_M(-p) = 0
\end{array}
\right. 
\qquad\quad
c\,p_0=\pm\,\hslash\omega_{\,\bf p}=\pm\sqrt{\mathbf{p}^{2}c^{\,2} + m^{2}c^{\,4}}
\]
which can be turned into a pair of energy eigenvalue problems
\begin{equation}
\widetilde H_{M}(\mathbf{p})\,\widetilde\psi_M(p) = \pm\,\hslash\omega_{\,\bf p}\,\widetilde\psi_M(p) 
\qquad\quad
c\,p^{\,k}\alpha^{\,k}_{M}  + mc^{2}\beta_{M}\equiv \widetilde H_{M}(\mathbf{p})
\end{equation}
where $ \widetilde H_{M}(\mathbf{p}) $ is the 1-particle Majorana Hamiltonian in the momentum space representation,
which satisfies the property $ \widetilde H^{2}_{M}(\mathbf{p}) = \omega_{\,\bf p}^{\,2}\,.$

\medskip
In order  to find explicit solutions in momentum space, let us first consider the following matrix-valued functions: namely,
$$
{\mathcal E}_M^{\,\pm}(\mathbf{p})=\left(\, mc\pm\p_M\right)/2mc
\qquad\qquad \mathrm{with}\quad c\,p_0=\pm\hslash\omega_{\,\bf p}=\pm\sqrt{\mathbf{p}^{2}c^{\,2} + m^{2}c^{\,4}}
$$
which do satisfy the following relationships
$$
[\,{\mathcal E}_M^{\,\pm}(\mathbf{p})\,]^{\,\ast}={\mathcal E}_M^{\,\mp}(\mathbf{p})
\qquad\quad
[\,{\mathcal E}_M^{\,\pm}(\mathbf{p})\,]^{\dagger}={\mathcal E}_M^{\,\pm}(-\,\mathbf{p})
$$
\[
[\,{\mathcal E}_M^{\pm}(\mathbf{p})\,]^{2}={\mathcal E}_M^{\,\pm}(\mathbf{p})
\qquad\quad
{\mathcal E}_M^{\,\pm}(\mathbf{p})\,{\mathcal E}_M^{\,\mp}(\mathbf{p})=0
\]
\[
\mathrm{tr}\,{\mathcal E}_M^{\,\pm}(\mathbf{p})=2\qquad\quad
{\mathcal E}_M^{\,+}(\mathbf{p})+{\mathcal E}_M^{\,-}(\mathbf{p})=\mathbb I
\]
The above matrix-valued functions can be fully understood as projectors, if we simultaneously perform the matrix Hermitean
conjugation and the full inversion, or parity transform, on the function variable $ \mathbf{p}\,\rightarrow\,-\,\mathbf{p}\,. $
The physical meaning of those projectors can be better featured and focused on the Majorana particle rest frame $ \mathbf{p}=0\,, $ 
in which $ \p_M=mc\beta_M\,. $
To this concern, we recall the spin operators $ S_{k}=\frac12\hslash \Sigma_{k}\ (\,k=1,2,3\,)$ in the Majorana representation where
\[
\Sigma_{M,1}=i\gamma^{\,2}_M\,\gamma_M^{\,3}=\left\lgroup
\begin{array}{cc}
0 & i\sigma_3
\\
-\,i\sigma_3 & 0
\end{array}
\right\rgroup
\qquad\quad
\Sigma_{M,\,3}=i\gamma^{\,1}_M\,\gamma_M^{\,2}=
\left\lgroup
\begin{array}{cc}
0 & -\,i\sigma_1
\\
i\sigma_1 & 0
\end{array}
\right\rgroup
\]
\[
\Sigma_{M,\,2}=i\gamma^{\,3}_M\,\gamma_M^{\,1}=
\left\lgroup
\begin{array}{cc}
\sigma_2 & 0
\\
0 & \sigma_2
\end{array}
\right\rgroup
\]
From the relationships
\[
\beta_{M}^{\,2}=1=\Sigma_{M,\,2}^{\,2}\qquad\quad[\,\beta_{M}\,,\,\Sigma_{M,\,2}\,]=0
\]
it appears that one can actually set up a complete orthogonal tetrad of  bispinors, 
which are common eigenvectors of the Hermitean commuting matrices $ \beta_{M} $
and $ \Sigma_{M,\,2}\,. $
It turns out that the spin states $\xi_{\,r}\ (\,r=\pm\,) $ are the pair of
degenerate eigenstates of the Majorana 1-particle Hamiltonian $ \widetilde H_{M}(\mathbf{p}) $
in the particle rest frame
${\bf p}=0\,,$ with positive eigenvalue $p_0=m$
and with opposite spin projections on the $OY$ axis, {\em viz.,}
\[
\beta_M\,\xi_{\,\pm}=\xi_{\,\pm}\qquad\quad
\left(\,\Sigma_{M,\,2}\mp1\,\right)\xi_{\,\pm}=0
\qquad\quad
\xi_{\,\mp}^{\,\dagger}\,\xi_{\,\pm}=0
\]
In the very same way we obtain
\[
\beta_M\,\eta_{\,\pm}= -\,\eta_{\,\pm}\qquad\quad
\left(\,\Sigma_{M,\,2}\mp1\,\right)\eta_{\,\pm}=0
\qquad\quad
\eta_{\,\mp}^{\,\dagger}\,\eta_{\,\pm}=0
\]
which means that the spin states $\eta_{\,r}\ (\,r=\pm\,) $ are in fact the two 
degenerate eigenstates of the Majorana 1-particle Hamiltonian in the particle rest frame
${\bf p}=0\,,$ with negative eigenvalue $p_0= -\, m$
and with opposite spin projections on the $OY$ axis.
Altogether we can write
\[
\xi_{\,r}^{\,\dagger}\,\xi_{\,s}
=2\delta_{\,rs}=\eta_{\,r}^{\dagger}\,\eta_{\,s}
\qquad\quad
\xi_{\,r}^{\,\dagger}\,\eta_{\,s} = 0
\qquad\quad
\forall\,r,s=\pm
\]
Notice that, from the relations $ \lbrace \gamma_{5},\gamma_{0} \rbrace=0= [\,\gamma_{5},\Sigma_2\,] $, it follows that the chiral matrix
$ \gamma_{5} $ maps the positive energy space into the negative one, and viceversa, in the Majorana particle rest frame $ \mathbf{p}=0\,. $
Hence, without loss of generality, we can always assume 
\[
\gamma_{M}^{\,5}\,\xi_{\,\pm} = \eta_{\,\pm}\qquad\quad\gamma_{M}^{\,5}\,\eta_{\,\pm} = \xi_{\,\pm}
\]
In general, the freedom in the choice of an orthogonal pair, in the positive and negative energy planes of the rest frame $ \mathbf{p}=0\,, $ 
is provided by the $ SO(2,\mathbb{R}) $ Abelian symmetry little group
\[
F_{\pm}(\theta)=\exp \lbrace \pm\,i\Sigma_{M,\,2}\,\theta\rbrace=\mathbf{I}\cos\theta \pm \gamma^{\,1}_{M}\gamma^{\,3}_{M}\sin\theta
\qquad\quad[\, F_{\pm}(\theta)\,]^{-1}=[\,F_{\pm}(\theta)\,]^{\top}=F_{\pm}(-\,\theta)
\]
because $ \beta^{\,\top}_{M}+\beta_{M}=0\,,\ \gamma^{\,k\top}_{M} =\gamma^{\,k}_{M}\,, $ in such a manner that 
\begin{equation}
\xi^{\,\prime}_{\,r}=\sum_{s=+,-}[\,F_{\,+}(\theta)\,]_{rs}\,\xi_{\,s}
\qquad\quad
\eta^{\,\prime}_{\,r}=\sum_{s=+,-}[\,F_{\,-}(\vartheta)\,]_{rs}\,\eta_{\,s}
\label{rotaspin} 
\end{equation}
To sum up, one can set up the positive and negative energy spaces projectors $ \varepsilon_{\,\pm} $, 
in the rest frame of the massive Majorana particle, which satisfy
\begin{eqnarray}
\varepsilon_{\,+} = \textstyle\frac12 ( \xi_{\,+} \otimes\xi_{\,+}^{\,\dagger} + \xi_{\,-} \otimes\xi_{\,-}^{\,\dagger} )
\qquad\quad
\varepsilon_{\,-} = \textstyle\frac12 ( \eta_{\,+} \otimes\eta_{\,+}^{\,\dagger} + \eta_{\,-} \otimes\eta_{\,-}^{\,\dagger} )
\\
\varepsilon_{\,\pm}^{\,2}=\varepsilon_{\,\pm}
\qquad\quad \varepsilon_{\,\pm}\,\varepsilon_{\,\mp} = 0
\qquad\quad \varepsilon_{\,+} + \varepsilon_{\,-} = 1
\qquad\quad \mathrm{tr}\,\varepsilon_{\,\pm}=2
\end{eqnarray}
Now, in order to move from the rest frame to an arbitrary inertial frame we can proceed as follows: consider the so called spin-states of
positive and negative energy, viz.,
\begin{equation}
\left\lbrace 
\begin{array}{c}
u_{\,r}(\mathbf{p}) \equiv 2m (2\omega_{\,\bf p}+2m)^{-\frac12}\,{\mathcal E}_M^{\,+}(\mathbf{p})\,\xi_{\,r}
\\ \\
v_{\,r}(\mathbf{p}) \equiv 2m (2\omega_{\,\bf p}+2m)^{-\frac12}\,{\mathcal E}_M^{\,-}(\mathbf{p})\,\eta_{\,r}
\end{array}
\right. 
\qquad(\,r=+,-\,)
\end{equation}
which again satisfy 
\[
\gamma_{M}^{\,5}\,u_{\,r}(\mathbf{p}) = v_{\,r}(\mathbf{p}) 
\qquad\quad
\gamma_{M}^{\,5}\,v_{\,r}(\mathbf{p}) = u_{\,r}(\mathbf{p}) 
\qquad\quad
(\,r=+,-\,)
\]
The above spin-states  do fulfill by construction 
\begin{equation}
\left\lbrace 
\begin{array}{c}
2m\,{\mathcal E}_M^{\,-}(\mathbf{p})\,u_{\,r}(\mathbf{p}) = (m - \p_{M})\,u_{\,r}(\mathbf{p}) = 0
\\ \\
2m\,{\mathcal E}_M^{\,+}(\mathbf{p})\,v_{\,r}(\mathbf{p}) = (m + \p_{M})\,v_{\,r}(\mathbf{p}) = 0
\end{array}
\right. 
\qquad\quad (\,p_{0} = \omega_{\bf p}\,,\ r=+,-\,)
\label{projector-} 
\end{equation}
and they are the two pair of degenerate eigenstates of the positive and negative energy projectors
\begin{equation}
{\mathcal E}_M^{\,+}(\mathbf{p})\,u_{\,r}(\mathbf{p}) = u_{\,r}(\mathbf{p})
\qquad\quad
{\mathcal E}_M^{\,-}(\mathbf{p})\,v_{\,r}(\mathbf{p}) = v_{\,r}(\mathbf{p})
\qquad\quad(\,r=+,-\,)
\label{projector+} 
\end{equation}
as well as of the Majorana 1-particle Hamiltonian in momentum space, viz.,
\begin{equation}
\widetilde H_{M}(\mathbf{p})\,u_{\,r}(\mathbf{p}) = \omega_{\,\bf p}\,u_{\,r}(\mathbf{p})
\qquad\quad
\widetilde H_{M}(-\,\mathbf{p})\,v_{\,r}(\mathbf{p}) = -\,\omega_{\,\bf p}\,v_{\,r}(\mathbf{p})
\qquad\quad(\,r=+,-\,)
\end{equation}
By direct inspection one can readily verify the orthogonality relations
\begin{equation}
\left\lbrace 
\begin{array}{cc}
u^{\,\dagger}_{\,r}(\mathbf{p})\,u_{\,s}(\mathbf{p}) = 2\omega_{\,\bf p}\,\delta_{\,rs}
& \quad\overline{u}_{\,r}(\mathbf{p})\,u_{\,s}(\mathbf{p}) = 2m\,\delta_{\,rs}
\\ \\
v^{\,\dagger}_{\,r}(\mathbf{p})\,v_{\,s}(\mathbf{p}) = 2\omega_{\,\bf p}\,\delta_{\,rs}
& \quad\overline{v}_{\,r}(\mathbf{p})\,v_{\,s}(\mathbf{p}) = -\,2m\,\delta_{\,rs}
\end{array}
\right. 
\qquad\quad(\,r,s=+,-\,)
\label{orthospin} 
\end{equation}
together with
\begin{equation}
u^{\,\dagger}_{\,r}(\mathbf{p})\,v_{\,s}(-\,\mathbf{p}) = \overline{u}_{\,r}(\mathbf{p})\,v_{\,s}(\mathbf{p}) = 0
\qquad\quad(\,r,s=+,-\,)
\end{equation}

\begin{footnotesize}
For example we have
\begin{eqnarray*}
&& v^{\,\dagger}_{\,r}(\mathbf{p})\,v_{\,s}(\mathbf{p}) =
(2\omega_{\,\bf p}+2m)^{-1}\,\eta^{\,\dagger}_{\,r}\,( m -\tilde{\p} )\,
( m - \p )\,\eta_{\,s}
\\
&=& \omega_{\,\bf p}\,(\omega_{\,\bf p}+m)^{-1}\,
\eta^{\,\dagger}_{\,r}\,( \omega_{\,\bf p} + \vec{p}\cdot\vec{\gamma}\,\beta_{M} - m\beta_{M} )\,\eta_{\,s}
= 2\omega_{\,\bf p}\,\eta^{\,\dagger}_{\,r}\,\eta_{\,s} = 2\omega_{\,\bf p}\,\delta_{\,rs}
\end{eqnarray*}
where use has been made of the reations $ \beta_{M}\eta_{\,s}=-\,\eta_{\,s} $ and $ \eta^{\,\dagger}_{\,r}\,\vec{\gamma}_{M}\,\eta_{\,s} \equiv 0\,. $
Furthermore we get
\begin{eqnarray*}
&&\overline{u}_{\,r}(\mathbf{p})\,u_{\,s}(\mathbf{p}) =
(2\omega_{\,\bf p}+2m)^{-1}\,\xi^{\,\dagger}_{\,r}\,( m +\tilde{\p} )\,\beta_{M}\,( m + \p )\,\xi_{\,s}
\\
&=& (2\omega_{\,\bf p}+2m)^{-1}\,\xi^{\,\dagger}_{\,r}\,( m + \p )^{2}\xi_{\,s}
=2m(2\omega_{\,\bf p}+2m)^{-1}\,\xi^{\,\dagger}_{\,r}\,( m + \p )\,\xi_{\,s} = 2m\,\delta_{\,rs}
\\ \\
&&\overline{u}_{\,r}(\mathbf{p})\,v_{\,s}(\mathbf{p}) =
(2\omega_{\,\bf p}+2m)^{-1}\,\xi^{\,\dagger}_{\,r}\,( m +\tilde{\p} )\,\beta_{M}\,( m - \p )\,\eta_{\,s}
\\
&=& (2\omega_{\,\bf p}+2m)^{-1}\,\xi^{\,\dagger}_{\,r}\,( m + \p )( m - \p )\,\eta_{\,s}\equiv 0
\\ \\
&& u^{\,\dagger}_{\,r}(\mathbf{p})\,v_{\,s}(-\,\mathbf{p}) =
(2\omega_{\,\bf p}+2m)^{-1}\,\xi^{\,\dagger}_{\,r}\,( m +\tilde{\p} )\,
( m - \tilde{\p} )\,\eta_{\,s} \equiv 0
\end{eqnarray*}
Finally, from the relation $ \beta_{M}\eta_{\pm} = - \eta_{\pm} $  we get for $ N=(2\omega_{\,\bf p}+2m)^{-1/2} $
\begin{eqnarray*}
v_{\,\pm}(\mathbf{p}) &=& -\,N ( m -\p_{M} ) \beta_{M}\,\eta_{\pm}
= N ( \omega_{\,\bf p} - m\beta_{M}  + p^{k}\alpha^{\,k}_{M} ) \,\eta_{\pm}
= N [\, \omega_{\,\bf p} - \widetilde H_{M}(-\,\mathbf{p})\,]\,\eta_{\pm}
\end{eqnarray*}
in such a manner that from the relationship $ \widetilde H^{2}_{M}(-\,\mathbf{p}) = \omega^{\,2}_{\,\bf p} $ we can immediately obtain
\begin{eqnarray*}
\widetilde H_{M}(-\,\mathbf{p})\,v_{\,r}(\mathbf{p}) = N\,\omega_{\,\bf p} [\,\widetilde H_{M}(-\,\mathbf{p}) - \omega_{\,\bf p}\,]\,\eta_{\,r}
= -\,\omega_{\,\bf p}\,v_{\,r}(\mathbf{p})\qquad\quad(\,r=+,-\,)
\end{eqnarray*}
as expected.
\end{footnotesize}

\medskip
Moreover, it is possible to prove the closure relations
\begin{equation}
\left\lbrace 
\begin{array}{c}
\sum_{r=+,-} u_{\,r}(\mathbf{p}) \otimes \overline{u}_{\,r}(\mathbf{p}) =\p + m
\\ \\
\sum_{r=+,-} v_{\,r}(\mathbf{p}) \otimes \overline{v}_{\,r}(\mathbf{p}) = \p - m
\end{array}
\right. 
\qquad\quad (\,p_{0} = \omega_{\bf p}\,)
\label{closurespinstates} 
\end{equation}

\medskip
\begin{footnotesize}
\texttt{Proof}. Consider for instance the closure relation for the negative energy spin-states. It turns out that
\begin{eqnarray*}
\sum_{r=+,-} v_{\,r}(\mathbf{p}) \otimes \overline{v}_{\,r}(\mathbf{p})\,v_{\,s}(\mathbf{p}) = -\,2m\,v_{\,s}(\mathbf{p}) 
\qquad\quad (\p-m) v_{\,s}(\mathbf{p}) = -\,2m\,\mathcal{E}_{M}^{-}(\mathbf{p})\,v_{\,s}(\mathbf{p}) = -\,2m\,v_{\,s}(\mathbf{p}) 
\end{eqnarray*}
\[
\sum_{r=+,-} v_{\,r}(\mathbf{p}) \otimes \overline{v}_{\,r}(\mathbf{p})\,u_{\,s}(\mathbf{p}) = 0 =  (\p-m) u_{\,s}(\mathbf{p}) 
\]
which means that the two matrix-valued functions on both sides of the above equality do share the same rank two, the same degenerate
eigenvalues $ (-\,2m, 0) $ and the corresponding very same eigenvectors.
A quite analogous conclusion holds obviously true even concerning closure relation for the positive energy spin-states. It follows that the matrix-valued 
functions in the equality (\ref{closurespinstates}) are proportional. However, if we multiply by $ \beta_{M} $ from the right both closure relations 
and take the trace, then we get the same result $ 4\omega_{\bf p}\,, $ which means that the proportionality constant is one, as required.

\medskip\noindent
As a corollary of the closure relations, after direct substitution, we come to the nontrivial general identity
\begin{equation}
( m\pm\p )\,\varepsilon_{\,\pm}\,( m \pm \p )\,=  ( \p \pm m )\,( \omega_{\bf p} + m )
\end{equation}
\end{footnotesize}

\medskip
Concerning the Majorana spinor wave functions, it turns out that a 
\textsf{complete orthonormal set of  complex plane wave bispinor solutions}
of the Majorana wave equation (\ref{MWE}) is provided by
\begin{equation}
\left\lbrace 
\begin{array}{c}
u_{\,{\bf p}\,,\,r}(x)\equiv
[\,(2\pi)^3 2\omega_{\,\bf p}\,]^{-\frac12}\,u_{\, r}(\mathbf{p})\,
\exp\{-\,i\omega_{\,\bf p}t + i{\bf p}\cdot{\bf x}\}
\\
v_{\,{\bf p}\,,\,r}(x)\equiv
[\,(2\pi)^3 2\omega_{\,\bf p}\,]^{-\frac12}\,v_{\, r}(\mathbf{p})\,
\exp\{\,i\omega_{\,\bf p}t - i{\bf p}\cdot{\bf x}\}
\end{array}
\right. 
\qquad\quad(\,r=+,-\,)
\label{spinornormalmodes} 
\end{equation}
which fulfill the energy eigenvalue problem
\begin{equation}
H_{M}\,u_{\,{\bf p}\,,\,r}(x) = \omega_{\,\bf p}\,u_{\,{\bf p}\,,\,r}(x)
\qquad\quad
H_{M}\,v_{\,{\bf p}\,,\,r}(x) = -\,\omega_{\,\bf p}\,v_{\,{\bf p}\,,\,r}(x)
\end{equation}
as well as the orthogonality and closure relations: namely,
\begin{equation}
\int\mathrm{d}\mathbf{x}\,u^{\,\dagger}_{\,{\bf p}\,,\,r}(x)\,u_{\,{\bf q}\,,\,s}(x)  = 
\delta_{\,rs}\,\delta(\mathbf{p} - \mathbf{q})
= \int\mathrm{d}\mathbf{x}\,v^{\,\dagger}_{\,{\bf p}\,,\,r}(x)\,v_{\,{\bf q}\,,\,s}(x)
\label{orthodag} 
\end{equation}
\begin{equation}
\begin{array}{c}
i\sum_{r=+,-}\int\mathrm{d}\mathbf{p}\, 
u_{\,{\bf p}\,,\,r}(x)\,\otimes\,\overline{u}_{\,{\bf p}\,,\,r}(y) = ( i\partial\!\!\!/_M+m )\,D^{(-)}(x-y)
\\ \\
i\sum_{r=+,-}\int\mathrm{d}\mathbf{p}\, 
v_{\,{\bf p}\,,\,r}(x)\,\otimes\,\overline{v}_{\,{\bf p}\,,\,r}(y) = ( i\partial\!\!\!/_M+m )\,D^{(+)}(x-y)
\end{array}
\label{closure} 
\end{equation}
where $ D^{(\pm)}(x-y) $ are the well known  tempered distribution
\[
D^{(\pm)}(x-y)=\frac{\pm1}{i(2\pi)^3}\int\mathrm{d}^{4}p\,\exp \lbrace \pm\,ip\cdot (x-y) \rbrace\, \delta( p^{2} - m^{2} )\, \theta(p_0)
\]

\medskip
\begin{small}
\texttt{Proof}. By direct replacement we find
\begin{eqnarray*}
&&i\sum_{r=+,-}\int\mathrm{d}\mathbf{p}\, 
u_{\,{\bf p}\,,\,r}(x)\,\otimes\,\overline{u}_{\,{\bf p}\,,\,r}(y) 
\\
&=& i\sum_{r=+,-}\int\mathrm{d}\mathbf{p}\, [\,(2\pi)^3 2\omega_{\,\bf p}\,]^{-1}
u_{\,r}(\mathbf{p}) \otimes \overline{u}_{\,r}(\mathbf{p})\,
\exp\{-\,i\omega_{\,\bf p} (x_0 - y_0) + i{\bf p}\cdot ( {\bf x} - \mathbf{y} )\}
\\
&=& i\,(\,i\gamma_{M}^{\,\mu} \partial_{\mu,x} + m\, )
\int\mathrm{d}\mathbf{p}\, [\,(2\pi)^3 2\omega_{\,\bf p}\,]^{-1} \exp\{-\,i\omega_{\,\bf p} (x_0 - y_0) + i{\bf p}\cdot ( {\bf x} - \mathbf{y} )\}
\\
&=&\frac{i}{(2\pi)^3}\,( m + i\partial\!\!\!/_M ) 
\int\mathrm{d}^{4}p\,\exp \lbrace -ip\cdot (x-y) \rbrace\, \delta( p^{2} - m^{2} )\, \theta(p_0)
=  ( i\partial\!\!\!/_M+m )\,D^{(-)}(x-y)
\\ \\
&&i\sum_{r=+,-}\int\mathrm{d}\mathbf{p}\, 
v_{\,{\bf p}\,,\,r}(x)\,\otimes\,\overline{v}_{\,{\bf p}\,,\,r}(y) 
\\
&=& i\sum_{r=+,-}\int\mathrm{d}\mathbf{p}\, [\,(2\pi)^3 2\omega_{\,\bf p}\,]^{-1}
v_{\,r}(\mathbf{p}) \otimes \overline{v}_{\,r}(\mathbf{p})\,
\exp\{\,i\omega_{\,\bf p} (x_0 - y_0) - i{\bf p}\cdot ( {\bf x} - \mathbf{y} )\}
\\
&=& -\,i\,(\,i\gamma_{M}^{\,\mu} \partial_{\mu,x} + m\, )
\int\mathrm{d}\mathbf{p}\, [\,(2\pi)^3 2\omega_{\,\bf p}\,]^{-1} \exp\{\,i\omega_{\,\bf p} (x_0 - y_0) - i{\bf p}\cdot ( {\bf x} - \mathbf{y} )\}
\\
&=&\frac{-\,i}{(2\pi)^3}\,( m + i\partial\!\!\!/_M ) 
\int\mathrm{d}^{4}p\,\exp \lbrace ip\cdot (x-y) \rbrace\, \delta( p^{2} - m^{2} )\, \theta(p_0)
=  ( i\partial\!\!\!/_M+m )\,D^{(+)}(x-y)
\end{eqnarray*}
\end{small}

\medskip
Now a key observation is in order, for the genius of Ettore Majorana to be fully recognized and appreciated.
It turns out that \textsf{only in the Majorana representation of the purely imaginary Dirac matrices},
two further complete orthonormal sets of positive and negative frequency solutions
of the field equation (\ref{MWE}) can be immediately obtained by complex conjugation of the previously obtained ones.
As a matter of fact we find
\begin{eqnarray*}
v^{\ast}_{\,r}(\mathbf{p}) \equiv 2m (2\omega_{\,\bf p}+2m)^{-\frac12}\,
{\mathcal E}_M^{\,+}(p)\,\eta_{\,r}^{\ast}
\qquad(\,r=\pm\,\vee\,p_0=\omega_{\,\bf p}\,)
\end{eqnarray*}
which manifestly satisfies 
\[
\widetilde H_{M}(\mathbf{p})\,v^{\ast}_{\,\pm}(\mathbf{p}) = \omega_{\,\bf p}\,v^{\ast}_{\,\pm}(\mathbf{p})
\]
owing to $ \beta_{M} \eta^{\ast}_{\pm} = \eta^{\ast}_{\pm} $ and $ \beta_{M} \xi^{\ast}_{\pm} = -\,\xi^{\ast}_{\pm}\,. $
In a quite analogous way, the spin states
\[
u^{\ast}_{\,r}(\mathbf{p}) \equiv 2m (2\omega_{\,\bf p}+2m)^{-\frac12}\,
{\mathcal E}_M^{\,-}(p)\,\xi^{\ast}_{\,r}
\qquad(\,r=\pm\,\vee\,p_0=\omega_{\,\bf p}\,)
\]
do satisfy
\[
\widetilde H_{M}(-\,\mathbf{p})\,u^{\ast}_{\,\pm}(\mathbf{p}) = -\,\omega_{\,\bf p}\,u^{\ast}_{\,\pm}(\mathbf{p})
\]
Notice \textit{en passant}  that, e.g., $ \gamma_{M}^{\,5}\,u^{\ast}_{\,\pm}(\mathbf{p})=-\,v^{\ast}_{\,\pm}(\mathbf{p})\,, $
whilst for the adjoint spin-states we find 
$ \overline{u}_{\,\pm}^{\,\ast} (\mathbf{p}) = [ \,u^{\dagger}_{\,\pm}(\mathbf{p})\,\beta_{M}\,] ^{\ast} = -\,u^{\top}_{\,\pm}(\mathbf{p})\,\beta_{M}\,.$
It turns out that any spin-state is truly orthogonal to its complex conjugate because, e.g.,
\[
\overline{u}_{\,r}(\mathbf{p})\,u^{\ast}_{\,s}(\mathbf{p}) =
\xi_{\,r}^{\,\dagger} (m+\tilde{\p})\,\beta_{M}\,(m-\p)\,\xi_{\,s}^{\,\ast}\,(\,2m+2\omega_{\,\bf p}\,)^{-1}\equiv0
\qquad\quad(\,r,s=+,-\,)
\]
while the closure relation can be directly obtained from the complex conjugation of eq.s (\ref{closurespinstates}) 
\begin{equation}
\sum_{r=+,-} u^{\ast}_{\,r}(\mathbf{p}) \otimes \overline{u}^{\,\ast}_{\,r}(\mathbf{p}) =
-\sum_{r=+,-} u^{\,\ast}_{\,r}(\mathbf{p}) \otimes u^{\,\top}_{\,r}(\mathbf{p})\,\beta_{M}
= -\,\p + m
\label{ccclosure} 
\end{equation}

\medskip
It follows therefrom that the \textbf{most general classical solutions} of the real Majorana wave field equation 
(\ref{MWE}) might appear in both real and complex forms: namely,
\begin{equation}
\psi_{M}(x) = \sum_{r=+,-}\int\mathrm{d}\mathbf{p}\
a_{\,{\bf p}\,,\,r}\,u_{\,{\bf p}\,,\,r}(x) + \mathrm{c.\,c.} =\psi^{\ast}_{M}(x)
\label{Majorana1} 
\end{equation}
\begin{equation}
\widehat{\psi}_{M}(x)=\sum_{r=+,-}\int\mathrm{d}\mathbf{p}\
b_{\,{\bf p}\,,\,r}\,v^{\,\ast}_{\,{\bf p}\,,\,r}(x) + \mathrm{c.\,c.} =\widehat{\psi}^{\,\ast}_{M}(x)
\label{Majorana2} 
\end{equation}
\begin{equation}
\Psi_{D}(x) = \sum_{r=+,-}\int\mathrm{d}\mathbf{p}\
[\,c_{\,{\bf p}\,,\,r}\,u_{\,{\bf p}\,,\,r}(x) + d_{\,{\bf p}\,,\,r}^{\,\ast}\,v_{\,{\bf p}\,,\,r}(x)\,]
\label{Dirac} 
\end{equation}
where all the coefficients $ a_{\,{\bf p}\,,\,\pm},b_{\,{\bf p}\,,\,\pm},c_{\,{\bf p}\,,\,\pm},d_{\,{\bf p}\,,\,\pm} $ are Grassmann valued complex numbers.
As already noticed, the dynamics and the symmetry properties related to the above real and complex solutions of the Majorana 
real wave field equations (\ref{MWE}) and (\ref{PsiMWE}) will be quite different and will be analyzed in the sequel, after the transition
to the quantum theory. The complex solutions of the Majorana real wave equation are usually called Dirac bispinors, or spinors, whilst the real 
solutions are the self-conjugated Majorana bispinors, or spinors.

\subsubsection{Canonical Quantum Theory}

The transition to the quantum theory is achieved by passing from the complex Grassmann valued c-number coefficients
$ a_{\,{\bf p}\,,\,\pm},b_{\,{\bf p}\,,\,\pm},c_{\,{\bf p}\,,\,\pm},d_{\,{\bf p}\,,\,\pm} $ to the corresponding creation-annihilation operators,
which satisfy the canonical anti-commutation relations, viz.,
\begin{eqnarray}
\lbrace\,a_{\,{\bf p}\,,\,r}\,,\,a^{\dagger}_{\,{\bf q}\,,\,s}\,\rbrace = \delta(\mathbf{p} - \mathbf{q})\,\delta_{\,rs}
\qquad\quad
\lbrace\,b_{\,{\bf p}\,,\,r}\,,\,b^{\,\dagger}_{\,{\bf q}\,,\,s}\,\rbrace = \delta(\mathbf{p} - \mathbf{q})\,\delta_{\,rs}
\\
\lbrace\,c_{\,{\bf p}\,,\,r}\,,\,c^{\,\dagger}_{\,{\bf q}\,,\,s}\,\rbrace = \delta(\mathbf{p} - \mathbf{q})\,\delta_{\,rs}
\qquad\quad
\lbrace\,d_{\,{\bf p}\,,\,r}\,,\,d^{\,\dagger}_{\,{\bf q}\,,\,s}\,\rbrace = \delta(\mathbf{p} - \mathbf{q})\,\delta_{\,rs}
\end{eqnarray}
all the other anti-commutators being null. In so doing, it appears that the Majorana quantum fields, which arise out of the real classical solutions
$ \psi_{M}(x) $ and $ \widehat{\psi}_{M}(x)\,, $ will become Hermitean and neutral quantum fields, whilst that one arising from
$ \Psi_{M}(x) $ will represent a charged quantum field.
Consider in fact the neutral, Hermitean, self-conjugated quantum spinor field
\begin{equation}
\psi_M(x) = \sum_{\,{\bf p}\,,\,r}\,\Big[\,
a_{\,{\bf p}\,,\,r}\,u_{\,{\bf p}\,,\,r}(x)
+ a_{\,{\bf p}\,,\,r}^{\,\dagger}\,u_{\,{\bf p}\,,\,r}^{\,\ast}(x)\,\Big]
= [\,\psi_M^{\,\dagger}(x)\,]^{\top} \equiv\ \psi^{\,c}_M(x) 
\end{equation}
\begin{equation}
\overline{\psi}_{M}(x) =\ \psi_{M}^{\,\dagger}(x)\,\beta_{M}=
\sum_{\,{\bf p}\,,\,r}\,\Big[\,
a^{\,\dagger}_{\,{\bf p}\,,\,r}\,\overline{u}_{\,{\bf p}\,,\,r}(x)
+ a_{\,{\bf p}\,,\,r}\,u_{\,{\bf p}\,,\,r}^{\,\top}(x)\,\beta_{M}\,\Big] = \psi_{M}^{\,\top}(x)\,\beta_{M}
\end{equation}
in which we have introduced the shorthand notation
\[
\sum_{\,{\bf p}\,,\,r}\equiv\int\mathrm{d}\mathbf{p}\sum_{r=+,-}
\]
Here the dagger symbol means simultaneously full Hermitean conjugation in the Fock space, as well as 
in the complex bispinor part. 
Let us first evaluate its charge or lepton number
\begin{eqnarray*}
&& \int\mathrm{d}\mathbf{x}\,\rho(x)\,=\, \int\mathrm{d}\mathbf{x}\,:\psi_{M}^{\,\dagger}(x)\,\psi_{M}(x):
\\ 
&=& \int\mathrm{d}\mathbf{x}\,
\sum_{\,{\bf p}\,,\,r}\sum_{\,{\bf q}\,,\,s} : \Big[\,
a^{\,\dagger}_{\,{\bf p}\,,\,r}\,{u}^{\,\dagger}_{\,{\bf p}\,,\,r}(x)
+ a_{\,{\bf p}\,,\,r}\,u_{\,{\bf p}\,,\,r}^{\,\top}(x)\,\Big]\,
\Big[\,a_{\,{\bf q}\,,\,s}\,u_{\,{\bf q}\,,\,s}(x)
+ a_{\,{\bf q}\,,\,s}^{\,\dagger}\,u_{\,{\bf q}\,,\,s}^{\,\ast}(x)\,\Big] :
\\
&=& \sum_{\,{\bf p}\,,\,r}\sum_{\,{\bf q}\,,\,s} 
\delta(\mathbf{p} - \mathbf{q})\,\delta_{\,rs}\,
: a^{\,\dagger}_{\,{\bf p}\,,\,r}\,a_{\,{\bf q}\,,\,s} + a_{\,{\bf p}\,,\,r}\,a_{\,{\bf q}\,,\,s}^{\,\dagger} :
\ = \sum_{\,{\bf p}\,,\,r}  \left( a^{\,\dagger}_{\,{\bf p}\,,\,r}\,a_{\,{\bf p}\,,\,r}  - a^{\,\dagger}_{\,{\bf p}\,,\,r}\,a_{\,{\bf p}\,,\,r}\right) 
\equiv 0
\end{eqnarray*}
which endorses the neutral nature of the Hermitean and self-conjugate Majorana spinor quantum field. Needless to say, the above equality is a formal one and 
deserves a regularization procedure - such as the point-splitting e.g. - but it is eventually and essentially true, thanks to the normal ordering prescription and the 
canonical anticommutation relations.

Let us determine the canonical anticommutators for the Hermitean Majorana quantum field, which is an operator valued tempered distribution. 
In order to be as clear as possible, it is better to recall that the Majorana spinor quantum field is a Hermitean operator valued tempered distribution,
which also lies in the four dimensional bispinor space $ \psi_{M,\,\alpha}(x)\ (\,\alpha=1,2,3,4\,)\,. $
More precisely, it turns out that the Majorana bispinor quantum field do actually live in a two dimensional subspace of the four dimensional linear bispinor space,
as it satisfies the constraints
\begin{eqnarray}
\mathfrak{E}_M^{\,-}\,\psi_{M}(x)=0\qquad\quad\mathfrak{E}_M^{\,+}\,\psi_{M}(x) =\ \psi^{\,c}_{M}(x)
\end{eqnarray}
\begin{equation}
\mathfrak{E}_M^{\,\pm} \equiv (\, m \pm i\partial\!\!\!/_M \,)/2m
\end{equation}
thanks to the basic relations (\ref{projector-}) and (\ref{projector+}). It is worthwhile to remark that the rank-two differential operators
$ \mathfrak{E}_M^{\,\pm}  $  \textsf{are real in the Majorana representation}, owing to the purely imaginary nature of the gamma matrices, so that
\[
\left[\, \mathfrak{E}_M^{\,\pm} \,\right] ^{\ast} = (\,m\mp i\gamma_{M}^{\,\nu\ast} \partial_{\nu}\,)/2m
= \mathfrak{E}_M^{\,\pm} 
\]
Then we can write
\begin{eqnarray}
&&\lbrace \psi_{M}(x)\,,\,\overline{\psi}_{M}(y)\rbrace = 
\lbrace \psi_{M}(x)\,,\,{\psi}^{\,\dagger}_{M}(y)\rbrace\,\beta_{M}
\nonumber\\
&=&\sum_{\,{\bf p}\,,\,r}\sum_{\,{\bf q}\,,\,s}\,
\left\lbrace \Big[\,a_{\,{\bf p}\,,\,r}\,u_{\,{\bf p}\,,\,r}(x)
+ a_{\,{\bf p}\,,\,r}^{\,\dagger}\,u_{\,{\bf p}\,,\,r}^{\,\ast}(x)\,\Big] ,
\Big[\,a^{\,\dagger}_{\,{\bf q}\,,\,s}\,{u}^{\,\dagger}_{\,{\bf q}\,,\,s}(y)
+ a_{\,{\bf q}\,,\,s}\,{u}_{\,{\bf q}\,,\,s}^{\,\top}(y)\,\Big]\right\rbrace \,\beta_{M}
\nonumber\\
&=&\sum_{\,{\bf p}\,,\,r}\sum_{\,{\bf q}\,,\,s}\,\left[ \,
\lbrace a_{\,{\bf p}\,,\,r}\,,\,a_{\,{\bf q}\,,\,s}^{\,\dagger} \rbrace\,u_{\,{\bf p}\,,\,r}(x) \otimes \overline{u}_{\,{\bf q}\,,\,s}(y)
+\lbrace a_{\,{\bf p}\,,\,r}^{\,\dagger}\,,\,a_{\,{\bf q}\,,\,s} \rbrace\,u^{\ast}_{\,{\bf p}\,,\,r}(x) \otimes {u}^{\,\top}_{\,{\bf q}\,,\,s}(y)\,\beta_{M}\,\right] 
\nonumber\\
&=&\sum_{\,{\bf p}\,,\,r} \left[ \,u_{\,{\bf p}\,,\,r}(x) \otimes {u}^{\,\dagger}_{\,{\bf p}\,,\,r}(y) 
+ u_{\,{\bf p}\,,\,r}^{\,\ast}(x) \otimes {u}^{\,\top}_{\,{\bf p}\,,\,r}(y)\,\right] \,\beta_{M}
\nonumber\\
&=& \int\mathrm{d}\mathbf{p}\,[\,(2\pi)^{3} 2\omega_{\,\bf p}\,]^{-1}
\left[ \, (\p+m)\,e^{-\,ip\cdot(x-y)} + (\p-m)\,e^{-\,ip\cdot(x-y)}\,\right] _{p_0=\omega_{\,\bf p}}
\nonumber\\
&=& (\, m + i\partial\!\!\!/_M \,)\int\mathrm{d}\mathbf{p}\,[\,(2\pi)^{3} 2\omega_{\,\bf p}\,]^{-1}
\left[ \,e^{-\,ip\cdot(x-y)} - e^{\,ip\cdot(x-y)}\,\right] _{p_0=\omega_{\,\bf p}}
\nonumber\\
&=& (-\,i)(\, m + i\partial\!\!\!/_M \,)\,D(x-y) \equiv S_{M}(x-y)
\end{eqnarray}
where we have set $ \partial\!\!\!/_M \equiv \gamma^{\,\mu}_{M}\,\partial_{\mu}\,, $ while use has been made of the closure relation (\ref{ccclosure}) 
whereas $ D(x-y) $ is the well known Pauli-Jordan tempered distribution
\begin{equation}
D(x-y)=\frac{i}{(2\pi)^{3}}\int\mathrm{d}^{4}k\,\exp \lbrace -\,ik\cdot (x-y) \rbrace\,\delta(k^{2} - m^{2})\,\mathrm{sgn}(k_0)
\end{equation}
As a matter of fact, the canonical spinor anticommutator $ \lbrace\psi(x)\,,\overline{\psi}(y)\rbrace =S(x-y) $ 
is generally and uniquely determined by 1. Lorentz covariance 2. the field equation
$(\, m - i\partial\!\!\!/\,)\,S(x-y)=0 $
and 3. the general \textbf{spinor  microcausality property}
\[
\gamma_{0} S(x-y) = \partial_{0} D(x-y)\qquad\quad (x-y)^{2}<0\quad
\Longrightarrow\quad S(0,\mathbf{x}-\mathbf{y})=\gamma_{0}\,\delta(\mathbf{x}-\mathbf{y})
\]
It is important to emphasize that the above mentioned noteworthy featured properties do not depend at all upon the specific Dirac matrices representations, 
so that they keep holding true for both Dirac and Majorana spinor quantum fields, \textsf{in spite of their deep difference in the internal symmetries},
the same conclusion being also valid, \textit{mutatis mutandis}, for the off-shell causal Green's functions or Feynman propagators.
However, it turns out that only for the Majorana spinor we get $ S_{M}(x-y) + S_{M}^{\,\ast}(x-y) = 0\,, $ thanks to the Majorana spinor self-conjugation
and the purely imaginary nature of the Dirac matrices in the Majorana representation.
\medskip
A further neutral, self-conjugate, Hermitean quantum Majorana spinor can be obtained out of the classical solution (\ref{Majorana2}), viz.,
\begin{equation}
\widehat{\psi}_M(x) = \sum_{\,{\bf p}\,,\,r}\,\Big[\,
b_{\,{\bf p}\,,\,r}\,v^{\,\ast}_{\,{\bf p}\,,\,r}(x)
+ b_{\,{\bf p}\,,\,r}^{\,\dagger}\,v_{\,{\bf p}\,,\,r}(x)\,\Big]
= [\,\widehat\psi_M^{\,\dagger}(x)\,]^{\top} \equiv \widehat{\psi}^{\,c}_M(x) 
\end{equation}
\begin{equation}
\widehat{\psi}^{\,\dagger}_M(x)\,\beta_{M} = \sum_{\,{\bf q}\,,\,s}\,\Big[\,
b^{\,\dagger}_{\,{\bf q}\,,\,s}\,v^{\,\top}_{\,{\bf q}\,,\,s}(x)\,\beta_{M}
+ b_{\,{\bf q}\,,\,s}\,\overline{v}_{\,{\bf q}\,,\,s}(x)\,\Big]
\end{equation}
which again satisfies the canonical anticommutation relations
\[
\lbrace \widehat{\psi}_{M}(x)\,,\,\widehat{\psi}^{\,\dagger}_{M}(y)\rbrace\,\beta_{M} = S_{M}(x-y)
\qquad\quad \mathrm{with}\qquad\quad
\lbrace \widehat{\psi}_{M}(x)\,,\,\overline{\psi}_{M}(y)\rbrace = 0
\]
Consider now the non-Hermitean combination
\begin{equation}
\Psi_{D}(x) = [\,\psi_{M}(x) + i\widehat{\psi}_{M}(x)\,]/\sqrt{2}
\end{equation}
which still satisfies the Majorana wave equation (\ref{MWE}). Its normal modes expansion reads
\begin{eqnarray*}
\Psi_{D}(x) &=& 2^{-1/2}\sum_{\,{\bf p}\,,\,r}\,\Big[\,
a_{\,{\bf p}\,,\,r}\,u_{\,{\bf p}\,,\,r}(x) + i b_{\,{\bf p}\,,\,r}\,v^{\,\ast}_{\,{\bf p}\,,\,r}(x)
+ a_{\,{\bf p}\,,\,r}^{\,\dagger}\,u_{\,{\bf p}\,,\,r}^{\,\ast}(x) + i b_{\,{\bf p}\,,\,r}^{\,\dagger}\,v_{\,{\bf p}\,,\,r}(x)\,\Big]
\\
&=& \sum_{\,{\bf p}\,,\,r}\,\Big[\,c_{\,{\bf p}\,,\,r}\,w_{\,{\bf p}\,,\,r}(x) + d^{\,\dagger}_{\,{\bf p}\,,\,r}\,z_{\,{\bf p}\,,\,r}(x)\,\Big]
\end{eqnarray*}
whence the identity follows
\begin{eqnarray}
a_{\,{\bf p}\,,\,r}\,u_{\,{\bf p}\,,\,r}(x) + i b_{\,{\bf p}\,,\,r}\,v^{\,\ast}_{\,{\bf p}\,,\,r}(x) = \sqrt{2}\,c_{\,{\bf p}\,,\,r}\,w_{\,{\bf p}\,,\,r}(x)
\\
 a_{\,{\bf p}\,,\,r}^{\,\dagger}\,u_{\,{\bf p}\,,\,r}^{\,\ast}(x) + i b_{\,{\bf p}\,,\,r}^{\,\dagger}\,v_{\,{\bf p}\,,\,r}(x)
 =\sqrt{2}\,d^{\,\dagger}_{\,{\bf p}\,,\,r}\,z_{\,{\bf p}\,,\,r}(x)\
\end{eqnarray}
which is nothing but a \textbf{local Bogoliubov transformation}, as we shall check here below.
As a matter of fact,
after taking inner products with some suitable spin-state wave functions, it is possible to prove that the creation/annihilation operators
$ c_{\,{\bf p}\,,\,r}\,,\,c^{\,\dagger}_{\,{\bf p}\,,\,r}\,,\,d_{\,{\bf q}\,,\,s}\,,\,d^{\,\dagger}_{\,{\bf q}\,,\,s} $ do still satisfy
the canonical anticommutation relations, which endorses the local nature of the above Bogoliubov transformation.

\medskip
\begin{footnotesize}
\texttt{Proof.} Consider the inner product
\[
\int\mathrm{d}\mathbf{x}\,u^{\,\dagger}_{\,\mathbf{q},\,s}(x)
\left[ \,a_{\,{\bf p}\,,\,r}\,u_{\,{\bf p}\,,\,r}(x) + i b_{\,{\bf p}\,,\,r}\,v^{\,\ast}_{\,{\bf p}\,,\,r}(x)\,\right]  
= \sqrt{2}\,c_{\,{\bf p}\,,\,r} \int\mathrm{d}\mathbf{x}\,u^{\,\dagger}_{\,\mathbf{q},\,s}(x)\,w_{\,{\bf p}\,,\,r}(x)
\]
We find
\begin{eqnarray*}
\int\mathrm{d}\mathbf{x}\,u^{\,\dagger}_{\,\mathbf{q},\,s}(x)\,v^{\,\ast}_{\,{\bf p}\,,\,r}(x) 
= (2\omega_{\,\bf p})^{-1}\,u^{\,\dagger}_{\, s}(\mathbf{p})\,v^{\,\ast}_{\, r}(\mathbf{p})  \,\delta(\mathbf{p} - \mathbf{q})
= \textstyle\frac12\,\xi^{\,\dagger}_{\,s}\,\eta^{\,\ast}_{\,r} \,\delta(\mathbf{p} - \mathbf{q})
\\
(2\omega_{\,\bf p})^{-1}\,u^{\,\dagger}_{\, s}(\mathbf{p})\,v^{\,\ast}_{\, r}(\mathbf{p})
= (2\omega_{\,\bf p})^{-1}\,(2m + 2\omega_{\,\bf p})^{-1}\,\xi^{\,\dagger}_{\,s} 
( m + \omega_{\,\bf p} \beta_{M} - p_{k} \gamma_{M}^{\,k} ) ( m + \omega_{\,\bf p} \beta_{M} + p_{\ell} \gamma_{M}^{\,\ell} ) \,\eta^{\,\ast}_{\,r}
\\
= (2\omega_{\,\bf p})^{-1}\,(2m + 2\omega_{\,\bf p})^{-1}\,\xi^{\,\dagger}_{\,s}\,
 [\,2\omega^{\,2}_{\,\bf p} + 2m\omega_{\,\bf p} \beta_{M} + \omega_{\,\bf p} \beta_{M}\,p_{\ell} \gamma_{M}^{\,\ell}  
- p_{k} \gamma_{M}^{\,k}\,\omega_{\,\bf p} \beta_{M}\,]\,\eta^{\,\ast}_{\,r}
\\
=  (2\omega_{\,\bf p})^{-1}(2m + 2\omega_{\,\bf p})^{-1}\,\xi^{\,\dagger}_{\,s}\, 
 [\,2\omega^{\,2}_{\,\bf p} + 2m\omega_{\,\bf p} + \omega_{\,\bf p}\,p_{\ell} \gamma_{M}^{\,\ell}  
- p_{k} \gamma_{M}^{\,k}\,\omega_{\,\bf p}\,]\,\eta^{\,\ast}_{\,r} = \textstyle\frac12\,\xi^{\,\dagger}_{\,s}\,\eta^{\,\ast}_{\,r} 
\end{eqnarray*}
because $ \xi^{\,\dagger}_{\,s} \beta_{M}=\xi^{\,\dagger}_{\,s}\,,\ \beta_{M} \eta^{\,\ast}_{\,r} = \eta^{\,\ast}_{\,r} $ 
and $ \xi^{\,\dagger}_{\,s} \gamma_{M}^{\,k} \eta^{\,\ast}_{\,r} = \xi^{\,\dagger}_{\,s} \lbrace\beta_{M}\,,\gamma_{M}^{\,k} \rbrace \,\eta^{\,\ast}_{\,r} \equiv 0\,. $
The two orthogonal pair $ \xi_{\,\pm}\,, \eta^{\,\ast}_{\,\pm} $ of the positive energy bispinors in the Majorana particle rest frame $ \mathbf{p}=0 $
are evidently connected by the $SO(2,\mathbb{R})$ transform (\ref{rotaspin})
so that $ \textstyle\frac12\,\xi^{\,\dagger}_{\,s}\,\eta^{\,\ast}_{\,r} = 
\frac12\xi^{\,\dagger}_{\,s}\,[\,F_{+}(\theta)\,]_{\,rt}\,\xi_{\,t} \equiv f_{\,rs}(\theta) $
in which $ f_{rs}^{-1}(\theta)=f_{sr}(\theta)=f_{rs}(-\,\theta)\ (\,r,s=+,-\,) $ and the summation over repeated indexes is understood.
Then we obtain
\begin{equation}
a_{\,{\bf p}\,,\,r} + i b_{\,{\bf p}\,,\,t}\,f_{\,tr}(\theta) = \sqrt{2}\,c_{\,{\bf p}\,,\,t}\,f_{\,tr}(\overline{\theta})
\qquad\quad
a^{\,\dagger}_{\,{\bf q}\,,\,s} - i b^{\,\dagger}_{\,{\bf q}\,,\,t}\,f_{ts}(\theta) = \sqrt{2}\,c^{\,\dagger}_{\,{\bf q}\,,\,t}\,f_{ts}(\overline{\theta})
\end{equation}
\begin{equation}
\sqrt{2}\,c_{\,{\bf p}\,,\,r} = a_{\,{\bf p}\,,\,t} \,f_{tr}(-\,\overline{\theta})+ i b_{\,{\bf p}\,,\,t}\,f_{\,tr}(\theta-\overline{\theta}) 
\qquad\quad
\sqrt{2}\,c^{\,\dagger}_{\,{\bf q}\,,\,s} = a^{\,\dagger}_{\,{\bf q}\,,\,u}\,f_{us}(-\,\overline{\theta}) - i b^{\,\dagger}_{\,{\bf q}\,,\,u}\,f_{\,us}(\theta-\overline{\theta}) 
\end{equation}
\begin{eqnarray*}
\lbrace c_{\,{\bf p}\,,\,r}\,,c^{\,\dagger}_{\,{\bf q}\,,\,s}\rbrace  &=& \textstyle\frac12\,
\lbrace a_{\,{\bf p}\,,\,t} \,f_{tr}(-\,\overline{\theta}) + i b_{\,{\bf p}\,,\,t}\,f_{\,tr}(\theta-\overline{\theta}) \,,\,
a^{\,\dagger}_{\,{\bf q}\,,\,u}\,f_{us}(-\,\overline{\theta}) - i b^{\,\dagger}_{\,{\bf q}\,,\,u}\,f_{\,us}(\theta-\overline{\theta}) \rbrace
\\
&=& \delta(\mathbf{p} - \mathbf{q})\,\delta_{\,rs}
\end{eqnarray*}
Analogous procedures drive to show the full validity of the canonical commutation relation 
$\lbrace d_{\,{\bf p}\,,\,r}\,,d^{\,\dagger}_{\,{\bf q}\,,\,s}\rbrace=\delta(\mathbf{p} - \mathbf{q})\,\delta_{\,rs}\,, $
as well as the vanishing of all the remaining ones.
\end{footnotesize}

\medskip
It follows that a Dirac spinor, which represents a charged and non-Hermitean quantum spinor field, can  always be expressed as
a complex combination of two independent neutral, self-conjugated and Hermitean Majorana quantum spinor fields, i.e.,
\begin{equation}
\psi_{D}(x)=[\,\psi_{M}(x) + i \widehat{\psi}_{M}(x)\,]/\sqrt{2}
\end{equation}
From the similarity transformation rule $ \gamma^{\,\mu}=S\gamma^{\,\mu}_{M}S^{-1} $ we can go back to the Weyl representation
of the gamma matrices and thereby
\begin{equation}
\psi_{D}(x)=\sum_{\,{\bf p}\,,\,r}\,\Big[\,c_{\,{\bf p}\,,\,r}\,S\,w_{\,{\bf p}\,,\,r}(x)\,S^{-1} + d^{\,\dagger}_{\,{\bf p}\,,\,r}\,S\,z_{\,{\bf p}\,,\,r}(x)\,S^{-1}\,\Big]
\end{equation}
The normal modes - in the Weyl representation of the gamma matrices - are provided by
\begin{equation}
\left\lbrace 
\begin{array}{c}
w_{\,{\bf p}\,,\,r}(x)\equiv
[\,(2\pi)^3 2\omega_{\,\bf p}\,]^{-\frac12}\,w_{\, r}(\mathbf{p})\,
\exp\{-\,i\omega_{\,\bf p}t + i{\bf p}\cdot{\bf x}\}
\\ \\
z_{\,{\bf p}\,,\,r}(x)\equiv
[\,(2\pi)^3 2\omega_{\,\bf p}\,]^{-\frac12}\,z_{\, r}(\mathbf{p})\,
\exp\{\,i\omega_{\,\bf p}t - i{\bf p}\cdot{\bf x}\}
\end{array}
\right. 
\qquad\quad(\,r=+,-\,)
\end{equation}
where
\begin{equation}
w_{\,r}(\mathbf{p})=(2m+2\omega_{\,\bf p})^{-1/2 }(m+\p)\,\varsigma_{\,r}
\qquad\quad
z_{\,r}(\mathbf{p})=(2m+2\omega_{\,\bf p})^{-1/2} (m-\p)\,\zeta_{\,r}
\end{equation}
where 
\[
\gamma_{0}\,\varsigma_{\,\pm}=\varsigma_{\,\pm}
\qquad\quad
\gamma_{0}\,\zeta_{\,\pm} = -\,\zeta_{\,\pm}
\qquad\quad
(\,\Sigma_{\,2} \mp 1\,)\,\varsigma_{\,\pm}=0
\qquad\quad
(\,\Sigma_{\,2} \pm 1\,)\,\zeta_{\,\pm}=0
\]
with $ \Sigma_{\,2}=i\gamma_{3}\gamma_{1}\,. $ The above spin states do satisfy
$ z_{\,\pm}(\mathbf{p})=i\gamma_{2}\,w^{\,\ast}_{\,\pm}(\mathbf{p}) $ and $ w_{\,\pm}(\mathbf{p})=i\gamma_{2}\,z^{\,\ast}_{\,\pm}(\mathbf{p})\,. $

\medskip
\begin{footnotesize}
\texttt{Proof}. After setting $ N=(2m+2\omega_{\,\bf p})^{-1/2 } $ we find
\begin{eqnarray*}
i\gamma_{2}\,w^{\,\ast}_{\,\pm}(\mathbf{p}) =
iN\gamma_{2} (m+\gamma_{0}\omega_{\,\bf p} + \gamma_{1}  p_{x} + \gamma_{3} p_z - \gamma_{2} p_y )\,\varsigma^{\,\ast}_{\,\pm}
= N ( m - \p )\,i\gamma_{2}\,\varsigma^{\,\ast}_{\,\pm}
\\
\gamma_{0}\,i\gamma_{2}\,\varsigma^{\,\ast}_{\,\pm} = -\,i\gamma_{2}\,\varsigma^{\,\ast}_{\,\pm} 
\qquad\quad
\Sigma_{2}\,i\gamma_{2}\,\varsigma^{\,\ast}_{\,\pm} = \mp\,i\gamma_{2}\,\varsigma^{\,\ast}_{\,\pm} 
\qquad\Longrightarrow\qquad i\gamma_{2}\,\varsigma^{\,\ast}_{\,\pm}\equiv \zeta_{\,\pm}
\qquad\Longrightarrow\qquad z_{\,\pm}(\mathbf{p})=i\gamma_{2}\,w^{\,\ast}_{\,\pm}(\mathbf{p})
\end{eqnarray*}
\end{footnotesize}

We have already introduced charge conjugation above. It is defined as the operation by 
which particles and antiparticles are exchanged, up to an arbitrary
overall phase factor. Thus let us set
\begin{equation}
{\mathcal C}\,c_{\,{\bf p}\,,\,r}\,{\mathcal C}^{\,-1} = \mathrm e^{i\eta}\,d_{\,{\bf p}\,,\,r}
\qquad
{\mathcal C}\,d_{\,{\bf p}\,,\,s}\,{\mathcal C}^{\,-1} = \mathrm e^{-\,i\eta}\,c_{\,{\bf p}\,,\,s}
\end{equation}
$$
\forall\,r,s=+,-\quad {\bf p}\in{\mathbb R}^3
$$
Repeated application of this operation yields
\begin{eqnarray*}
 {\mathcal C}\,\Big({\mathcal C}\,c_{\,{\bf p}\,,\,r}\,{\mathcal C}^{\,-1}\Big)\,{\mathcal C}^{\,-1}
= \mathrm e^{i\eta}\,{\mathcal C}\,d_{\,{\bf p}\,,\,r}\,{\mathcal C}^{\,-1}=c_{\,{\bf p}\,,\,r}
\end{eqnarray*}
Therefore we can always assume the following properties:

\[
{\mathcal C}^{\,2}\ =\ {\mathbb I}\qquad\quad\Longrightarrow\qquad\quad
{\mathcal C}={\mathcal C}^{\,\dagger}={\mathcal C}^{\,-1}
\]
It turns out that the above defined spin-states 
do fulfill the remarkable relationship
\[
z_{\,\pm}({\bf p}) = i\,\gamma_{2}\,w_{\,\pm}^{\,\ast}({\bf p})
\qquad\quad
w_{\,\pm}({\bf p}) = i\,\gamma_{2}\,z_{\,\pm}^{\,\ast}({\bf p})
\]
Hence, the charge conjugation transformation law can be rewritten as
\begin{eqnarray}
\psi_{D}^{\,c}(x) &=& \sum_{{\bf p}\,,\,r}\ \mathrm e^{i\eta}\,\Big[\,
d_{\,{\bf p}\,,\,r}\,w_{\,{\bf p}\,,\,r}(x) +
c_{\,{\bf p}\,,\,r}^{\,\dagger}\,z_{\,{\bf p}\,,\,r}(x)\,\Big]\nonumber
\\
&=&  e^{\,i(\eta-\pi/2)}\,\sum_{{\bf p}\,,\,r}\ \Big[\,
d_{\,{\bf p}\,,\,r}\,\gamma^{\,2}\,z^{\,\ast}_{\,{\bf p}\,,\,r}(x)
+ c_{\,{\bf p}\,,\,r}^{\,\dagger}\,\gamma^{\,2}\,w^{\,\ast}_{\,{\bf p}\,,\,r}(x)\,\Big]\nonumber
\\
&=& e^{\,i(\eta-\pi/2)}\,\gamma^{\,2}\,\left(\,\psi_{D}^{\,\dagger}(x)\,\right)^\top
\end{eqnarray}
Thus, if we choose $\eta=\pi/2$ then we obtain the transformation rule
\begin{equation}
 \psi_{D}^{\,c}(x) = \gamma^{\,2}\,\left(\,\psi_{D}^{\,\dagger}(x)\,\right)^\top
\end{equation}
which precisely corresponds to the classical transformation
$ \psi^{\,c}(x)=\gamma^{\,2} \psi^{\,\ast}(x)\,.$

\subsubsection{Observables}

Let us now analyze and discuss the observables of a Majorana quantum field and the related properties of the Majorana particles, 
if eventually detected to any concern in some forthcoming experiment.

First of all, it is worthwhile to notice that from the orthogonality relations
\begin{eqnarray*}
\left(\,u_{\,{\bf p}\,,\,r}\,,\,u_{\,{\bf q}\,,\,s}\,\right)
=\int\mathrm{d}\mathbf{x}\ 
\overline{u}_{\,{\bf p}\,,\,r}(x)\,\beta_M\,
u_{\,{\bf q}\,,\,s}(x)=\delta_{\,rs}\;\delta({\bf p}-{\bf q})
\\
\left(\,u_{\,{\bf p}\,,\,r}\,,\,u^{\,\ast}_{\,{\bf q}\,,\,s}\,\right)
=\int\mathrm{d}\mathbf{x}\ 
\overline{u}_{\,{\bf p}\,,\,r}(x)\,\beta_M\,
u^{\,\ast}_{\,{\bf q}\,,\,s}(x)=0
\\
\left(\,u^{\,\ast}_{\,{\bf p}\,,\,r}\,,\,u^{\,\ast}_{\,{\bf q}\,,\,s}\,\right)
=\int\mathrm{d}\mathbf{x}\ 
u^{\,\top}_{\,{\bf p}\,,\,r}(x)\,\beta_M\,
u^{\,\ast}_{\,{\bf q}\,,\,s}(x) = -\,\delta_{\,rs}\;\delta({\bf p}-{\bf q})
\end{eqnarray*} 

one can easily derive all the observable quantities involving a Majorana massive quantum spinor field. For example the energy-momentum four-vector takes the form

\begin{eqnarray*}
P_{\mu} &=& 
\frac{i}{4}\int\mathrm{d}\mathbf{x}\, :\overline{\psi}_M(x)\,\beta_M \parl_\mu\,\psi_M(x):
\\
&=& \frac{i}{4}\int\mathrm{d}\mathbf{x}\, :\sum_{\,{\bf q}\,,\,s}\,
\Big[\,a^{\,\dagger}_{\,{\bf q}\,,\,s}\,\overline{u}_{\,{\bf q}\,,\,s}(x)
+ a_{\,{\bf q}\,,\,s}\,\overline{u}_{\,{\bf q}\,,\,s}^{\,\ast}(x)\,\Big]
\\
&\times& \beta_M \parl_\mu
\sum_{\,{\bf p}\,,\,r}\,\Big[\,
a_{\,{\bf p}\,,\,r}\,u_{\,{\bf p}\,,\,r}(x)
+a_{\,{\bf p}\,,\,r}^{\,\dagger}\,u_{\,{\bf p}\,,\,r}^{\,\ast}(x)\,\Big]:
\\
&=& \frac12\sum_{{\bf p}\,,\,r}\ p_{\,\mu} :
a^{\,\dagger}_{\,{\bf p}\,,\,r}\,a_{\,{\bf p}\,,\,r}
-  a_{\,{\bf p}\,,\,r}\,a^{\,\dagger}_{\,{\bf p}\,,\,r} :
\\
&=& \sum_{{\bf p}\,,\,r}\ p_{\,\mu}\,
a^{\,\dagger}_{\,{\bf p}\,,\,r}\,a_{\,{\bf p}\,,\,r}
\qquad\quad(\,p_{\,0}=\omega_{\,\bf p}\,)
\end{eqnarray*}
where the normal ordering prescription has been customarily adopted.

\medskip
Concerning the total angular momentum operator, owing to the rotation invariance of the theory and without loss of generality,
one can always arrange the Observer reference frame in such a manner that the direction of the motion of a Majorana particle lies
along the $Oy-$axis.
In such a circumstances, for example, if
$\partial_{x}\psi_M=\partial_{z}\psi_M=0$ that implies $\psi_M(t,y)=\psi_M(t,0,y,0)\,,$ 
then we obtain for the energy-momentum tensor
\[
2i\,T^{\,13}(t,y)=\overline{\psi}_M(t,y)\,\gamma_M^{1}\parl_z\,\psi_M(t,y)=0
\]
\[
2i\,T^{\,31}(t,z)=\overline{\psi}_M(t,y)\,\gamma_M^{\,3}\parl_x\,\psi_M(t,y)=0
\]
and thereby
\[
\partial_{\,\mu}\,M_{\,13}^{\,\mu}=\partial_{\,\mu}\,S_{\,13}^{\,\mu}=0
\]
whence it follows that the helicity - the projection of the spin angular momentum on the direction of the motion - is conserved in time.
After insertion of the normal modes expansion one gets
\begin{eqnarray*}
\mathrm{h} &=& \int_{-\infty}^\infty\mathrm{d} y\,
:{\textstyle\frac12}\,\psi_M(t,y)\,\Sigma_{M,\,2}\,\psi_M(t,y):
\\
&=& \int_{-\infty}^\infty\mathrm{d}y\,
:\sum_{{p}\,,\,r}\ \Big[\,a_{\,{p}\,,\,r}\,u_{\,{p}\,,\,r}(t,y)
+ a_{\,{p}\,,\,r}^{\,\dagger}\,u_{\,{p}\,,\,r}^{\,\ast}(t,y)\,\Big]
\\
&\times& {\textstyle\frac12}\,\Sigma_{M,\,2}\,\sum_{{q}\,,\,s}\ \Big[\,
a_{\,{q}\,,\,s}^{\dagger}\,u_{\,{q}\,,\,s}^{\,\ast}(t,y)
+ a_{\,{q}\,,\,s}\,u_{\,{q}\,,\,s}(t,y)\,\Big]:
\end{eqnarray*}
in which we have set 
\[
{\bf p}=(0,p,0)\qquad\quad{\bf q}=(0,q,0)\qquad\quad
\omega_{\,p}=\sqrt{p^2+m^2}
\]
\[
u_{\,{p}\,,\,r}(t,y) = [\,4\pi\omega_{\,p}\,]^{\,-1/2}\,
u_{\,r}(p)\,\exp\{ipy-it\omega_{\,p}\}\qquad(\,r=+,-\,)
\]
the normalization being now consistent with the occurrence that the spinor plane waves 
are independent of the transverse spatial coordinates $x_\perp=(x^1\,,\,x^3)\,.$
From the commutation relation
\[
[\,\omega_{\,p}\gamma^{\,0}_M - 
p\,\gamma^{\,2}_M\,,\,\Sigma_{M,\,2}\,]=0
\]
together with the definition 
\[
u_{\,\pm}(p)\equiv 
(2\omega_{\,p}+2m)^{\,-1/2}\;
\left(m+\omega_{\,p}\gamma^{\,0}_M - p\,\gamma^{\,2}_M\right)\xi_{\,\pm}
\]
it can be readily derived that
\[
(\,\Sigma_{M,\,2}\mp1\,)\,\xi_{\,\pm}=0\qquad\Rightarrow\qquad
(\,\Sigma_{M,\,2}\mp1\,)\,u_{\,\pm}(p)=0
\]
which yields in turn
\[
\mathrm{h}={\textstyle\frac12}\hslash
\int_{-\infty}^\infty \mathrm{d}p\ \Big[\,
a^{\,\dagger}_{\,{p}\,,\,+}\,a_{\,{p}\,,\,+} - a^{\,\dagger}_{\,{p}\,,\,-}\,a_{\,{p}\,,\,-}\,\Big]
\]
It follows therefrom that the 1-particle states $a^{\,\dagger}_{\,{\bf p}\,,\,\pm}\,\vert\,0\,\rangle$ 
do represent charge self-conjugated, or neutral,  Majorana massive particles with energy-momentum
$p^{\,\mu}=(\omega_{\,\bf p},{\,\bf p})$ and two positive/negative helicity states, that has to be eventually detected 
with \textbf{equal probability.}

\medskip
{
Soon after the discovery by Super-Kamiokande \cite{Fukuda} of the neutrino's flavor oscillations,
it has been promptly suggested that, were the neutrinos described by massive Majorana particles, then the lepton number conservation
would be no longer true, in such a manner that the neutrino's flavor oscillations could be accounted for and actually explained. 
However, it turns out that the renowned Goldhaber \emph{et al.} experiments \cite{Goldhaber} clearly proved that:

\begin{itemize}
\item 
\textbf{neutrinos and antineutrinos are distinguishable chiral particles and charge conjugated one to each other};
\item 
their \textbf{chirality and helicity are strongly correlated}, in such a manner that neutrinos are merely left-handed with solely negative helicity,
while antineutrinos are strictly right-handed and solely with positive helicity.
\end{itemize}

The Goldhaber \emph{et al.} experiments together with the famous 1957 experiment by Madame Chien-Shiung Wu do actually firmly endorse
the parity violation of the weak interactions as well as the so called V-A theory, which is the low-energy limit of the present day Standard Model
where the neutrino and antineutrino particles and fields are described by chiral massless Weyl fermions.
As we have thoroughly and at length discussed in full detail in the present review, the above featured experimental facts are compatible
\textbf{only }with the assumption that in the relativistic quantum field theory neutrinos and antineutrinos are described by Weyl massless spin$ \frac12 $ fermions.
These very same experimental facts \textbf{strongly disagree} with the phenomenological \textit{ad hoc} hypothesis that neutrinos and
antineutrinos could be described by massive Majorana parity invariant and charge self-conjugated, i.e. neutral, particles and fields 
with both chiralities and two equivalent helicity states.
As a matter of fact, for the quanta of a massive Majorana spinor field one can always select the particle rest frame in which the particle spin must be up or down 
with equal probability, if all the interactions are Lorentz invariant and \textit{a fortiori} rotation invariant with no privileged direction and versus.
As a consequence
\begin{itemize}
\item
in any arbitrary inertial frame the massive Majorana quanta must always exhibit, if eventually detected,
both chirality and helicity quantum numbers with equal probability,
no matter how small the mass could be;
\item
it is not possible to distinguish between particle and antiparticle: the very deep reason why the lepton number is not conserved is that
it cannot even be \textbf{operatively} defined for massive Majorana quanta;
\item
the zero mass limit for a massive Majorana spinor yields a massless Majorana fermion and never any Weyl spinor, which exhibits quite
different mathematical and physical properties; moreover, the zero mass limit is \textbf{singular}, because a new symmetry arises solely
for massless Majorana spinors - see below.\footnote{Quite often in the literature it is argued that a tiny value for the Majorana mass does eventually
provide a small perturbation to the neutrino sector of the Standard Model, which allows a simple explanation of the neutrino's flavor oscillations without any drastic
modification of any other physical setting and properties of the SM itself. As we have thoroughly analyzed in this review, this is wrong, false and badly misleading.}

\end{itemize}
Hence the phenomenological \textit{ad hoc} hypothesis of a massive Majorana spinor to describe neutrinos and antineutrinos is actually ruled out by the 
experimental evidence and the basic features of the renormalizable relativistic quantum field theory.
}

\medskip
\begin{footnotesize}
\texttt{Atiah-Singer index.}
We recall that, only for massless self-conjugated spinors, there exists an accidental U(1) internal symmetry 
for the Majorana quantum spinor field, 
the corresponding conserved 
pseudo-scalar charge being expressed by
\[
Q_{5}=\pm\,\textstyle\frac12\int\mathrm{d}\mathbf{x}\,:\overline{\psi}_M(t,\mathbf{x})\beta_M\gamma^{\,5}_M \psi(t, \mathbf{x}):
\quad\qquad\dot{Q}_5=0
\]
the overall sign being conventional and irrelevant. After insertion of the normal mode expansion we get
\begin{eqnarray*}
Q_5 &=& \pm\int\mathrm{d}\mathbf{x}:
\sum_{\,{\bf q}\,,\,s}\,\Big[\,
a^{\,\dagger}_{\,{\bf q}\,,\,s}\,\overline{\upsilon}_{\,{\bf q}\,,\,s}(x)
+a_{\,{\bf q}\,,\,s}\,\overline{\upsilon}_{\,{\bf q}\,,\,s}^{\,\ast}(x)\,\Big]
\\
&\times& {\textstyle\frac12}\,\beta_M\gamma^{\,5}_M
\sum_{\,{\bf p}\,,\,r}\,\Big[\,
a_{\,{\bf p}\,,\,r}\,\upsilon_{\,{\bf p}\,,\,r}(x)
+a_{\,{\bf p}\,,\,r}^{\,\dagger}\,\upsilon_{\,{\bf p}\,,\,r}^{\,\ast}(x)\,\Big]
\end{eqnarray*}
where we have denoted by 
\begin{eqnarray*}
\upsilon_{r}(\mathbf{p})=p\!\!/_{\!M}\;\upsilon_{r}/\sqrt{2\wp}
\qquad\quad(\,\wp\equiv\vert\,\mathbf{p}\,\vert\,,\ r=+,-\,)
\\
\upsilon_{+}=\left\lgroup
\begin{array}{c}
1\\-i\\1\\-i
\end{array}
\right\rgroup
\qquad\quad
\upsilon_{-}=\left\lgroup
\begin{array}{c}
i\\-1\\i\\-1
\end{array}
\right\rgroup
\\
\upsilon_{r,\,\mathbf{p}}(x)=[\,(2\pi)^{3}2\wp\,]^{-\frac12}\,\upsilon_{r}(\mathbf{p})\,\exp \lbrace-\,it\wp + i\mathbf{p}\cdot\mathbf{x}\rbrace
\end{eqnarray*}
the spin-states and plane wave-functions respectively of the massless  self-conjugated spinor field.
The spin-states satisfy the conventional orthogonality and closure relations as they correspond to the regular limit 
$ \lim_{\, m\rightarrow0} w_{r}(\mathbf{p}) = \upsilon_{r}(\mathbf{p})\,. $ Now we have
\begin{eqnarray*}
\gamma^{\,5}_M=\left\lgroup
\begin{array}{cccc}
0 & 0 & 0 & i
\\
0 & 0 & -i & 0
\\
0 & i & 0 & 0
\\
-i & 0 & 0 & 0
\end{array}
\right\rgroup
\end{eqnarray*}
so that
\[
\begin{array}{c}
\gamma^{\,5}_M\,\upsilon_{\pm}=\pm\,\upsilon_{\pm}
\qquad
\quad
\gamma^{\,5}_M\,\upsilon_{\pm}(\mathbf{p})=\mp\,\upsilon_{\pm}(\mathbf{p})
\end{array}
\]
which means that the spin-states in the massless case are eigenstates of the chiral matrix in the Majorana representation.

It turns out that in the 1d case, \textit{i.e.} the dynamics along \textit{e.g.} the $ Oy- $axis, which is the relevant situation for the construction of the
helicity operator, we actually have 
\[
\upsilon_{\pm}(p_y)=(\,\beta_{M}\,\wp - \gamma^{\,2}_{M}\,p_y\,)\, \upsilon_{\pm}/\sqrt{2\wp}
\qquad\quad[\,\wp=p_y\mathrm{sgn}(p_y)\,]
\]
in such a manner that the above spin-states are common eigenstates of the chiral and spin matrices because
$$
[\,\beta_{M}\,\wp - \gamma^{\,2}_{M}\,p_y,\,\Sigma^{\,2}_{M}\,]=0
\qquad\quad
[\,\Sigma^{\,2}_{M},\,\gamma^{\,5}_M\,]=0
$$
Now we obtain
\begin{eqnarray*}
\int\mathrm{d}\mathbf{x}\,\upsilon^{\,\dagger}_{r,\,\mathbf{p}}(,\mathbf{x})\,
\gamma^{\,5}_{M}\,\upsilon^{\,\ast}_{s,\,\mathbf{q}}(t,\mathbf{x}) =
\frac{1}{2\wp}\;\upsilon^{\,\dagger}_{r}(\mathbf{p})\,\gamma^{\,5}_M\,\upsilon^{\,\ast}_{s}(-\,\mathbf{p})\,
\exp \lbrace 2i\wp t\rbrace
\\
\upsilon^{\,\dagger}_{r}(\mathbf{p})\,\gamma^{\,5}_M\,\upsilon^{\,\ast}_{s}(-\,\mathbf{p}) =
-\;\frac{ \tilde{p}^{\,2} }{ 2\wp }\;\upsilon^{\,\dagger}_{r}\gamma^{\,5}_M\,\upsilon^{\,\ast}_{s}\,\equiv 0
\qquad\quad(\,\forall\,r,s=+,-\,\vee\,\tilde{p}^{\,\nu}=p_{\nu}\,)
\end{eqnarray*}
Moreover we find
\[
\int\mathrm{d}\mathbf{x}\,\upsilon^{\,\dagger}_{r,\,\mathbf{p}}(,\mathbf{x})\,
\gamma^{\,5}_{M}\,\upsilon_{s,\,\mathbf{q}}(t,\mathbf{x}) = \textstyle\frac12\,
\upsilon^{\,\dagger}_{r}\gamma^{\,5}_M\,\upsilon_{s} = \frac12\,\delta_{rs}\,(\,\delta_{r -} - \delta_{s +}\,)
\qquad\quad(\,\forall\,r,s=+,-\,)
\]
whence it follows, after a suitable choice of the overall sign, that the above introduced pseudo-scalar charge $ Q_5 $
is nothing but the quantum counterpart of the Atiah-Singer index for a Majorana spinor field
\[
Q_5=\nu_{M}\equiv\int\mathrm{d}\mathbf{p}\,\left( 
a^{\,\dagger}_{\,\mathbf{p},\,+}\,a_{\,\mathbf{p},\,+} - a^{\,\dagger}_{\,\mathbf{p},\,-}\,a_{\,\mathbf{p},\,-}
\right) 
\]
\textit{i.e.} the difference between the number of quanta with positive and negative helicity.
\end{footnotesize}

\medskip
We can therefore conclude  that the 1-particle states $ a^{\,\dagger}_{\,\mathbf{p},\,\pm}\vert\,0\,\rangle $ do represent neutral, i.e. charge self-conjugated, 
spin $ \frac12 $ Majorana particles,
massive or massless, with energy-momentum $ p_{\mu}=(\omega_{\,\mathbf{p}},\mathbf{p} )$ and two equally probable positive/negative helicity states.

\medskip\noindent

To complete the review of Majorana spinor fields we  write down the appropriate causal distribution and Green's function for a neutral Majorana spinor field. From the Majorana Lagrangian and the related equations of motion one can see that they are:
\begin{eqnarray*}
\left\lbrace \psi_{M}(x),\,\overline{\psi}_{M}(y) \right\rbrace  = S(x-y;m)
\\
\langle\,0\,\vert\,T\,\psi_{M}(x)\,\overline{\psi}_{M}(y)\,\vert\,0\,\rangle = S^{\,F}(x-y;m)
\end{eqnarray*}
the massless limit being smooth.

\section{Dirac, Majorana and Weyl fermions in 4d }
\label{s:MW}

In this Section we briefly summarize and highlight the main properties of the Dirac, Weyl and Majorana fermion fields that we have thoroughly discussed 
at length in the previous Sections.

Let us start from a few basic definitions and properties of  {spinors 
on a 4d Minkowski space}. 
Under a Lorentz transformation a four-component Dirac fermion $\psi$ transforms as
\be
\psi(x) \rightarrow \psi'(x')=\exp\left[-\frac12
\lambda^{\mu\nu}\Sigma_{\mu\nu}\right]\psi(x)\,,\label{Lorentzpsi}
\ee
for $x'^{\mu}=\Lambda^\mu{}_\nu \, x^\nu $. {Here $ \lambda^{\mu\nu}+\lambda^{\nu\mu}=0 $ are six real canonical coordinates 
for the Lorentz group, $ \Sigma_{\mu\nu} $ are the generators in the 4d reducible representation of Dirac bispinors, 
while $ \Lambda^\mu{}_\nu  $ are the Lorentz matrices in the irreducible vector representation  $ D(\frac12,\frac12) $.}
The invariant {kinetic} Lagrangian for a free Dirac field is
\be
i \overline{\psi}\gamma^\mu\partial_\mu \psi=i \overline{\psi}\,\gamma\cdot\partial\,\psi\label{freeDirac}
\ee
where $ \overline{\psi}= \psi^\dagger \gamma_0\,.$
 A Dirac fermion admits a Lorentz invariant mass term $m \overline{\psi}\psi\,.$

A Dirac {bispinor} can be written as the {direct} sum of two Weyl {spinors}\footnote{Here we use the $ (L,R) $ indexes to understand chiral Weyl bispinors with a nonvanishing 
(left,right)-handed 2-component Weyl spinors.}
\be
\psi_L = P_L \psi, \quad\quad \psi_R = P_R \psi, \quad\quad {\rm where}
\quad\quad P_L=\frac {1-\gamma_5}2,
\quad\quad  P_R=\frac {1+\gamma_5}2\0
\ee
with opposite chiralities
\be
\gamma_5 \psi_L= -\psi_L, \quad\quad \gamma_5 \psi_R= \psi_R.\0
\ee 
A left-handed Weyl fermion admits a Lagrangian kinetic term
\be
i(\psi_L, \gamma\cdot\partial\,\psi_L)\equiv\overline\psi_L\gamma^\mu
\partial_\mu \psi_L \label{freeWeyl}	 
\ee
but not a mass term, because $(\psi_L,\psi_L)=0$, since
$\gamma_5\gamma^0+\gamma^0\gamma_5=0$. 
{Therefore, \textsf{a Weyl fermion is massless and this
property is protected by chirality conservation.}}

For classical Majorana spinors we need the notion of {\sl Lorentz
covariant charge conjugate} spinor $\psi^{\,c}$
\be
 \psi^{\,c} =\mathcal{C}\,\psi\,\mathcal{C}^{-1}= \eta_C \gamma^{\,2}\psi^*
 \label{CC}
\ee
where $\eta_C$ is a phase which, for simplicity, in the sequel we set equal to 1. 
It is easy to prove that if $\psi$ transforms like (\ref{Lorentzpsi}), then
\be
\psi^{\,c} (x) \rightarrow \exp\left[-\frac12
\lambda^{\mu\nu}\Sigma_{\mu\nu}\right]\psi^c(x)\,.\label{Lorentzpsihat}
\ee
Therefore one can impose on $\psi$ the condition 
\be
\psi=\psi^{\,c} 
\label{Majo}
\ee

because both sides transform in the same way. Then, as before a  spinor satisfying (\ref{Majo} is defined to be a Majorana spinor. As shown before a Majorana spinor admits both kinetic and mass term.

The Lorentz group spinor representation properties are as follows: the chiral matrix $\gamma_5$ commutes with Lorentz transformation matrix 
$\exp\left[-\frac12\lambda^{\mu\nu}\Sigma_{\mu\nu}\right]$ and so do $P_L$ and $P_R\,.$ This implies that the Dirac representation is reducible. Multiplying the spinors by  $P_L$ and
$P_R$ we select two Weyl irreducible representations. In more precise terms,
Weyl representations are irreducible representations of the group $SL(2,C)$, which is the covering
group of the {\it proper ortochronous} Lorentz group. They are usually denoted
$(\frac 12,0)$ and $(0,\frac 12)$ in the $SU(2)\times SU(2)$ notation of the
$SL(2,C)$ irreducible representations. As we have seen above
Lorentz transformations commute also with the charge
conjugation operation \eqref{CC}. 

It follows that Dirac spinors are reducible and suggests another possible reduction:
the condition (\ref{Majo}) defines another irreducible representation, the
Majorana one. From the literature, \cite{Racah,Cornwell}, we learn that the Majorana representation is the minimal irreducible representation of a (one out of eight) covering of the {\it complete} Lorentz group. In view of the next sections of this review we must stress a fact that is well known and self-evident from our presentation: \textsf{the Majorana and Weyl
representations in 4d are incompatible. }

Let us consider next the charge conjugation and parity discrete symmetries and recall the relevant properties of a Weyl spinor. We have
\be
\mathcal{C}\,\psi_L\,\mathcal{C}^{-1} =P_L  \mathcal{C}\,\psi\,\mathcal{C}^{-1} = P_L\psi^{\,c}= 
(\psi^{\,c})_L\label{CpsiL}
\ee
The charge conjugate of a Majorana field is, by
definition, itself. While the Action of a Majorana field is
invariant under charge conjugation, for a Weyl spinor we have [\,$ \gamma^{\,\mu}\partial_{\mu}\equiv\gamma\cdot\partial $\,]
\be
{\mathcal{C}}\left(\int \overline{\psi_L }\,i\gamma\cdot\partial\,\psi_L\right)  {\mathcal{C}}^{-1}
= \int \overline{(\psi^{\,c})_L}\,i\gamma\cdot\partial\,\psi_L^{\,c}
=\int \overline{\psi_R }\,i\gamma\cdot\partial\,\psi_R\label{CactionLH}
\ee
{i.e.} a Weyl fermion is, so to say, maximally non-invariant under charge conjugation.

The parity operation is defined by
\be
\mathcal{P}\,\psi_L (t, \vx)\,\mathcal{P}^{-1} =\eta_P \gamma_0 \psi_R (t,
-\stackrel{\rightarrow}{x})\label{parityop}
\ee
where $\eta_P$ is a phase, which we set to 1. As for the Action we have
\be
\mathcal{P}\left(\int\overline{\psi_L}\,i\gamma\cdot\partial\,\psi_L  \right) \mathcal{P}^{-1}=
\int\overline{\psi_R}\,i\gamma\cdot\partial\,\psi_R\label{parityaction}
\ee
On the contrary, the Action for a Majoran spinor is parity invariant.

This suggests a useful representation for a Majorana fermion. 
Let $\psi_R = P_R\,\psi $ be a generic Weyl spinor: as seen above, we have $P_R\,\psi_R=\psi_R$ and it is easy to see that $P_L (\psi_R)^{c} =  (\psi_R)^{c}  $, i.e. $ (\psi_R)^{c} $ is left-handed. 
Therefore the sum $\psi_M= \psi_R+ (\psi_R)^c$ is a Majorana spinor because it satisfies \eqref{Majo} and any Majorana fermion can be represented in this way.

Let us consider next the CP discrete symmetry. From above follows that the Action of a Majorana spinor is obviously invariant under it. 

For a Weyl spinor we have
\be
\mathcal{CP}\,\psi_L(t,\vx) ( \mathcal{CP} )^{-1}= \gamma_0\,P_R\,\psi^{\,c}(t,-\vx)= \gamma_0 (\psi^{\,c})_R(t,-\vx)\label{CP}
\ee  
Let us apply now the CP transformation to the Action for a Weyl spinor. We get 
\begin{eqnarray}
&&\mathcal{CP}\int\overline{\psi_L }\,i\gamma\cdot\partial\,\psi_L\,( \mathcal{CP} )^{-1}
= {\int \overline{(\psi^{\,c})_R} (t,-\vx)\,i\gamma^{\,\dagger}\cdot\partial\,(\psi^{\,c})_R(t,-\vx)}
\nonumber\\
&=&\int\overline{(\psi^{\,c})_R} (t,\vx)\,i\gamma\cdot\partial\,(\psi^c)_R(t,\vx)
\label{CPactionLH}
\end{eqnarray}
But we can prove as well that
\be
\int\overline{(\psi^{\,c})_R} (t,\vx)\,i\gamma\cdot\partial\,(\psi^{\,c})_R(t,\vx)
=\int\overline {\psi_L}(x)\,i\gamma\cdot\partial\,\psi_L(x)
\label{CPinvariance}
\ee
Therefore that the Action for a Weyl spinor is CP invariant. Moreover, as we have seen in the previous Section, it is also, separately,
T invariant and so CPT invariant. 

The transformation properties of the Weyl and Majorana spinor fields are summarized as follows:

{
\be
\begin{matrix}\quad &{\rm Majorana}&\quad& {\rm Weyl} \\
{\rm P}:\quad&\mathcal{P}\,\psi (t, \vx)\,\mathcal{P}^{-1} = \gamma_0 \psi (t,
-\stackrel{\rightarrow}{x})&\quad &\mathcal{P}\,\psi_L (t, \vx)\,\mathcal{P}^{-1} = \gamma_0 \psi_R (t,-\stackrel{\rightarrow}{x})
\\
{\rm  C}:\quad& \mathcal{C}\,\psi\,\mathcal{C}^{-1} = \gamma^{\,2}\psi^*=\psi&\quad&\mathcal{C}\,\psi_L\,\mathcal{C}^{-1}  
= P_L\psi^{\,c} = (\psi^c)_L
\\
{\rm CP}:\quad& \mathcal{CP}\,\psi(t,\vx)\,(\mathcal{CP})^{-1} = \psi(t,-\vx) &\quad& \mathcal{CP}\,\psi_L(t,\vx)\,( \mathcal{CP} )^{-1} 
= \gamma_0 (\psi^{\,c})_R(t,-\vx)
\end{matrix}
\ee
}

In the conventional Standard Model of the fundamental interactions the neutrino and antineutrino fields are treated as massless chiral Weyl bispinors:
according to our conventions and notations we can set
\begin{equation}
\psi(x) = \nu_{L}(x) + \nu_{R}(x) =
\int\mathrm{d}\mathbf{p}\,\left[ \, c_{-,\,\mathbf{p}}\, u_{-,\,\mathbf{p}}(x)
+ d^{\,\dagger}_{+,\,\mathbf{p}}\, v_{+,\,\mathbf{p}}(x) \,\right] + \nu_{R}(x)
\label{neutrinoquantumfield} 
\end{equation}
where $u_-,v_+$ are fixed and independent left- and right-handed spinors
that, without loss of generality, we can identify with those ones of eq.~(\ref{Weylspinorwavefunctions}).
It is worthwhile to stress that, in spite of being a spin $ \frac12 $ field, the left-handed neutrino field necessarily exhibits
positive frequency plane waves with only negative helicity and negative frequency plane waves with solely positive helicity.
Contrary to some popular belief, this characteristic striking constraint for the Weyl spinors $ \psi_{\mp} $ is not merely due to their masslessness, because
Majorana and Dirac massless spinors do behave quite differently.

We can interpret the Fourier expansion (\ref{neutrinoquantumfield}) of the Standard Model neutrino field  as follows: $c_{-,\,\mathbf{p}}$ annihilates a left-handed
particle with negative helicity, while $d^{\,\dagger}_{+,\,\mathbf{p}}$ creates a right-handed anti-particle with positive helicity.
Needless to say, it appears to be evident that in this game right-handed neutrinos $ \nu_{R} $ and left-handed anti-neutrinos 
play only the role of sterile and non-interacting spectators,
this role being confined to provide a consistent causal Green's function: 
it turns out that
the latter have never been detected neither in laboratory experiments, nor in astrophysical or cosmological observations.

\section{Regularizations for Weyl spinors}.

\medskip\noindent
The classical Lagrange density for a Weyl (left) spinor in the four component formalism 
\[
\psi(x)=\chi_{L}(x)=\left\lgroup
\begin{array}{c}
\chi(x) \\ 0
\end{array}
\right\rgroup
\]
reads
\[
\mathcal{K}(x)=\overline{\psi}(x)\,i\partial\!\!\!/\,\psi(x)=\chi_{L}^{\dagger}(x)\,\alpha^{\nu}i\partial_{\nu}\chi_{L}(x)
\]
It follows that the corresponding matrix valued Weyl differential operator
\[
w_{L}\equiv\alpha^{\nu}i\partial_{\nu}P_{L}\qquad\quad
[\,P_{L}=\textstyle\frac12(\mathbb{I}-\gamma_5)\,]
\]
is singular and does not possess any rank-four inverse. After minimal coupling with a real massless vector field $ A^{\mu}(x) $
we come to the classical Lagrangian
\[
\mathcal{L}=\chi_{L}^{\dagger}\,\alpha^{\nu}i\partial_{\nu}\chi_{L}
+ gA^{\nu}\,\chi_{L}^{\dagger}\,\alpha_{\nu}\chi_{L} -\textstyle\frac14\,F^{\,\mu\nu}\,F_{\mu\nu}
\]
where $ 0\le g< 1 $ is a non-negative small parameter, while $ F_{\mu\nu}=\partial_{\mu}A_{\nu}-\partial_{\nu}A_{\mu}\,. $
It turns out that the classical Action 
\[
S=\int\mathrm{d}^{4}x\,\mathcal{L}
\]
is invariant under the Poincar\'{e} group,  as well as under  the internal U(1) phase transformations
$ \chi_L(x)\mapsto\,e^{\,ig\theta}\chi_{L}(x)\,. $ The Action is invariant under the so called scale or
dilatation transformations, viz.,
\begin{eqnarray*}
x^{\,\prime\mu}=e^{-\varrho}x^{\,\mu}\qquad \chi_L^{\,\prime}(x)=e^{\,\frac32\varrho}\chi_L(e^{\,\varrho}x)
\qquad A^{\prime\mu}(x)=e^{\,\varrho}A^{\mu}(e^{\,\varrho}x)
\end{eqnarray*}
with $ \varrho\in\mathbb{R} $, as well as with respect to the local phase or gauge transformations
\[
\chi^{\,\prime}_L(x)=\,e^{\,ig\theta(x)}\chi_{L}(x)\qquad\quad A^{\prime}_{\nu}(x)=A_{\nu}(x)+\partial_{\nu}\theta(x)
\]
which amounts to the ordinary U(1) phase transform in the limit of constant phase. It follows therefrom that there are twelve conserved charges
in this model at the classical level and, in particular, owing to scale and gauge invariance, no mass term is allowed for both spinor and vector
fields. The question naturally arises if those symmetries hold true after the transition to the quantum theory and, in particular, if they are
protected against loop radiative corrections within the perturbative approach. Now, in order to develop perturbation theory, one has to face
the problem of the lack of an inverse for both the Weyl and gauge fields, owing to chirality and gauge invariance. In order to solve this problem, it is 
expedient to add to the Lagrangian non-interacting terms, which are fully decoupled from any physical quantity. 
\textbf{These terms break  chirality and
gauge invariance}, albeit in a harmless way, just to allow the setting up of a Feynman propagator, or causal Green's functions,
for both the Weyl and gauge quantum fields. The simplest choice, which keeps Poincar\'{e} and internal U(1) phase change symmetries,
is provided by
\[
\mathcal{L}^{\,\prime}=\varphi_{R}^{\dagger}\,\alpha^{\nu}i\partial_{\nu}\varphi_{R} 
-\textstyle\frac12(\partial\cdot A)^{2}
\]
where 
\[
\psi(x)=\varphi_{R}(x)=\left\lgroup
\begin{array}{c}
0 \\ \varphi(x)
\end{array}
\right\rgroup
\]
is a left-chirality breaking right-handed Weyl spinor field. Notice \textit{en passant} that the modified Lagrangian 
$ \mathcal{L}+\mathcal{L}^{\prime} $ does exhibit the further U(1) internal symmetry under the so called chiral phase transformations
\[
\psi^{\prime}(x)=(\cos\theta+ig\sin\theta\,\gamma_{5})\psi(x)
\qquad\quad\psi(x)=\left\lgroup
\begin{array}{c}
\chi(x) \\ \varphi(x)
\end{array}
\right\rgroup
\]
so that the modified theory involves another conserved charge at the classical level.
From the modified Lagrange density we get the Feynman propagators for the massless Dirac field $ \psi(x) $,
as well as for the massless vector field in the so called Feynman gauge: namely,
\[
S(p)=\frac{ip\!\!/}{p^{2}+i\varepsilon}\qquad\quad D_{\mu\nu}(k)=\frac{-ig_{\mu\nu}}{k^{2}+i\varepsilon}
\]
and the vertex $ ig\gamma^{\nu}P_L\,,  $ with $ k+p-q=0 $, which involves a vector particle of momentum $ k $
and a Weyl pair of particle and anti-particle of momenta $ p $ and $ q $ respectively and of opposite helicity.\footnote{Customarily, 
the on-shell 1-particle states of a left Weyl spinor field are a left-handed particle with negative helicity 
$ -\frac12\hslash $ and  a right-handed antiparticle of positive helicity $ \frac12\hslash $.}

The lowest order 1-loop correction to the Weyl kinetic term $ p\!\!/P_L $ is formally provided by the Feynman rules
in the Minkowski space, viz.,
\[
\Sigma_2(p\!\!/) =
 -\,ig^{\,2}\int\frac{\mathrm{d}^{4}\ell}{(2\pi)^{4}}\;
\gamma^{\,\mu}\,D_{\mu\nu}(\,p-\ell\,)\,S(\ell)\,\gamma^{\,\nu}P_L
\]
By na\"{\i}ve power counting the above 1-loop integral turns out to be UV divergent. Hence a regularization procedure is
mandatory to give a meaning and evaluate the radiative correction $ \Sigma_2(p\!\!/) $ to the Weyl kinetic operator. 
Here in the sequel we shall examine in detail 
the dimensional, Pauli-Villars and UV cut-off regularizations.

\medskip
\subsubsection{Dimensional regularization}

\medskip
In a $ 2\omega- $dimensional space-time the radiative correction to the Weyl kinetic term  takes the form 
\[
\mathtt{reg}\,\Sigma_2(p\!\!/) 
= -\,ig^{\,2}\mu^{\,2\epsilon}\int\frac{\mathrm{d}^{2\omega}\ell}{(2\pi)^{2\omega}}\;D_{\mu\nu}(\ell)\,
\gamma^{\,\mu}\,S(\ell+p)\,\gamma^{\,\nu}P_L
\]
where $ \epsilon=2-\omega>0 $ is the shift with respect to the Minkowski space. Since the above expression  
is traceless and has the canonical engineering dimension of a mass in natural units, it is quite apparent that the
latter cannot generate any mass term, which would be proportional to the unit matrix. 
Hence, mass is forbidden and it is convenient to set and evaluate
\begin{eqnarray*}
\mathtt{reg}\,\Sigma_2(p\!\!/) &\equiv& f(\,p^{\,2})\,p\!\!/P_L\qquad\quad
\mathrm{tr}\,[\,p\!\!/\mathtt{reg}\,\Sigma_2(p\!\!/)]=\textstyle\frac12\,2^{\,\omega}p^{\,2}f(\,p^{\,2})
\\
\mathrm{tr}\,[\,p\!\!/\mathtt{reg}\,\Sigma_2(p\!\!/)]
&=& g^{\,2}\mu^{\,2\epsilon}\,{(2\pi)^{-\,2\omega}}
\int{\mathrm{d}^{2\omega}\ell}\;
\frac{(-\,i\,)\,\mathrm{tr}\left(\,p\!\!/\gamma^{\,\lambda}\ell\!\!/\gamma_{\lambda}P_L\,\right) }
{\left[\,(\ell-p)^2+i\varepsilon\,\right]\left(\,\ell^{\,2}+i\varepsilon\,\right)}
\end{eqnarray*}
Concerning dimensional regularization, we collect here below the
definitions and key properties for the $2^\omega\times
2^\omega$ $\gamma-$matrices in a $2\omega-$dimensional space-time with
a Minkowski signature
\begin{eqnarray*}
&&
\gamma^{\,\mu} = \left\lbrace
\begin{array}{cc}
\bar\gamma^{\,\mu} \qquad & \mu = 0,1,2,3
\\
\hat\gamma^{\,\mu} \qquad & \mu = 4,\ldots,2\omega-4
\end{array}\right.
\\
&&\left\{\bar\gamma^{\,\mu},\bar\gamma^\nu\right\}=2\bar g^{\,\mu\nu}\,
{\mathbb I}\qquad
\left\{\hat\gamma^{\,\mu},\hat\gamma^\nu\right\}=2\hat g^{\,\mu\nu}\,
{\mathbb I}\qquad
\left\{\bar\gamma^{\,\mu},\hat\gamma^\nu\right\}=0
\label{A4}\\
&&\left\|\,\bar g\,\right\|={\rm diag}\,(+,-,-,-)\qquad 
\left\|\,\hat g\,\right\|=\ -\ \hat{\mathbb I}
\label{A5}\\
&&\gamma_5\equiv i\bar\gamma^0\bar\gamma^1\bar\gamma^2\bar\gamma^3\qquad
\gamma_5^2={\mathbb I}\qquad
\left\{\bar\gamma^{\,\mu},\gamma_5\right\}=0=\left[\,\hat\gamma^{\,\mu},\gamma_5\,\right] 
\end{eqnarray*}
Taking all the above listed equations into account, it is not
difficult to check the following trace formulas: namely,
\begin{eqnarray*}
\mathrm{tr}\left(\gamma^\mu\gamma^\nu\right)&=&
g^{\mu\nu}\,\mathrm{tr}\,{\mathbb I}\ =\ 2^{\,\omega}\,g^{\,\mu\nu}\label{traccia2}
\\
2^{-\omega}\mathrm{tr}\left(\gamma^\kappa\gamma^\lambda\gamma^\mu\gamma^\nu\right)&=&
g^{\kappa\lambda}\,g^{\mu\nu}\ -\
g^{\kappa\mu}\,g^{\lambda\nu}\ +\ g^{\kappa\nu}\,g^{\lambda\mu}\label{traccia4}
\\
\mathrm{tr}\left(\bar\gamma^\kappa\bar\gamma^\lambda\hat\gamma^\mu\hat\gamma^\nu\right) &=&
2^{\,\omega}\,\bar g^{\kappa\lambda}\,\hat g^{\mu\nu}
\\
\mathrm{tr}\left(\gamma_5\bar\gamma^\mu\bar\gamma^\lambda\bar\gamma^\rho
\bar\gamma^\nu\right)&=& -\ i\,2^{\,\omega}\epsilon^{\mu\lambda\rho\nu}
\label{A7}
\end{eqnarray*}
Traces involving an odd number of Dirac matrices do vanish. Then we get
\[
\mathrm{tr}\left(\,p\!\!/\gamma^{\,\lambda}\ell\!\!/\gamma_{\lambda}\,P_L\,\right) 
=2^{\,\omega}(\epsilon-1)\,p\cdot\ell 
\]
and thereby
\[
p^2\,f(\,p^{\,2})  = ig^{\,2}\mu^{\,2\epsilon}\;\frac{\epsilon-1}{(2\pi)^{2\omega}}
\int\frac{2p\cdot\ell\ \mathrm{d}^{2\omega}\ell}
{\left[\,(\ell - p)^2+i\varepsilon\,\right]\left(\,\ell^{\,2}+i\varepsilon\,\right)}
\]
Turning to the Feynman parametric representation we obtain
\[
p^2\,f(\,p^{\,2}) = ig^{\,2}\mu^{\,2\epsilon}\;\frac{\epsilon-1}{(2\pi)^{2\omega}}
\int_0^1\mathrm{d} x\int\frac{2p\cdot\ell\ \mathrm{d}^{2\omega}\ell}
{\left[\,\ell^{\,2} - 2x\,p\cdot \ell + xp^{\,2}  + i\varepsilon\,\right]^2}
\]
Completing the square in the denominator and after shifting the momentum
$\ell^{\,\prime}\equiv \ell-xp\,,$ dropping the linear term in $\ell^{\,\prime}$ in the numerator owing
to symmetric integration, we have
\[
f(\,p^{\,2}) = 2ig^{\,2}\mu^{\,2\epsilon}\;\frac{\epsilon-1}{(2\pi)^{2\omega}}
\int_0^1\mathrm{d} x\,x\int\frac{\mathrm{d}^{2\omega}\ell}
{\left[\,\ell^{\,2}+x(1-x)\,p^{\,2}+i\varepsilon\,\right]^2}
\]
When $ p^{\,2}<0 $ one can perform the Wick rotation and readily get the result
\begin{eqnarray*}
f(\,p^{\,2}) &=& -\,2 g^{\,2}\mu^{2\epsilon}\;\frac{\epsilon-1}{(4\pi)^{\omega}}
\int_0^1\mathrm{d} x\,x\int_{0}^{\infty}\mathrm{d}\tau\,\tau^{\,\epsilon-1}\,e^{-\tau x(1-x)\,p_{E}^{\,2}}
\\
&=& 2\left( \frac{g}{4\pi}\right)^{2}[\,\Gamma(\epsilon) - \Gamma(1+\epsilon)\,]
\left( -\,\frac{4\pi\mu^{2}}{p^{\,2}}\right)^{\epsilon}
\int_{0}^{1}\mathrm{d}x\,x^{\,1-\epsilon}(1-x)^{-\epsilon}
\\
&=& 2\left( \frac{g}{4\pi}\right)^{2}[\,\Gamma(\epsilon) - \Gamma(1+\epsilon)\,]
\left( -\,\frac{4\pi\mu^{2}}{p^{\,2}}\right)^{\epsilon} B(2-\epsilon,1-\epsilon)
\end{eqnarray*}
Expansion around $ \epsilon=0 $ yields
\begin{eqnarray*}
f(p^2) &=& \left( \frac{g}{4\pi}\right)^{2} \left[\,\frac{1}{\epsilon} - \mathbf{C} - 1 +\cdots\,\right] 
\left[ \,1 + \epsilon\ln\left( -\,\frac{4\pi\mu^{2}}{p^{\,2}}\right) +\cdots\,\right] 
\\
&\times& \left\lbrace 1 - \epsilon\,[ \,\psi(2)+\psi(1)-2\psi(3)\,]+\cdots\right\rbrace 
\\
&=&\left( \frac{g}{4\pi}\right)^{2} \left[\,\frac{1}{\epsilon} + 1
+3\mathbf{C} + \ln\left( -\,\frac{4\pi\mu^{2}}{p^{\,2}}\right) \,\right] + \mathrm{evanescent}
\end{eqnarray*}

\medskip
\subsubsection{Pauli-Villars regularization}

\medskip
Let us now repeat the very same calculation for the left Weyl spinor self-energy in the Pauli-Villars regularization.
The latter is simply implemented by the following replacement of the massless Dirac propagator
\[
\mathtt{reg}\,\Sigma_2(p\!\!/) =
 -\,ig^{\,2}\int\frac{\mathrm{d}^{4}\ell}{(2\pi)^{4}}\;
\gamma^{\,\mu}\,D_{\mu\nu}(\,p-\ell\,)\,
\sum_{s\,=\,0}^S C_s\,S(\ell\,,\,M_s)\,
\,\gamma^{\,\nu}P_L
\]
where $M_0 = 0\,,\ C_0=1$ while 
$\{\,M_s\equiv \lambda_s\,M\,|\,\lambda_s\gg 1\ (\, s=1,2,\ldots,S\,)\,\}$
is a collection of very large auxiliary masses. The set of constants
$C_s$ will be suitably selected,
as we shall see in the sequel, in such a manner 
to obtain a specific and mathematically meaningful form for the 
ultraviolet divergences that will manifest themselves in the limit
$\lambda_{s} \rightarrow\infty\,.$
Since we have
\[
S(\ell\,,\,M_s)=\frac{ i(\ell\!\!/ + M_{s}) }{ \ell^{\,2} - M_{s}^{2} + i\varepsilon }
\qquad\quad(\,s=1,2,\ldots,S\,)
\]
it is now convenient to set
\[
\mathtt{reg}\,\Sigma_2(p\!\!/) \equiv f(p^{\,2})\,p\!\!/P_L - M\upsilon(\,p^{\,2})
\]
in such a manner that we can write
\begin{eqnarray*}
\mathrm{tr} \left[ \,\mathtt{reg}\,\Sigma_2(p\!\!/)\,\right]  &=& -\,4M\upsilon(\,p^{\,2})
\\
&=& -\,ig^{\,2}\int\frac{\mathrm{d}^{4}\ell}{(2\pi)^{4}}\;\sum_{s\,=\,1}^S C_s\,
\frac{ 8M_s }{ [\,(\ell-p)^{2}+i\varepsilon\,] \left( \ell^{\,2} - M_{s}^{2} + i\varepsilon \right)  }
\\
\mathrm{tr} \left[ \,p\!\!/\mathtt{reg}\,\Sigma_2(p\!\!/)\,\right]  &=& 2p^{\,2}f(\,p^{\,2})
\\
&=& ig^{\,2}\int\frac{\mathrm{d}^{4}\ell}{(2\pi)^{4}}\;\sum_{s\,=\,0}^S C_s\,
\frac{ 4\,p\cdot\ell }{ [\,(\ell-p)^{2}+i\varepsilon\,] \left( \ell^{\,2} - M_{s}^{2} + i\varepsilon \right)  }
\end{eqnarray*}
Once again, if we take profit of the Feynman parametric formula we obtain
\begin{eqnarray*}
\upsilon(\,p^{\,2})&=&
\int\frac{\mathrm{d}^{4}\ell}{(2\pi)^{4}}\;\sum_{s\,=\,1}^S C_s\int_{0}^{1}\mathrm{d}x\,
\frac{ 2ig^{\,2}\lambda_s }{\left[\,\ell^{\,2} - 2x\,p\cdot \ell + xp^{\,2}  - (1-x)M_{s}^{2} + i\varepsilon\,\right]^2}
\\
&=&\int_{0}^{1}\mathrm{d}x\int\frac{\mathrm{d}^{4}\ell^{\,\prime}}{(2\pi)^{4}}\;\sum_{s\,=\,1}^S 
\frac{ 2ig^{\,2}C_s\lambda_s }{\left[\,\ell^{\,\prime 2}  + x(1-x)p^{\,2}  - (1-x)M_{s}^{2} + i\varepsilon\,\right]^2}
\\
&=&-\,2g^{\,2}\int_{0}^{1}\mathrm{d}x\int\frac{\mathrm{d}^{4}\ell_E}{(2\pi)^{4}}\;\sum_{s\,=\,1}^S 
\frac{ C_s \lambda_s }{\left[\,\ell_{E}^{\,2}  + x(1-x)p_{E}^{\,2}  + (1-x)M_{s}^{2} \,\right]^2}
\end{eqnarray*}
In order to proceed further on it is convenient to consider the generating integral
\[
I_{\,n}\,(z_E) \equiv
(-1)^{n}\int\frac{\mathrm{d}^4 \ell_E}{(2\pi)^{4}}
\sum_{s=1}^S C_s\lambda_{s}\,\frac{\exp\lbrace i\ell_E\cdot z_E\rbrace}
{(\ell_E^{\,2}+\Delta_{s}^{2})^{n}}
\]
where $ \Delta_{s}^{2}=(1-x)[\,xp_{E}^{\,2}  + M^{2}\lambda_{s}^{2}\,]>0\,,\ \forall s=1,2,\ldots,S\,. $
Explicit calculation yields
\begin{equation}
I_{\,n}\,(z_E) = \frac{(-1)^{\,n}}{8\pi^2\,\Gamma(n)}\,
\sum_{s=1}^S C_s\lambda_{s}\,\left(\frac{2\,\Delta_{s}}{|\,z_E\,|}\right)^{2-n}\,
K_{\,2-n}(\,\Delta_{ s}\,\vert\,z_E\,\vert\,)
\end{equation}
where $ z_{E}=(\mathbf{z},z_{4}) $ while $ \vert\,z_E\,\vert=\sqrt{\mathbf{z}^{2}+z_4^{2}}\,, $
whence we immediately obtain 
$$
I_{\,2}\,(z_E) =
\frac{1}{8\pi^2}\,
\sum_{s\,=\,1}^S C_s\lambda_{s}\,K_{0}(z_{s})\qquad\quad(\,z_{s}=\Delta_{ s}\,\vert\,z_E\,\vert\,)
$$
where $K_{0}$ is the modified Bessel function of the third kind and of order zero, the series representation of which
is provided by
$$
K_{0}(z)=\sum_{k\,=\,0}^\infty
\frac{1}{(k!)^2}\left(\frac{z}{2}\right)^{2k}
\left[\,\psi(k+1)-\ln\frac{z}{2}\,\right]=\ln\frac{2}{z} -\,\mathbf{C}+O(z^{2}\ln z)
$$
Then we can write
\[
\frac{1}{4M}\,\mathrm{tr} \left[ \,\mathtt{reg}\,\Sigma_2(p\!\!/)\,\right] = -\,\upsilon(\,p^{\,2}) = 2g^{\,2}\int_{0}^{1}\mathrm{d}x\,
\lim_{\,z_{E}\to\,0}I_{\,2}\,(z_E) 
\]
whence it is clear that the limit exists, so that the regularization works, iff the Pauli-Villars condition is fulfilled, viz.,
\[
\sum_{s\,=\,1}^S C_s\lambda_{s}=0
\]
which provides in turn $ \upsilon=0 $, i.e., no mass term is allowed in the 1-loop Weyl kinetic term, both
in dimensional  and Pauli-Villars regularizations, as expected.

Let us turn now to the other form factor
\begin{eqnarray*}
&&\mathrm{tr} \left[ \,p\!\!/\mathtt{reg}\,\Sigma_2(p\!\!/)\,\right]  = 2p^{\,2}f(\,p^{\,2})
\\
&=& ig^{\,2}\int\frac{\mathrm{d}^{4}\ell}{(2\pi)^{4}}\;\sum_{s\,=\,0}^S C_s\,
\frac{ 4\,p\cdot\ell }{ [\,(\ell-p)^{2}+i\varepsilon\,] \left( \ell^{\,2} - M_{s}^{2} + i\varepsilon \right)  }
\\
&=&\int\frac{\mathrm{d}^{4}\ell}{(2\pi)^{4}}\;\sum_{s\,=\,0}^S C_s\int_{0}^{1}\mathrm{d}x\,
\frac{ 4ig^{\,2}p\cdot\ell }{\left[\,\ell^{\,2} - 2x\,p\cdot \ell + xp^{\,2}  - (1-x)M_{s}^{2} + i\varepsilon\,\right]^2}
\\
&=&4ig^{\,2}\int_{0}^{1}\mathrm{d}x\int\frac{\mathrm{d}^{4}\ell^{\,\prime}}{(2\pi)^{4}}\;\sum_{s\,=\,0}^S 
\frac{ C_s\,xp^{\,2} }{\left[\,\ell^{\,\prime 2}  + x(1-x)p^{\,2}  - (1-x)M_{s}^{2} + i\varepsilon\,\right]^2}
\\
&=& -\,2g^{\,2}p^{\,2}\int_{0}^{1}\mathrm{d}x\,x\int\frac{\mathrm{d}^{4}\ell_E}{(2\pi)^{4}}\;\sum_{s\,=\,0}^S 
\frac{ C_s  }{\left[\,\ell_{E}^{\,2}  + x(1-x)p_{E}^{\,2}  + (1-x)M_{s}^{2} \,\right]^2}
\end{eqnarray*}
whence we obtain
\begin{eqnarray*}
f(p^{\,2}) &=& -\,g^{\,2}\int_{0}^{1}\mathrm{d}x\,x\int\frac{\mathrm{d}^{4}\ell_E}{(2\pi)^{4}}\;\sum_{s\,=\,0}^S 
\frac{ C_s  }{\left[\,\ell_{E}^{\,2}  + \Delta_{s}^{2} \,\right]^2}
\\
&=& -\,g^{\,2}\int_{0}^{1}\mathrm{d}x\,x\int\frac{\mathrm{d}^{4}\ell_E}{(2\pi)^{4}}\;\sum_{s\,=\,0}^S  C_s  
\int_{0}^{\infty}\mathrm{d}\tau\,\tau\,\exp\left\lbrace -\,\tau\,\ell_{E}^{\,2}  - \tau\,\Delta_{s}^{2} \right\rbrace 
\\
&=&-\,\frac{g^{\,2}}{16\pi^{2}}\int_{0}^{1}\mathrm{d}x\,x  
\int_{0}^{\infty}\frac{\mathrm{d}\tau}{\tau}\sum_{s\,=\,0}^S C_s 
\exp \left\lbrace -\,\tau(1-x)[\,xp_{E}^{\,2}  + M^{2}\lambda_{s}^{2}\,]\right\rbrace 
\\
&=&-\,\frac{g^{\,2}}{16\pi^{2}}\int_{0}^{1}\mathrm{d}x\,x  
\int_{0}^{\infty}\frac{\mathrm{d}\tau}{\tau}
\exp \left\lbrace \tau x(1-x)\,p^{\,2}\right\rbrace
\\
&\times&\sum_{s\,=\,0}^S C_s \left[\, 1-\sum_{n=1}^{\infty}\frac{\tau^{\,n}}{n!}(1-x)^{n}M^{2n}\lambda_{s}^{2n}\,\right]
\end{eqnarray*}
Integrability requires the further Pauli-Villars condition
\[
\sum_{s\,=\,1}^S C_s =-1
\]
and consequently
\begin{eqnarray*}
f(p^{\,2}) &=& -\,\frac{g^{\,2}}{16\pi^{2}}\int_{0}^{1}\mathrm{d}x\,x  
\sum_{n=1}^{\infty}\frac{1}{n!}(1-x)^{n}M^{2n}\sum_{s\,=\,1}^S C_s\lambda_{s}^{2n}
\\
&\times&\int_{0}^{\infty}\frac{\mathrm{d}\tau}{\tau}\,\tau^{\,n}
\exp \left\lbrace \tau x(1-x)\,p^{\,2}\right\rbrace
\\
&=& -\,\frac{g^{\,2}}{16\pi^{2}}\sum_{s\,=\,1}^S C_s\int_{0}^{1}\mathrm{d}x\,x  
\sum_{n=1}^{\infty}\frac{1}{n}\left( \frac{\lambda_{s}^{2}M^{2}}{xp^{\,2}}\right)^{n}
\\
&=&\frac{g^{\,2}}{16\pi^{2}}\sum_{s\,=\,1}^S C_s\int_{0}^{1}\mathrm{d}x\,x \,
\ln\left( 1+\frac{\lambda_{s}^{2}M^{2}}{-\,xp^{\,2}}\right)
\\
&=&\frac{g^{\,2}}{16\pi^{2}}\int_{0}^{1}\mathrm{d}x\,x\left\lbrace 
\sum_{s\,=\,1}^S C_s\,\ln\left( \lambda_{s}^{2} -\,\frac{xp^{\,2}}{M^{2}}\right)  
- \ln\left( -\,\frac{xp^{\,2}}{M^{2}}\right) \right\rbrace 
\\
&=&\frac{g^{\,2}}{16\pi^{2}}\left\lbrace 
-\,\frac12\,\ln\left( -\,\frac{p^{\,2}}{M^{2}}\right) - \int_{0}^{1}\mathrm{d}x\,x\,\ln x
+ \frac12\sum_{s\,=\,1}^S C_s\,\ln\lambda_{s}^{2}\right. 
\\
&+& \left. \int_{0}^{1}\mathrm{d}x\,x\,\sum_{s\,=\,1}^S C_s\,\ln\left( 1 -\,\frac{xp^{\,2}}{M_{s}^{2}}\right)  
\right\rbrace 
\\
&=&\left( \frac{g}{4\pi}\right)^{2} \left[\,
\sum_{s\,=\,1}^S C_s\,\ln\lambda_{s} + \frac14 + \frac12\ln\left( -\,\frac{M^{\,2}}{p^{\,2}}  \right)\, 
\right] + \mathrm{evanescent}
\\
&\sim& \left( \frac{g}{4\pi}\right)^{2} \left[\,\frac{1}{\epsilon} + 1
+3\mathbf{C} + \ln\left( -\,\frac{4\pi\mu^{2}}{p^{\,2}}\right) \,\right] + \mathrm{evanescent}
\end{eqnarray*}
where, in the last line, a comparison with the result in dimensional regularization has been exhibited.
It follows that by identifying 
\[
\frac{1}{\epsilon}=\sum_{s\,=\,1}^S C_s\,\ln\lambda_{s} 
\]
in which the two Pauli-Villars conditions\footnote{This means the the minimal choice is $ S=2 $.} hold true, viz.,
\[
\qquad\sum_{s\,=\,1}^S C_s =-1\qquad\sum_{s\,=\,1}^S C_s\lambda_{s}=0
\]
the coefficient of the divergent part turns out to be the same for both regulators, as expected.

\medskip
\subsubsection{Cut-off regularization} 

\medskip\noindent
Let us repeat once more the one loop calculation
of the Weyl spinor self-energy with a physical very large cut-off regulator, e.g.  $ K=l_{P}^{-1} $ with $ l_{P}=\sqrt{\hslash G_N/c^{\,3}} $
the Planck length. Of course, once again, no mass term is allowed for the trace of an odd number of gamma matrices is null.
Moreover, for the previously introduced coefficient of the Weyl kinetic term we find with
$ \ell^{\,\mu}=(\ell_{0},	\vec{\ell}\;) $
\begin{eqnarray*}
p^2\,f(\,p^{\,2})  &=& \frac{-\,ig^{\,2}}{(2\pi)^{4}}
\int\frac{2p\cdot\ell\ \theta\left(K^{2}-\vec{\ell}^{\;2}\,\right)\,\mathrm{d}^{4}\ell}
{\left[\,(\ell - p)^2+i\varepsilon\,\right]\left(\,\ell^{\,2}+i\varepsilon\,\right)}
\\
&=& \frac{-\,ig^{\,2}}{(2\pi)^{4}}\int_{0}^{1}\mathrm{d}x
\int\frac{2p\cdot\ell\ \theta\left(K^{2}-\vec{\ell}^{\;2}\,\right)\,\mathrm{d}^{4}\ell}
{\left[\,\ell^{\,2} -2x p\cdot\ell + x p^{\,2}+i\varepsilon\,\right]^{2}}
\\
&=& \frac{-\,ig^{\,2}}{(2\pi)^{4}}\int_{0}^{1}\mathrm{d}x\int\mathrm{d}\vec{\ell}\ \theta\left(K^{2}-\vec{\ell}^{\;2}\,\right)
\\
&\times& \lim_{\eta\,\rightarrow\,0}\,\frac{\mathrm{d}}{\mathrm{d}\eta}\int_{-\infty}^{\infty}\mathrm{d}\ell_{0}\;
\frac{ 2p_{0}\ell_{0}-2\vec{p}\cdot\vec{\ell} }
{\ell^{\,2} - 2x p\cdot\ell + x p^{\,2} -\eta +i\varepsilon}
\end{eqnarray*}
After translation of the integration variable $ \ell\mapsto\ell-xp $, for a suitably large cut-off $ K $ and space-like momentum
$ p^{\,2}<0 $ we get
\begin{eqnarray*}
p^2\,f(\,p^{\,2})  &=& 
\frac{-\,ig^{\,2}}{(2\pi)^{4}}\int_{0}^{1}\mathrm{d}x\int\mathrm{d}\vec{\ell}\ \theta\left(K^{2}-(\vec{\ell}+x\vec{p}\,)^{2}\,\right)
\\
&\times& \lim_{\eta\,\rightarrow\,0}\,\frac{\mathrm{d}}{\mathrm{d}\eta}\int_{-\infty}^{\infty}\mathrm{d}\ell_{0}\;
\frac{ 2p_{0}\ell_{0}-2\vec{p}\cdot\vec{\ell} +2xp^{\,2} }
{\ell^{\,2} + x(1-x) p^{\,2} -\eta +i\varepsilon}
\\
&=& 
\frac{-\,ig^{\,2}}{(2\pi)^{4}}\int_{0}^{1}\mathrm{d}x\int\mathrm{d}\vec{\ell}\ \theta\left(4K^{2}-\vec{\ell}^{\;2}\,\right)
\\
&\times& \lim_{\eta\,\rightarrow\,0}\,\frac{\mathrm{d}}{\mathrm{d}\eta}\int_{-\infty}^{\infty}
\frac{ \mathrm{d}\ell_{0}\;2xp^{\,2} }
{\ell_{0}^{\,2} - \vec{\ell}^{\;2} + x(1-x) p^{\,2} -\eta +i\varepsilon}
\\
&=&\frac{-\,g^{\,2}p^{\,2}  }{(2\pi)^{3}}\int_{0}^{1}\mathrm{d}x\,x\int\mathrm{d}\vec{\ell}\ \theta\left( 4K^{2}-\vec{\ell}^{\;2}\,\right)
\\
&\times& 
\lim_{\eta\,\rightarrow\,0}\,\frac{\mathrm{d}}{\mathrm{d}\eta}
\left[ \,\vec{\ell}^{\;2} - x(1-x)\, p^{\,2} -\eta \,\right] ^{-1/2}
\\
f(\,p^{\,2}) &=&\frac{g^{\,2}}{4\pi^{2}}\int_{0}^{1}\mathrm{d}x\,x\int_{0}^{2K}\mathrm{d}\ell\,\ell^{\,2}
\left[\, \ell^{\,2}-x(1-x)\,p^{\,2}\,\right] ^{-3/2}
\\
&=&\frac{g^{\,2}}{4\pi^{2}}\int_{0}^{1}\mathrm{d}x\,x\int_{0}^{1}\mathrm{d}y\,y^{\,2}
\left[\, y^{\,2}-x(1-x)\,\frac{p^{\,2}}{4K^{2}}\,\right] ^{-3/2}
\end{eqnarray*}
Consider $ u=\sqrt{a+y^{2}}\,,\,a=x(1-x)(-p^{2}/4K^{2})>0 $, then from \cite{GR}, \textbf{2.272} 4. p. 86 we get
the elementary result
\begin{eqnarray*}
\int_{0}^{1}\mathrm{d}y\;\frac{y^{2}}{u^{3}}= \left[ \,-\,\frac{y}{u} +\ln(y+u)\,\right] _{0}^{1}
=\frac{-1}{\sqrt{a+1}} + \ln\left( 1+\sqrt{a+1}\,\right) - \ln \sqrt{a}
\end{eqnarray*}
For a very large cut-off $ K\rightarrow\infty $ we obtain
\begin{eqnarray*}
f(\,p^{\,2}) &=&\frac{g^{\,2}}{4\pi^{2}}\int_{0}^{1}\mathrm{d}x\,x\left\lbrace 
-1+\ln2+\frac12\ln\left( -\,\frac{4K^{2}}{p^{\,2}}\right) -\frac12\ln x(1-x)\right\rbrace +\cdots
\\
&=& \left( \frac{g}{4\pi}\right) ^{2}\left[ \,\ln\left( -\,\frac{4K^{2}}{p^{\,2}}\right) + \ln4 \,\right] 
+\mathrm{evanescent}
\end{eqnarray*}

\medskip
To sum up, we have verified that the 1-loop correction to the (left) Weyl spinor self-energy has the general form,
which is \textbf{universal, i.e. regularization independent}: namely,
\begin{eqnarray*}
\mathtt{reg}\,\Sigma_2(p\!\!/) &\equiv& f(\,p^{\,2})\,p\!\!/P_L
\\
f(\,p^{\,2})&:=&\left( \frac{g}{4\pi}\right)^{2} \left[\,\frac{1}{\epsilon} + 1
+3\mathbf{C} + \ln\left( -\,\frac{4\pi\mu^{2}}{p^{\,2}}\right) \,\right]\qquad  (\,\mathrm{DR}\,)
\\
&:=&\left( \frac{g}{4\pi}\right)^{2} \left[\,
\sum_{s\,=\,1}^S C_s\,\ln\lambda_{s} + \frac14 + \frac12\ln\left( -\,\frac{M^{\,2}}{p^{\,2}}  \right)\, 
\right]\qquad (\,\mathrm{PV}\,)
\\
&:=&\left( \frac{g}{4\pi}\right)^{2} \left[\,\ln\left( -\,\frac{4K^{2}}{p^{\,2}}\right) + \ln4 \,\right] 
\qquad\quad(\,\mathrm{CUT-OFF}\,)
\end{eqnarray*}

\medskip\noindent
Some final remarks and comments are in  order.
\begin{enumerate}
\item In the present model of a left Weyl spinor minimally coupled with a gauge vector potential,
\textbf{no mass term can be generated by the radiative corrections} in any regularization scheme.
The left-handed part of the classical kinetic term does renormalize, while its right-handed part
does not undergo any radiative correction and keeps on being free. The latter has to be necessarily 
introduced in order to define a Feynman propagator for the massless spinor field, like the gauge
fixing term to invert the kinetic term of the gauge potential.
The (one loop) renormalized  Lagrangian for a Weyl fermion minimally coupled with a gauge vector potential 
has the universal - i.e. regularization independent - form
\begin{eqnarray*}
\mathcal{L}_{1-\rm loop} &=& \mathcal{L}=\chi_{L}^{\dagger}\,\alpha^{\nu}i\partial_{\nu}\chi_{L}
+ gA^{\nu}\,\chi_{L}^{\dagger}\,\alpha_{\nu}\chi_{L} -\textstyle\frac14\,F^{\,\mu\nu}\,F_{\mu\nu}
\\
&+& \varphi_{R}^{\dagger}\,\alpha^{\nu}i\partial_{\nu}\varphi_{R}  
-\textstyle\frac12(\partial\cdot A)^{2} - (Z_{3}-1)\textstyle\frac14\,F^{\,\mu\nu}\,F_{\mu\nu}
\\
&+& (Z_{2}-1)\chi_{L}^{\dagger}\,\alpha^{\nu}i\partial_{\nu}\chi_{L}
+ (Z_{1}-1)gA^{\nu}\,\chi_{L}^{\dagger}\,\alpha_{\nu}\chi_{L} 
\\
(Z_{2}-1)&=& -\left( \frac{g}{4\pi}\right)^{2} \left[\,\frac{1}{\epsilon} + F_{2}(\epsilon,p^{\,2}/\mu^{2})\,\right] 
+ \cdots
\\
&=& -\left( \frac{g}{4\pi}\right)^{2} \left[\,
\sum_{s\,=\,1}^S C_s\,\ln\lambda_{s} + \widetilde F_2(\lambda_{s},p^{\,2}/M^{2})\,\right] 
+ \cdots
\\
&=& -\,\left( \frac{g}{4\pi}\right)^{2} \left[\,\ln\left( -\,\frac{4K^{2}}{p^{\,2}}\right) + \ln4 + \widehat{F}_{2}(K^{2}/p^{\,2})\,\right] 
+\cdots
\end{eqnarray*}
where conventional notations have been employed. Notice that, as usual, the arbitrary finite parts $ F_2,\widetilde F_2, \widehat{F}_2 $ 
of the countertems  are analytic for $ \epsilon\rightarrow0 $ and $ \lambda_{s},K\rightarrow\infty $, respectively,  and have to be 
uniquely fixed by the renormalization prescription.
\item The interaction definitely preserves left chirality and scale invariance of the counterterms  in the transition from the classical to the 
(perturbative) quantum theory: no mass coupling between the left-handed (interacting) Weyl spinor $ \chi_{L} $
and right-handed (free) Weyl spinor $ \varphi_{R} $ can be generated by radiative loop corrections.

\item While the cut-off and dimensional regularized theory does admit a local formulation 
in  $ D=4 $ or $ D=2\omega $ space-time dimensions, there
is no such local formulation for the Pauli-Villars regularization. The reason is that the PV spinor propagator
\[
\sum_{s\,=\,0}^S C_s\,S(\ell\,,\,M_s)
\]
where $M_0 = 0\,,\ C_0=1$ while 
$\{\,M_s\equiv \lambda_s\,M\,|\,\lambda_s\gg 1\ (\, s=1,2,\ldots,S\,)\,\}$,
cannot be the inverse of any local differential operator of the Calderon-Zygmund type.
Hence, there is no local Action involving a bilinear spinor term that can produce, after a suitable causal inversion,
the Pauli-Villars regularized spinor propagator.

{
In order to be fully explicit and exhaustive, consider the simplest example: namely,
\begin{equation}
\mathtt{reg}\,S(p,m)= \frac{i(\p+m)}{p^{2} - m^{2} + i\varepsilon} - \frac{i(\p+M)}{p^{2}-M^{2}+i\varepsilon}
\label{PVpropagator} 
\end{equation}
where $ m $ is the physical fermion mass while $ M  $ is a very large $ M>>m $ regularization mass parameter.
As a first comment we notice that the minus sign in front of the second propagator means that in the regularized theory
an unphysical ghost fermion has been introduced that provides negative norm states in the Fock space.
The expression (\ref{PVpropagator}) can be readily rewritten in the more suggestive form
\begin{eqnarray*}
i\,\mathtt{reg}\,S(p,m) = (M-m) [\,\p(m+M) + p^{2} +mM\,] [\,(\,p^{2} - m^{2} + i\varepsilon)(\,p^{2} - M^{2} + i\varepsilon)\,]^{-1}
\end{eqnarray*}
whence it appears to be evident that the ultraviolet behavior of the regularized Feynman propagator has been improved by one power
in momentum space, as expected. Now we easily find
\begin{eqnarray*}
[\,\p (m+M) + p^{2} + mM\,]^{-1} = [\,\p (m+M) - p^{2} - mM\,] [\,p^{2}(m+M)^{2} - (\,p^{2} + mM)^{2}\,]_{\,\ast}^{-1}
\end{eqnarray*}
where the symbol $ [\,\ldots\,]_{\,\ast}^{-1} $ means up to some suitable prescription to shift the poles.
As a consequence we eventually get
\begin{equation}
[\,i\,\mathtt{reg}\,S(p,m)\,]^{-1} = \frac{1}{M-m}\cdot [\,(\,p^{2} - m^{2})(\,p^{2} - M^{2})\,]
\cdot \frac{ \p (m+M) - p^{2} - mM }{ \left[ \,p^{2}(m+M)^{2} - (\,p^{2} + mM)^{2}\,\right] _{\,\ast}  }
\end{equation}
Now it appears to be quite evident that even the inverse of the simplest PV-regularized  propagator does correspond to a
\textbf{highly non local} expression in coordinate space, owing to the presence of the denominator
$ \left[ \,p^{2}(m+M)^{2} - (\,p^{2} + mM)^{2}\,\right] _{\,\ast}  $ in momentum space. To this concern we recall that,
for example, the far much simpler Nakanishi integral-differential inverse of the D'Alembert operator does actually involve
a \textbf{non local integral kernel}: namely,
\begin{eqnarray*}
\mathcal{D} \equiv \textstyle\frac12 (\nabla^{2})^{-1} (x_{0}\partial_{0} - \mathrm{C})
\end{eqnarray*}
where $ \nabla^{2} $ is the Laplace differential operator and C an arbitrary constant, which actually satisfies
\[
\square\,\mathcal{D} \ast f(x) = f(x)\qquad\quad\mathrm{iff}\qquad\quad \square f(x)=0
\]
The above basic elementary example clearly show that the Pauli-Villars regularized propagator can never be derived from any local
classical Action integral as the inverse of any local differential operator of the Calderon-Zygmund type, as above claimed.}
\end{enumerate}

\medskip\noindent
\textbf{One such pitfall (that is confusing the inverse of a sum of differential operators with the sum of their inverses)  is not rare in the literature and can be found in several approaches nominally utilizing a PV regularization. They have nothing to do with the Pauli-Villars regularization in the present calculation, notwithstanding their misleading denomination}.

\medskip
Now the main point concerning the chiral anomaly and the unitarity of the collision operator. 
As we have seen in details, in the present model of a left-handed Weyl spinor
interacting with a gauge vector potential, all the symmetries of the classical Action do survive the divergent and finite parts of the
counter-terms, for any choice of the regulators and renormalization prescription. As it is well-known since a long time, this is not 
so for the famous chiral left-handed triangle diagram, which turns out to be UV finite and yields
\begin{eqnarray*}
&&\left.\tilde\Gamma^{(L)}_{\mu_1\mu_2\mu_3}(k_1,k_2,k_3)\right|_{\rm 1-loop}
\\
&=& \frac{g^{\,3}}{12\pi^2}\,\varepsilon_{\sigma\mu_1\mu_2\mu_3}	\,
\left(k_1^\sigma -k_2^\sigma\right)\,k_3^2\, I_{12}(k_1,k_2,k_3) +
{\rm cyclic\ permutations}
\\
&+& \frac{g^{\,3}}{4\pi^{2}}\,\varepsilon_{\sigma\tau\mu_1\mu_2}\,k_1^\sigma k_2^\tau k_3^{\mu_3}\,
I_{12}(k_1,k_2,k_3)+{\rm cyclic\ permutations}
\\
&&(\,k_1+k_2+k_3=0\,)
\\
\\
&&I_{rs}(k_1,k_2,k_3)
\\
&=& -\int_0^1 \mathrm{d} x_1\int_0^1	\mathrm{d} x_2\int_0^1\mathrm{d} x_3\,
\frac{2x_r\,x_s\,\delta(1-x_1-x_2-x_3)}{k_1^2 x_2 x_3+k_2^2 x_3 x_1+k_3^2 x_1 x_2}
\\
&&(\,r,s=1,2,3\,)
\end{eqnarray*}
The one loop chiral left-handed triangular amplitude does satisfy the anomalous Ward identity
\[
k_1^\sigma\,\tilde\Gamma^{(L)}_{\sigma\mu_2\mu_3}(k_1,k_2,k_3)=
-\,\frac{g^{\,3}}{12\pi^{2}}\,\varepsilon_{\sigma\tau\mu_2\mu_3}\,k_2^\sigma\,k_3^\tau
\]
and related cyclic permutations. 
\\
\noindent
\textbf{The above expression for the 1-loop triangular amplitude is finite, it does not
depend at all upon regulators, or counterterms or even renormalization prescriptions and it does not undergo any higher loop corrections
(Adler-Bardeen theorem). }However, it turns out that, owing to the anomalous Ward identity, the left-handed Weyl current
does no longer satisfy the continuity equation, at variance with the classical case, viz.,
\[
\partial_{\nu}\chi_{L}^{\dagger}\,\alpha^{\,\nu}\chi_{L} = \frac13\left( \frac{g}{4\pi}\right) ^{2}F^{\,\mu\nu}_{\ast}F_{\mu\nu}
\qquad\qquad F^{\,\mu\nu}_{\ast}=\textstyle\frac12\,\varepsilon^{\,\mu\nu\rho\sigma}\,F_{\rho\sigma}
\]
This finite and universal quantum breaking of the U(1) symmetry under phase transformations on the left Weyl spinor field
does actually put in jeopardy the unitarity issue. As a matter of fact the Lorentz invariant quantum theory of a gauge vector field unavoidably
involves a Fock space of states with indefinite norm. Now, in order to select a physical Hilbert subspace of the
Fock space a subsidiary condition is necessary. In the Abelian case, when the fermion current satisfies the continuity equation then 
the equations of motion lead to  $ \square(\partial\cdot A)=0 $, so that a subspace of states of non-negative norm can be selected
through the auxiliary condition
\[
\partial\cdot A^{(-)}(x)\,\vert\,\mathrm{phys}\,\rangle=0
\]
$A^{(-)}(x)$ being the destruction, positive frequency part of a d'Alembert quantum field.
Conversely, in the present chiral model we find
\[
-\,g^{-1}\square(\partial\cdot A)=\partial_{\nu}\chi_{L}^{\dagger}\,\alpha^{\,\nu}\chi_{L} 
= \frac13\left( \frac{g}{4\pi}\right) ^{2}F^{\,\mu\nu}_{\ast}F_{\mu\nu}\neq0
\]
in such a manner that nobody knows how to select a physical subspace of states with non negative norm, if any, where a unitary
restriction of the collision operator $ S $ could be defined.

\section{ Weyl and Dirac propagators}

Equipped with the previous definitions, clarifications and results, in the rest of the paper we cope with  some critical issues one comes across in quantum  field theories involving Weyl fermions. We start with the fermion functional integral.

We denote by $\slashed D$ the 
Dirac operator: namely,  the massless  matrix-valued differential 
operator applied in general to Dirac spinors on the 4d curved space with 
Minkowski signature $ (+,-,-,-) $
\be
\slashed D= i \left(\slashed \partial + \slashed V\right)\label{dirac}
\ee 
where $V_\mu$ is any real vector potential, valued in a some Lie algebra. When we are in presence of a non-trivial background metric $\slashed{V}$ contains also a spin connection, which we understand here. Dirac fermions will be represented as  four component spinors. The functional integral, i.e. the effective Action for a 
quantum Dirac spinor in the presence of a classical background vector potential 
\be
{\cal Z}[V]=\int {\cal D}\psi {\cal D}\bar \psi\; e^{i\int d^4x \, \sqrt{g}\, 
\bar \psi \slashed D \psi}, \label{pathint}
\ee
is \textbf{formally } understood to be the determinant of $\slashed D\, :$ 
$ 
\det \,(\slashed D) = \prod^\infty_i \lambda_i\,.
$

From an operational point of view, the latter defined in two different ways: either by means of perturbation theory, 
i.e. as the sum of an infinite number of 1-loop Feynman diagrams, some of which 
containing UV divergences by naive power counting, 
or by means of a non-perturbative approach,
i.e. as the suitably regularized infinite product of the eigenvalues of 
$\slashed D$ via analytic continuation. 
In the perturbative approach, in order to assign a meaning to a
finite number of UV divergent 1-loop diagrams, one needs a regulator (dimensional, cutoff, PV or others).  In the non-perturbative framework the
complex power construction and the analytic continuation tool, if available,  
may provide themselves the necessary set up for the 
infinite product of eigenvalues, without need of any 
further regulator, usually via the Heat-Kernel expansion or similar \cite{Klaus,Dima}.

Any variation of \eqref{pathint} with respect to a parameter contained in the Dirac operator  requires the existence of an inverse of the kinetic operator. Formally
\be
\delta \det ({\slashed D})=  \det ({\slashed D}) \,\tr ({\slashed D}^{-1} \delta {\slashed D})\label{Dvariation}
\ee
In other words the existence of the inverse of $\slashed D$ is essential for the definition itself of the functional integral. It turns out that, in the case of Dirac fermions, an inverse of $\slashed D$ does exist. If full causality is imposed in forwards and backwards  time evolution on e.g. Minkowski space,  this inverse coincides with the Feynman propagator or Schwinger distribution $\slashed S$, which is 
unique and characterized by the well-known Feynman prescription and satisfies
\be 
\slashed D_x \slashed S(x-y)= \delta(x-y),\quad\quad \slashed D \slashed 
S=1,\label{DP}
\ee
The second equation is a synthetic notation\footnote{For simplicity we understand factors of $\sqrt{g}$, which 
are necessary, see \cite{DeWitt,dewitt1,dewitt2}, but inessential in this discussion.}.

\vskip 1cm 
An example is the way we
extract the trace anomaly from the functional integral. It is the response of the latter under a Weyl 
transform $\delta_\omega g_{\mu\nu} = 2 \omega 
g_{\mu\nu}\,:$
\be
 \delta_\omega \log {\cal Z} = \int d^4x\, \omega(x)\, g_{\mu\nu}(x)\, 
\langle\,T^{\mu\nu}(x)\rangle\label{deltaomega}
\ee
which is nothing but a particular case of \eqref{Dvariation}.  Here $g_{\mu\nu}(x) \langle T^{\mu\nu}(x)\rangle$ is the quantum trace of the 
energy-momentum tensor. Again, the latter can be calculated in various ways with perturbative or 
non-perturbative methods, that is with the Feynman diagram technique or with analytic functional methods, respectively. 
The latter include the Schwinger's proper-time method \cite{Schwinger}, the heat kernel method \cite{Klaus,Dima}, the Seeley-DeWitt \cite{DeWitt,Seeley1,Seeley2}, 
the zeta-function regularization \cite{Hawking}. Also the Fujikawa technique may be included in this category.

The results concerning the trace anomaly for Dirac (and Majorana) fermions as well as other anomalies,  are well established. 
It is worth recalling that one basic requirement for the viability of these calculations is the feasibility  of the Wick rotation, i.e. turning to the Euclidean 
formulation. This is allowed thanks to the above mentioned Feynman 
prescription to get the Schwinger causal Green's function or Feynman propagator 
for the quantum Dirac spinor field. In the absence of such a transition to a Euclidean formulation, the whole construction becomes mathematically meaningless and unreliable.

For Weyl fermions the situation drastically changes. Let us denote a 
left-handed Weyl fermion by  $\psi_L= P_L \psi$, where $P_L= \frac12(
{1-\gamma_5})$, so that the classical Action on the 4d Minkowski space reads
\be
S_L=\int \mathrm{d}^4x\, \bar \psi_L \slashed D \psi_L \label{weylaction}
\ee
The Dirac operator, acting on left-handed spinors maps them to right-handed 
ones. Hence, {\bf the eigenvalue problem itself is not even well posed}. Therefore
the Weyl determinant cannot be defined at all. This is trivially apparent in the fact that the inverse of
\be
{\slashed D}_L = {\slashed D}P_L = P_R {\slashed D}\label{DL}
\ee
does not exist, since it is the product of an invertible operator times a 
projector. The full propagator of a Weyl fermion does not exist in this 
naive form (this problem can be circumvented in a more sophisticated approach, 
see below). It is incorrect to pretend that the propagator is  $\slashed 
S_L=\slashed S P_R = P_L \slashed S$. First because such an inverse does not 
exist, second because, even formally,
\be
{\slashed D}_L \slashed S_L= P_R,\quad\quad {\rm and}\quad\quad{\slashed S}_L 
{\stackrel {\leftarrow}{ \slashed D_L}}= P_L \label{wrong}
\ee
The inverse of the Weyl kinetic operator is not the inverse of the Dirac operator multiplied by a chiral projector.  Therefore the propagator for a Weyl fermion is not the Feynman propagator for a Dirac fermion multiplied by the same projector. The impossibility to define eigenvalues of the kinetic operator or the lack of an inverse for the chiral Weyl kinetic term is a very severe problem.
It is clear that the corresponding usual formulas for the chiral Weyl quantum theory do not exist at all.

How can we cope with these problem? In perturbation theory, in order to construct Feynman diagrams, one usually employs the ordinary {\bf free} Feynman propagator for Dirac fermions. The use of a free Dirac propagator is formally justified, because one can add a  free right-handed fermion 
to allow the inversion of the kinetic operator, as was done above. This is an easy way out, but it immediately raises several questions: in this way one explicitly breaks gauge and Lorentz invariance, and, if a background metric is minimally coupled to the Weyl fermion also diffeomorphisms and local Lorentz invariances are explicitly broken; moreover how can one preserve in this way the notion of chirality, which is a distinctive feature of Weyl spinors? There  is no intrinsic way to guarantee that it is preserved all the way through the nontrivial steps of the calculations.  In any case one must check {\it  a posteriori} that diffeomorphism invariance is satisfied\footnote{In 4d we know that there is no room for consistent diffeomorphisms or local Lorentz anomalies: however, spurious (trivial) anomalies of this kind may arise, depending on the regularization procedure. 
Such spurious anomalies should be eliminated by subtracting counterterms and thus by modifying for consistency the effective Action.}.

The reason why one uses such an expedient is because one hopes that the 
information about chirality is preserved by the fermion-boson-fermion vertex in \eqref{weylaction}, 
which contains the $P_L$ projector. As it has been shown in the example of Section 4, this method is believed to work correctly as long as it does not come across anomalies. Therefore absence of anomalies is a warranty that we are not introducing unwanted (opposite-handed) degrees of freedom. But this precisely cast doubts on the validity of results when anomalies are present. How can we trust an anomaly calculation when we know that the calculation is reliable only when anomalies are absent?  {We face a logical obstacle}. We cannot trust our calculations precisely when we wish to compute anomalies. It sounds very much like the liar's paradox: we are asking true information to somebody we know may be a liar. This logical {hurdle} is not surprising because the procedure we have outlined has a bag at the very beginning: when we add a free right-handed fermion to the left-handed one, the partition function of which we aim to compute, the resulting Dirac operator has well defined eigenvalues, as peculiar as they may be. Hence, the use of the path integral definition as a product of eigenvalues is justified, but here lies the contradiction: this is not what we are looking for, because what we want to compute is the determinant of an operator whose eigenvalues are not even defined.

When we pass to nonperturbative approaches it is not possible to simply replace the missing Weyl propagator with the Dirac one.  
The reason is that the latter does not contain the correct information about chirality. To follow closely the perturbative approach we should consider the inverse of 
\be
 i \left(\slashed \partial +{\slashed V} P_L\right)\label{diracweyl}
\ee 
The operator \eqref{diracweyl}, as it stands,  just treats  the two chiralities asymmetrically. We will return to this point later on, but what matters here is to remark that the differential operator \eqref{diracweyl} is not the inverse of \eqref{dirac} multiplied by $P_L$.  If one  pretends to replace the full Weyl propagator with the full Dirac propagator, one loses  any information concerning the chirality, and trying to recover it by multiplying the Dirac propagator by $P_L$ {is simply not serious.} 

{Even though \eqref{diracweyl} can be used to compute consistent gauge anomalies yielding satisfactory results (for reasons that will be explained further on), it shares the same drawbacks as the just above outlined perturbative method.  One should always and firmly keep in mind that all these approaches  are based on the above manipulations, where we assume that the inverse of the relevant kinetic operator exists and if, by any chance, the kinetic operator is not invertible we pretend that it is and  go on undeterred with the formalism. This attitude, based on the time-honored faith on practical calculations, may however hide unpleasant implications such as the logical loophole pointed out above. The inevitable conclusion is that all the results (even ours own) obtained with such procedure - that understands the invertibility of the kinetic operator when this is not the case - are intrinsically contradictory and should be considered with a good amount of skepticism.}

 There is however a way to avoid the logical trap pointed out before. 
A little thought shows what is critical in the previous procedures: there is a discontinuous jump between say \eqref{diracweyl}  and the operator \eqref{DL} which we should be computing. It is somewhat like studying a function only at the point where it is singular. There is no continuous parameter that connects one to the other. An elegant way out to bypass the difficulties with Weyl fermions is precisely following this suggestion; it consists in enlarging the space of potentials, by adding a spectator axial vector field, as was first done by W.A.Bardeen \cite{bardeen1969}. After all the most practical way to overcome the liar's paradox is to ask many people.

Limiting ourselves for simplicity to the Abelian gauge case the Action integral for a Dirac fermion coupled to a vector $V_\mu$ and an axial potential $A_\mu$ is 
\be
S[V,A]= \int d^4x \,i\, \overline \psi \left( \slashed{\partial}-i\slashed{V} -i \slashed {A} \gamma_5
\right)\psi\label{SDiracAV}
\ee
It is invariant under the following vector and axial-vector transformations
\be
&&\delta_\alpha V_\mu=\partial_\mu \lambda, \quad\quad \delta_\alpha A_\mu =0\0\\
&&\delta_\beta V_\mu=0, \quad\quad \delta_\beta A_\mu= \partial_\mu \rho\label{VAVtransf}
\ee
One can now do perturbative quantization of this theory without any difficulty since the fermion is Dirac and its propagator is the standard one. {Simultaneously the definition of the functional integral as the product of eigenvalues makes sense because the eigenvalue problem is well defined for it, as complicated as these eigenvalues may be}. After carrying out all the necessary calculations one can extract the results relevant for the  Weyl fermion case by taking the limit $V\to V/2, A\to -V/2$. For in this limit in \eqref{SDiracAV} one recovers the kinetic operator \eqref{diracweyl}. This yields the {\it bonafide} correct results for a theory of Weyl fermions. 

This method can be easily extended to non-Abelian potentials. {The calculation of the effective Action generated out of 
\eqref{SDiracAV} can be carried out along the traditional lines. For the perturbative approach everything needed for a correct computation 
is well defined: the propagator is the one for Dirac fermions, the vertices are of two types: for fermion-fermion-vector potential and  for fermion-fermion-axial potential. 
The approach is the appropriate and traditional one for an operator that has a well defined spectrum of eigenvalues, 
like the Dirac operator discussed at the beginning of this Section.  No poetic license is needed and any logical paradox is avoided. {Altogether this method is heuristic, that is it does not follow from first principles, but it is well motivated and so far it has never been contradicted. In particular it yields the same results as the family's index theorem of Atiyah ad Singer as far as consistent anomalies are concerned.}
Now it is also evident why the very common (and basically contradictory) approach based on the simple replacement of the (missing) Weyl propagator with the Dirac one, may yield correct results. It is because the calculation steps are the same as in the \eqref{SDiracAV} case as long as one imitates the substitution $A_\mu=0$ for the covariant case and $V\to V/2, A\to -V/2$ for the consistent case at the right points: in most cases a risky but often successful approach.}

The extension to calculations with a non-trivial  background metric requires a further step. This is easily understandable by looking at the kinetic operator  \eqref{diracweyl} which is supposed to be applied to Dirac fermions. In this case we must add a spin connection $\omega_\mu$, i.e. $\slashed V$ is replaced by $\slashed V + \slashed \omega$, and $\gamma^\mu = \gamma^a e_a^\mu$ where $e_a^\mu$ is the inverse tetrad. It is immediately  evident that the corresponding theory is not invariant under diffeomorphisms. If, on the other hand,  we replace $\slashed V P_L$ with $\slashed \omega + \slashed V P_L$ diffeomorphisms are preserved in the classical Action, but we loose the effect the chirality can have on the metric dependence. The surplace can be overcome by introducing a second spectator metric. The corresponding construction is conceptually simple but requires a laborious extension of the formalism. We give up the idea of presenting it here, and refer to the literature. It was first formulated in \cite{BCDGPS}. The interested reader can found an extensive review with applications in \cite{BG24}. One thing is however worth being reminded: in this formalism a central role is played by diffeomorphism invariance. There are unfortunately in the literature examples of calculations where such invariance has simply been explicitly broken, thus compromising the validity of the results.

\subsection{Weyl and massless Majorana fermions}

The second criticality is related to the (wrong) identification between Weyl and Majorana massless spinors. If this identification were possible the problem discussed in the previous Section would not exist,  because we could simply replace the Weyl kinetic operator with the massless Majorana one, 
and the missing Weyl propagator with the existing Majorana propagator (identical to the Dirac one).
We have already explained in detail above that Weyl and Majorana spinors belong to different representations of the Lorentz group. A Weyl spinor is a complex spinor with definite chirality (an eigenstate of $\gamma_5$) which corresponds to its helicity, constituting a minimal representation of the Lorentz group. A Majorana spinor is a self-conjugate bispinor, which can be cast in real form with a suitable choice of the gamma matrices. A Majorana spinor can be represented as a superposition of two Weyl spinors, conjugate to each other and thus with two opposite chirality. 
Therefore a Majorana spinor has undefined chirality, so that the relation with its helicity is also undefined. Moreover a parity operation maps the Majorana Action into itself, while it maps the Weyl Action (\ref{freeWeyl}) into the same Action for the opposite chirality.
The same holds for the charge conjugation operator. 

There is no possibility of confusing a Weyl spinor with a Majorana massless spinor. Therefore one may wonder how such misunderstanding may have arisen and spread. It certainly comes from a superficial knowledge of the mathematical properties of spinors in 4d, in which a false analogy with 2d, where Weyl and Majorana spinors coexist, may have played a role. But, above all, we suspect, the common properties of the two types of spinors may be at the origin of the misunderstanding. Beside being both massless they have other properties in common. For instance we can decompose any (massless) Dirac fermion into the sum of two Weyl spinors of opposite chirality or into the sum of two independent Majorana spinors. In this sense in model building it is equivalent to use Weyl or Majorana, provided the end result consists of Dirac spinors.  Another reason why they are sometimes considered the same object may be due to the fact that we can establish a one-to-one correspondence between the
components of a Weyl spinor and those of a Majorana spinor
in such a way that the Lagrangian, in two-component notation, looks
the same. If, in the chiral representation,
we write $\psi_L$ as $\left(\begin{matrix} \omega\\ 0\end{matrix}\right)$,
where $\omega$ is a two component spinor, then (\ref{freeWeyl}) above becomes
\be
i \omega^\dagger \bar \sigma^\mu \partial_\mu \omega \label{MjoranaWeyl}
\ee
which has the same form {(up to an overall factor)}  as a massless Majorana Action (see \eqref{MajoWeyl}). 
But the value of an Action is not everything in a theory. The action must be accompanied by a set of specifications, which we usually understand. What is essential in an action for Weyl and Majorana fermions is not its formal value but the way they respond to a variation of the Weyl and Majorana fields. One yields the Weyl, the other the Majorana equation of motion. Precisely this is the delicate issue: when onr takes the variation of an Action with respect to a field in order to derive equation of motion, one must make sure that the {variation}
respects the symmetries and the properties of the field\footnote{For instance, in gravity theories, the metric variation $\delta
g_{\mu\nu}$ is generic, but a symmetric infinitesimal tensor.}. 
If we wish the eom to preserve chirality, which is the distinctive feature of Weyl fermions, we must use variations that preserve chirality, i.e. must be
eigenfunctions of $\gamma_5$. If instead we wish the eom appropriate for 
Majorana fermions we have to use variations that transform suitably, i.e.
must be eigenfunctions of the charge conjugation operation. If we respect this minimal rule we
obtain two different results, which are irreducible to each other, no matter what Action we use.

{ At this stage it is perhaps useful to state a conclusion about two-component and four-component formalisms for spinors. As we have shown above, the representation of fermions by means of the two-componnet formalism alone is not always reliable, notably in what concerns chirality. Only the four componet formalism is reliable in all respects.}

At this point it is perhaps not useless to spend a few words concerning the already mentioned U(1) continuous symmetry of Weyl fermions, because thee latter is often confused with an axial ${\mathbb R}$ symmetry of Majorana fermions and may also be the origin of the illegal identification of Weyl and massless Majorana fermions. It is useful to start with a free massless Dirac fermion $\psi$. Its free Action is clearly invariant under the transformation $\delta\psi =i (\lambda +\gamma_5\rho)\psi$, where $\lambda$ and $\rho$ are real numbers. This symmetry can become local by minimally coupling $\psi$ with a vector potential $V_\mu$ and an axial potential $A_\mu$, i.e. couling it with the combination $V_\mu +\gamma_5 A_\mu$, in such a way that $\lambda$ and $\rho$ become arbitrary real functions. We choose, for convenience,  the Majorana representation for gamma matrices (see above). Thus all of them, including $\gamma_5$, are imaginary. 
If $\psi$ is a Majorana fermion, its four components will be real and only the symmetry parametrized 
by $\rho$ makes sense in the Action (let us dub it $\rho$ symmetry). 
If instead $\psi$ is Weyl, say $\psi=\psi_L$, then, since $\gamma_5 \psi_L=\psi_L$, 
the symmetry transformation will be  $\delta\psi_L =i (\lambda -\rho)\psi_L$. 

This is the possible source of confusion, for it looks like we can merge the two parameters $\lambda$ and $\rho$ into a unique one and identify it with the $\rho$ of the Majorana axial $\rho$ symmetry. But this is wrong because for a right handed Weyl fermion  the symmetry transformation is $\delta\psi_R =i (\lambda + \rho)\psi_R$. Dropping $\rho$,
the Majorana fermion does not transform. Dropping $\lambda$, both Weyl and Majorana fermions transform, but Weyl fermions transform with opposite signs for opposite chiralities. 
This distinction reflects on anomalies. It is a well-known fact that the axial ${\mathbb R}$ Majorana symmetry is anomalous, giving rise to the well-known covariant or ABJ anomaly $\sim \int d^4x \, \beta F_A \wedge F_A$ ($F_A$ is the curvature of $A$).  Instead for a left-handed (right-handed) Weyl fermion the anomalous symmetry has parameter $\lambda-\rho$ ($\lambda+\rho$). This anomaly satisfies the Wess-Zumino consistency conditions. In the Abelian case we are considering for simplicity, it coincides (but not in the non-Abelian case) in form with the covariant anomaly, but it has a coefficient different than the covariant one and with opposite sign for opposite chirality. As a Dirac fermion can be represented as a sum of two Weyl fermions with opposite chiralities, the anomalies triggered by the $\lambda$ transformation cancel out in a Dirac fermion, while the anomalies triggered by $\rho$ add up. This matches with the well-known fact that for a Dirac fermion the  $U(1)$ $\lambda$ symmetry is not anomalous, while the axial $\rho$ symmetry is anomalous and corresponds to twice the anomaly of a Majorana fermion. From this example it is very clear that symmetries and anomalies of Majorana fermions are different from the symmetries and anomalies of Weyl fermions.

Notwithstanding their common properties, the difference between Weyl and massless Majorana spinors becomes crucial in certain instances. One of these is the functional measure in the path integral. The functional measure is well defined for Dirac or Majorana fermions, while it is undefined for Weyl fermions. Here the problem of course arises in nonperturbative calculations. 
K. Fujikawa invented a method \cite{Fujikawa} to compute ABJ anomalies (i.e. anomalies that appear in theories with  Dirac or Majorana fermions). 
It is a heuristic method based on the calculation of the symmetry variation of the functional measure. 
It works very well for Dirac and Majorana fermions. But the functional measure is computed relying on the decomposition of the fermion fields into a complete set of  eigenvectors of the relevant Dirac operator. We know that the eigenvalue problem is not even defined in the case of Weyl fermions, 
thus it is mandatory to resort to other methods.  

There are other more flexible methods, beside Fujikawa's one. They are the functional method alluded to above, which we denoted with the term functional or `heat kernel' methods. They translate into a mathematically reliable language the idea of determinant of the kinetic operator. Since a formulation of these methods with  a linear Dirac operator is not known (yet), one resorts to a quadratic elliptic operator (for instance the square of the Dirac operator) and bases all calculations on the square root of the resulting path integral.  For the method to be valid, however, such a quadratic operator must preserve all the properties of the linear one. Now, it is not difficult to concoct an elliptic operator out of the Weyl-Dirac one, 
but it is impossible to do it while preserving all the critical information, 
i.e. while preserving the chirality of the model and its classical symmetries\footnote{Even the symmetry whose anomaly is being computed. The anomaly must result from the clash with the regulator, not because of the explicit breaking of the symmetry at the level of the quadratic operator.}. More comments on this below. 

There are in fact insurmountable difficulties. As a matter of fact, first of all, in order to turn the square of the Dirac operator  into an elliptic normal differential and matrix-valued operators,  the transition to the Euclidean formulation is necessary. The latter is legitimate and 
viable for Dirac fermions, whilst it does not exist at all for Weyl fermions. {
One reason why we believe it is deep and sharp and has its roots in group theory: the Euclidean 4d symmetry group is $ O(4,\mathbb{R}) $ which is locally isomorphic to the direct  product $ O(3,\mathbb{R})\times O(3,\mathbb{R}) $, in such a manner that any item in this theory must be invariant with respect to the exchange of any spin representation of the two identical and equivalent orthogonal trivial factors, no room being left to the very concept of chirality, which requires two non-equivalent irreducible 2d representations of the Lorentz group. 

We recall that a complete treatment of the invertibility of a (linear) Dirac operator in a Euclidean metric case is provided by the family's index theorem of Atiyah and Singer, and will be summarily presented in section 8.}

The solution to these series of problems exists without straining any fundamental principle or lore, and has been anticipated at the end of the previous section (and will be further commented in the next section). One has to escape from the trap of circular, self-delusive arguments and accept the idea of enlarging the space of potentials.

To close the discussion about the difference between Weyl and massless Majorana fermions  let us briefly focus on its physical implications. A Majorana spinor describes neutral spin 1/2 objects - not yet detected in Nature. As a consequence there is no phase transformation (U(1) continuous symmetry)  applicable to self-conjugated 
Majorana spinors (and this is independent of the presence or not of a mass term). Hence, 
e.g., its particle states do not allow for antiparticles of opposite charge, simply 
because for charge self-conjugated
spinors a charge does not exist at all (this was the surprising discovery of Ettore Majorana, 
after the appearance of the Dirac equation and the positron detection).
The general solution of the equation of motion of a free Majorana spinor 
always implies the existence of  two polarization states with opposite helicity.
On the contrary, a chiral Weyl spinor, describing massless quarks and leptons (before the electroweak symmetry breaking\footnote{Thanks to chiral anomaly cancellation quarks and leptons pair up, except possibly for neutrinos, to form Dirac spinors which acquire masses after symmetry breaking.})  in the Standard Model, admits only one polarization or helicity state. It always implies antiparticles of opposite helicity {and} it always 
carries a conserved internal quantum number such as the lepton number, which is opposite for particles and antiparticles.

\subsection{Dirac and Majorana mass terms}

Another particularly confusing concept seems to be connected with mass terms for fermions. Let us start from a basic element, the mass term $\bar \psi \psi$ for a Dirac spinor, which is uncontroversial. It  can be also
rewritten by projecting the latter into its chiral components
\be
\bar\psi \psi= \overline{\psi_L} \psi_R+ \overline {\psi_R}\psi_L
\label{masschiral}
\ee
If $\psi$ is a Majorana spinor this term can be rewritten  
\be
\bar\psi \psi^c = \overline{(\psi^c)_L} \psi_R+ \overline {\psi_R}(\psi^c)_L
,\label{massMajo}
\ee
which is therefore well defined and Lorentz invariant by construction. {Now,
using the Lorentz covariant conjugate we can rewrite (\ref{massMajo}) as
\be
(\psi_L)^T C^{-1} \psi_L + \psi_L^\dagger C(\psi_L)^*,\label{massMajoL}
\ee
which is expressed solely in terms of $\psi_L$. The above expression (\ref{massMajoL}) for the mass term may create the false 
illusion that also for Weyl fermions there may exist a mass term. But this is not so. Should we add this term to the kinetic term (\ref{freeWeyl}) the resulting
Action gives rise to a field equation for $\psi_L$ which is not Lorentz covariant: the kinetic and mass
terms in the field equation would belong to two different representations. 
In fact carrying out the complete Lagrangian derivation, i.e. taking the variation with respect to $\overline {\psi_L} $, one gets
\be
i \gamma^\mu \partial_\mu \psi_L -m( \psi^c)_R=i \gamma^\mu \partial_\mu \psi_L -m \psi_R=0,\label{wrong}
\ee
recalling that $\psi =\psi^c$.
That is, in order to be able to write the equation of motion for $\psi_L$ one needs also $\psi_R$.
In fact this equation is nothing but the right-handed part of the Dirac equation. It is not possible to write down a massive equation of motion for $\psi_L$ without involving $\psi_R$.

As a consequence, in particular,}  there does not exist such a
thing as a ``massive Weyl propagator'', that is a massive propagator involving only one chirality. This is something that, in particular, renders the use of the Pauli-Villars regularization problematic. In fact it would be natural to regularize the theory by means of massive Weyl ghost fields. But this is impossible.  Sometimes in place of the missing massive Weyl propagator a Dirac or Majorana propagator is used. But this is a leap in the dark: since in these propagators both left and right chiralities appear, one  has no warranty that chirality is preserved throughout the calculations and thus that the result one obtains is the correct one. A minimal precaution, in such a case, would be to crosscheck the results obtained with this regularization by comparing them with those obtained with others. In Sectiom 5, for instance, we have shown that the masslessness result of the Weyl kinetic term  is preserved at one loop with three different regularization, including a PV one.
These results, which are clearly visible in the four component formalism used
so far, are much less visible in the two-component formalism.

\medskip
{ It turns out that, on the one hand, the renowned and famed Goldhaber \textit{et al.} series of experiments \cite{Goldhaber} unequivocally proves 
that neutrinos are  \textbf{always} left-handed and of negative helicity,
while anti-neutrinos are distinguished and charge conjugated, since they always appear to be right-handed and of positive helicity: it means, beyond any reasonable doubt,
that \textbf{neutrinos and anti-neutrinos are truly Weyl fermions}. On the other hand, soon after the discovery of the neutrino flavor oscillations it has been promptly suggested
to understand the latter in terms of a Majorana mass term which provides, as we have seen (\ref{leptonnumberviolation}), a breakdown of the lepton number
conservation and an open door towards flavor changes. Unfortunately, as we have thoroughly discussed and elucidated in the present review, 
the hypothesis of a Majorana massive neutrino badly and strongly contradicts the Goldhaber \textit{et al.} long ago well established and solid results: 
it is impossible to simultaneously keep a full barrel and a drunk wife. The deep reasons to explain and understand neutrino oscillations are elsewhere but in the
Majorana mass, at least if we believe in the principles of quantum field theory and the Standard Model construction and if we trust in \cite{Goldhaber}.}

\vskip 1cm

Finally, going to nonperturbative approaches, we have already noted that the fermion functional integral measures are different in the two cases. Actually this is a rather heuristic way of phrasing the issue. One can express the same idea and calculations in terms of more rigorous functional or heat-kernel  like methods. 
In fact in the literature one meets various different formulations of this problem, which oscillate between rigorous and heuristic approaches. 
Anyhow, the main central object is the quadratic operator that should  represent the square of the kinetic operator, the Dirac operator in our case. 
Since, as we have already remarked, a version of the heat-kernel involving the linear Dirac operator  is not available, one considers a quadratic version, i.e. a functional integral in which the linear Dirac operator is replaced by its square, and then takes the square root of it. 
This is the logical extension to continuous operators of the determinant of a matrix represented by the product of its eigenvalues. 
The square of the matrix is represented by the product of the squared eigenvalues. 

Once this procedure has been accepted, the most important move is the choice of the quadratic operator. Let us explain this point in some detail.
 In the case of a Dirac fermion, by applying twice the Dirac operator makes sense because, for instance, for a gauge transformation $\delta \psi = i \alpha \psi$ with
$\alpha=\alpha^a T^a$, where $T^a$ are the Hermitean generators of a Lie algebra, we have
\be
\delta \slashed D \psi = i \alpha\,  \slashed D \psi\0
\ee
Thus it makes sense to apply twice $\slashed D$ to $\psi$, because $ \slashed D^2 \psi$ under a gauge transformation has the same properties as $\slashed D \psi$. 
This was the meaning of our statement in the previous Section whereby the quadratic operator must respect the symmetries and features of the original kinetic operator. 
In many cases one meets in the literature this is not the case.

Let us consider a simple example. The functional integral for a free Dirac fermion
(\ref{freeDirac}) ${\slashed D}=i{\slashed \partial+\slashed{V}}$ (where $V$ denotes any
potential), is the (suitably regularized) product of its eigenvalues. This is true  for a massless Majorana fermion, while for a Weyl
fermion it is not so straightforward. Since the Dirac operator anticommutes with
$\gamma_5$, it maps a left-handed spinor to a right-handed one. Therefore the
eigenvalue problem is not even defined for $ {\slashed D}_L = {\slashed D} P_L$.
We may replace the looked for $\det {\slashed D}_L$ with $\left(\det\left(
{\slashed D}^\dagger_L{\slashed D}_L\right)\right)^{\frac 12}$. If $\slashed D$ is self-adjoint 
(after a Wick rotaton) this reduces to $\left(\det\left(
{\slashed D}^2 P_L\right)\right)^{\frac 12}$. But this is deceivingly simple. It is true that the symmetries are preserved but the most important structure, i.e. the interaction vertex, is lost. Even if 
the quadratic operator contains a chiral projector the latter is totally ineffective. It is analogous to stating that the inverse of the Dirac-Weyl operator is the inverse of the Dirac operator multiplied by chiral projector. To protect the interaction vertex from disappearing one should use \eqref{diracweyl} as basic kinetic operator. But, as already noticed, \eqref{diracweyl} breaks gauge and diffeomorphism covariance. The puzzle can be solved with the already mentioned Bardeen's approach. {The real important point to be retained from this discussion is that the squared Dirac operator one is obliged to use in the functional methods must preserve the same symmetries as the linear Dirac operator which appear in the defining Action, and chirality in the Weyl-Dirac case.}

In most of the nonperturbative functional methods, after turning to a Euclidean setting, one uses the dimensional regularization or a regularization based on analytic continuation such as the zeta function one. 
At variance with these standard treatments, among the heat-kernel-like approaches there have been also attempts to use masses as regulators, much as in the PV regularization method. We have already remarked that the presence of massive terms introduce in general two chiralities into the game, so that the PV regularization
does not seem the most suitable method for calculations in which the chirality must be preserved. In the perturbative case, however, an educated use of PV can lead to the correct result. But, due to the original sin on which it is based, the result must always be confirmed with other regularizations.

The nonperturbative use of masses as regulators consists in choosing instead of a massless quadratic operator, the square of a massive operator, which can only be a  Dirac-Majorana massive operator. This method is sometime sold as PV. 
Now, first of all, as already noticed several times, the quadratic operator inevitably breaks the conservation of chirality, since it involves both chiralities in a balanced form and excludes any possibility to capture, for instance, consistent chiral anomalies.  Moreover, 
the perturbative 1-loop corrections to the classical Weyl Action, as obtained from the original true Pauli-Villars regularization method, 
by no means can generate any dynamical mass term, neither Dirac nor Majorana, since a PV regularization preserves gauge invariance 
and U(1) internal chiral phase transformation invariance, so that it protects classical scale invariance and forbids any mass term at the quantum level. 
Beside other less formal reserves, \textsf{it is roughly inappropriate to call this type of approach a PV regularization.}

To sum up, the use of a Majorana-Dirac operator inevitably drives to a chirally symmetric operator, which cannot produce any left-right asymmetry, i.e. in particular no chiral split anomaly. i.e. with opposite sign for opposite chiralities. Such methods, in particular, are unfit to intercept consistent chiral anomalies.

{In Table II below we have summarized the main properties of Weyl and Majorana fermions.

\eject

\vskip 0,3cm
 \centerline {Table II. Summary of Weyl and Majorana properties in 4d}
\vskip 0.2cm
 \noindent\fbox{
    \parbox{\textwidth}{
\begin{itemize}
\item Weyl spinors and Majorana spinors belong to different representation of the Lorentz group: Weyl spinors belongs to the minimal representation of the Lorentz group which is eigenstate of the chirality matrix, Majorana spinors belong to the minimal representation of the Lorentz group which is eigenstate of the charge conjugation operator.
\item Weyl spinors have 2 complex independent degrees of freedom, Majorana spinors have 4 degrees of freedom which can be chosen real.
\item Weyl spinor have definite chirality, the chirality of Majorana spinors is indefinite.
\item Weyl spinors are massless, Majorana spinors admit a mass term.
\item In perturbative field theory Weyl spinors do not have propagators, Majorana spinors have the same propagator as Dirac spinors.
\item The Lagrangian of free Weyl spinors has a (rigid) $U(1)$ symmetry. Massive Majorana spinors do not have any such symmetry, but in the massless case they admit an axial $U(1)$ symmetry.
\item Weyl spinors can have chiral consistent anomalies (obstructive anomalies), Majorana spinors do not have this kind of anomalies.
\item Majorana spinors, like Dirac spinors, can have chiral anomalies of the ABJ type (non-obstructive).
\item For Majorana fermions the path integral measure is well defined, while for Weyl fermions it is not.

\end{itemize}
}} 
}
\vskip 0.3 cm

\section{Euclidean Fermion Field Theories}

As explained in the introduction, the technique of the Wick rotation to regularize the divergent integrals appearing in Feynman diagrams 
cannot be applied to the original classical Action. Below we explain why, but to be able to explain it we need a short introduction to the Euclidean field theory for fermions. To start with, a warning: the problem of the transition from a Minkowski fermion theory to a corresponding Euclidean one 
\cite{schwinger1958,schwinger1959,OS72,OS73,OS73b,FO74} does not admit a unique solution. 
Here we present a simple approach.

The initial point is the theory of a massive Dirac fermion and its
spinor causal Green's function, or Schwinger function\footnote{The $n-$point Green's functions in a Euclidean spacetime are usually
named $n-$point Schwinger functions.}. We perform the standard substitutions
\begin{equation}
i p_0=p_4\qquad\quad i x_0=x_4\label{E1} 
\end{equation}
so as to obtain a positive definite denominator in the Fourier transform, that is 
\begin{eqnarray*}
S^{\,F}_{\,\alpha\beta}(\, -ix_4,{\bf x} )\ =
i\,(2\pi)^{-4}
\int \mathrm{d}\mathbf{p}\int_{-\infty}^{\,\infty}\! \mathrm{d}p_4  \,
 \frac{\exp\{ip_4x_4+i{p}_{\jmath}{ x}_{\jmath}\}}{p_4^2+{\bf p}^2+M^2}\
(\,\gamma^0 p_4 - i\,\gamma^{\, k}  p^{\, k} + i M\,)_{\,\alpha\beta}
\end{eqnarray*}
Next, we define the {\sf  Euclidean Dirac Matrices}
\begin{eqnarray}
\overline{\gamma}_4\;\equiv\;\gamma^0,\qquad
\overline{\gamma}_k\;\equiv\;i\,\gamma_k=\,-\,i\,\gamma^{\, k}
\qquad\quad
(\, k=1,2,3\,)\label{E2} 
\\
\overline{\gamma}_{\mu}=\left(\overline{\gamma}_k\,,\,\overline{\gamma}_4\right)\;=\;\overline{\gamma}_\mu^{\,\dagger},
\qquad\quad
\left\{\overline{\gamma}_\mu\,,\,\overline{\gamma}_\nu\right\}=2\,\delta_{\mu\nu}
\end{eqnarray}
so that we can write
\begin{eqnarray}
S^{\,F}_{\,\alpha\beta}(\, -i x_4,{\bf x})
&=& \frac{i}{(2\pi)^4}\int \mathrm{d}^{4}p_E\ 
\exp\{i p_E\cdot x_E\}    \
\left\lgroup  \frac{\overline{\gamma}_\mu\, p_{E\mu}+i M} {p_E^2+M^2} \right\rgroup_{\alpha\beta} 
\nonumber\\
&=& \frac{i}{(2\pi)^4} \int \mathrm{d}^4 p_E\ \exp\{i p_E\cdot x_E\}\
\left\lgroup \frac{1}{{\slashed p}_E-i M}\right\rgroup_{\alpha\beta}
\nonumber\\
&\equiv& -\,S^{\,E}_{\,\alpha\beta}(x_E)
\label{eucldiracprop}
\end{eqnarray}
where we understand
\[\mathbf p=(p^{1},p^{2},p^{3})=(p_{x},p_{y},p_{z})=\vec p_E=(\bar p_{1},\bar p_{2},\bar p_{3})\]
together with
\begin{eqnarray*}
p_{4}\overline{\gamma}_4\;+\overline{\gamma}_{k}\,\bar p_{k}\,=\,\overline{\gamma}_\mu\,p_{E\mu}\,\equiv p\!\!/_{\!E}
\\
\partial\!\!\!/_E\,\equiv\,\overline{\gamma}_\mu\, \partial_{E\mu}\,=\,
\overline{\gamma}_k\, \partial_{k} +\overline{\gamma}_4\, \partial_{4}
\end{eqnarray*}

The previous substitutions, however, have a price. The Schwinger function \eqref{eucldiracprop} conflicts with hermiticity. The way out was suggested by Osterwalder and Schrader, who took as a  basic requirement reflection positivity. This suggests to assume that the appropriate classical variables for a  Euclidean formulation of the Dirac spinor field theory, be not one but two distinct Euclidean bispinors. We may denote them again by $\psi_E$ and $\bar\psi_E$, but they obey
\begin{equation}
\{\psi_E(x)\,,\,\psi_E(y)\}=\{\bar\psi_E(x)\,,\,\bar\psi_E(y)\}=
\{\psi_E(x)\,,\,\bar\psi_E(y)\}=0\label{euclcc}
\end{equation}
for any two points $x$ and $y$ of the four-dimensional Euclidean space ${\mathbb R}^4\,$. On the other hand  the equal time canonical bracket between $\psi_E$ and the Hermitean conjugate of $\overline \psi_E$ is non-trivial. It must be noted that the last of the relation \eqref{euclcc} is  crucial, for it implies that $\bar\psi_E$ is not the Hermitean conjugate of $\psi_{E}$ multiplied by some matrix $\gamma_4\,.$ It follows that, if our aim is a meaningful Euclidean formulation for the Dirac spinor field theory, then we must treat $\psi_E$ and $\bar\psi_E$ as \textit{totally independent
classical Grassmann-valued fields}. The main novelty of the Euclidean fermion field theory is this independence, the rest of the construction being straightforward.

As an example, in order to  to derive the $O(4)$ transformation law for $\psi_E$ we use the definition of the Euclidean matrices $ \overline{\gamma}_{\mu} $ in the usual way. We obtain the six Hermitean generators
\begin{eqnarray}
\Sigma_{\mu\nu}=\frac{i}{4}\,\left[ \overline{\gamma}_{\mu},\overline{\gamma}_{\nu} \right] 
\qquad \Sigma_{\imath}\equiv\varepsilon_{\imath\jmath k} \Sigma_{\jmath k}
\quad (\,\imath,\jmath,k=1,2,3\,)
\\
\Sigma_{23} \equiv \Sigma_{1} = {\textstyle\frac12}\left\lgroup
\begin{array}{cc}
\sigma_{1} & 0
\\
0 & \sigma_{1}
\end{array}
\right\rgroup
\qquad
\Sigma_{31} \equiv \Sigma_{2} = {\textstyle\frac12} \left\lgroup
\begin{array}{cc}
\sigma_{2} & 0
\\
0 & \sigma_{2}
\end{array}
\right\rgroup
\\
\Sigma_{12} \equiv \Sigma_{3} = {\textstyle\frac12} \left\lgroup
\begin{array}{cc}
\sigma_{3} & 0
\\
0 & \sigma_{3}
\end{array}
\right\rgroup
\qquad
\Sigma_{k4} \equiv \overline{\Sigma}_{k} = {\textstyle\frac12}\left\lgroup
\begin{array}{cc}
\sigma_{k} & 0
\\
0 & -\,\sigma_{k}
\end{array}
\right\rgroup
\end{eqnarray}

\medskip
\begin{footnotesize}
$ so(4)- $\texttt{Lie algebra}. The six $ O(4,\mathbb{R}) $ generators satisfy the following Lie algebra commutation relations, \textit{viz.,}
\begin{eqnarray*}
[\,\Sigma_{\jmath}\,,\,\Sigma_{k}\,]=i \varepsilon_{\jmath k\ell}\,\Sigma_{\ell} = [\,\overline{\Sigma}_{\jmath}\,,\,\overline{\Sigma}_{k}\,]
\qquad\quad
[\,\Sigma_{\jmath}\,,\,\overline{\Sigma}_{k}\,]= i \varepsilon_{\jmath k\ell}\,\overline{\Sigma}_{\ell}
\end{eqnarray*}
We can recast them in the four-vector notations and get
\begin{eqnarray}
[\,\Sigma_{\mu\nu}\,,\,\Sigma_{\kappa\lambda}\,]= i\delta_{\mu\kappa}\,\Sigma_{\nu\lambda}
- i\delta_{\lambda\mu}\,\Sigma_{\nu\kappa} + i\delta_{\nu\lambda}\,\Sigma_{\mu\kappa} - i\delta_{\kappa\nu}\,\Sigma_{\mu\lambda} 
\label{O4Liealgebra} 
\end{eqnarray}
\end{footnotesize}

\medskip
Therefore for the two-component spinors and four-component bispinors of the Euclidean formulation we have the following orthogonal transformation rules , 
\begin{eqnarray}
\psi_{\uparrow}^{\,\prime}(x_E^{\,\prime})=\exp \lbrace (i/2)\,\vec{\sigma}\cdot (\vec{\alpha} + \vec{\eta}\, )\rbrace\, \psi_{\uparrow}(x_E)
\\
\psi_{\downarrow}^{\,\prime}(x_E^{\,\prime})=\exp \lbrace (i/2)\,\vec{\sigma}\cdot (\vec{\alpha} - \vec{\eta} \,)\rbrace\, \psi_{\downarrow}(x_E)
\end{eqnarray}
with $ x_E^{\,\prime} = R\,x_E\,,\, R\in O(4) $\footnote{The orthogonal group $O(4)$ of the
rotations in the Euclidean space ${\mathbb R}^4$ is a semi-simple Lie group
$O(4)$  isomorphic to $O(3)\times O(3)\,.$} while $ 0\le \vert\,\vec{\alpha}\,\vert<2\pi $ and $ 0\le \vert\,\vec{\eta}\,\vert<2\pi\,, $ that yields
$$
\psi_E(x_E)=\left\lgroup
\begin{array}{c}
\psi_{\uparrow}(x_E) \\ \psi_{\downarrow}(x_E)
\end{array}
\right\rgroup
$$
\begin{eqnarray}
\psi^{\,\prime}(x_E^{\,\prime}) = \exp \left\lbrace (i/2) \Sigma_{\mu\nu} \theta_{\mu\nu} \right\rbrace \psi_E(x_E)
\\
\theta_{k4} = -\,\eta_{k}\qquad\quad \theta_{\imath\jmath} = -\,\varepsilon_{\imath\jmath k} \,\alpha_{k}
\end{eqnarray}

Let us highlight a crucial difference: the upper and lower spinor components of a bispinor transform according to equivalent fundamental representations $ D_{\frac12} $ of SU(2), unlike the left and right Weyl spinors in Minkowski spacetime. It follows that the Euclidean
bispinor $ \psi_E(x_E) $ belongs to the 4d reducible representation $ D_{\frac12}\oplus D_{\frac12}\,. $
Thus,  in this Euclidean formulation of the Dirac spinor field theory the concept itself of non-equivalent Weyl spinors of different chiralities, characteristic of Minkowski spacetime, is lost.
Moreover,  by definition, \textit{$ \overline{\psi}_E$ transforms just
like the Hermitean conjugated of $\psi_E\,$. }
As a matter of fact, from the unitary property
\begin{eqnarray}
\sfU(\theta)=\exp \lbrace (i/2)\,\overline{\Sigma}_{\mu\nu}\,\theta_{\mu\nu}\rbrace
\qquad\quad
\sfU^{\dagger}(\theta)=\sfU^{-1}(\theta)=\sfU(-\,\theta)
\label{unitarity} 
\end{eqnarray}
we have
\begin{eqnarray}
\left( \overline{\psi}^{\,\prime}_{E}\,,\psi^{\,\prime}_{E} \right) \!\!=\!\!
\int\!\!\mathrm{d}^{4}x_E^{\,\prime}\,\overline{\psi}^{\,\prime}_{E}(x_E^{\,\prime})\,\psi^{\,\prime}_{E} (x_E^{\,\prime})\!
=\!\!\int\!\mathrm{d}^{4}x_E\,\overline{\psi}_E(x_E)\,\sfU^{\dagger}(\theta)\,\sfU(\theta)\,\psi_{E}(x_E)\!=\!(\,\overline{\psi}_E\,,\psi_E\,)\0
\end{eqnarray}
with $ x_E^{\,\prime} = R \,x_E\,,\ R^{\top}=R^{-1}\,. $

In this regard, let us focus on the conversion of the Minkowski space chiral matrix 
$$ 
\gamma_5=i\gamma_0\gamma^{1}\gamma^{2}\gamma^{3}=\gamma_5^{\,\dagger} 
\qquad\quad 
\lbrace \gamma_5,\,\gamma^{\,\mu}\rbrace = 0 
$$
The standard  definition in the Euclidean $ D- $dimensional Clifford algebra is
\begin{equation}
\overline{\gamma}_5 \equiv \overline{\gamma}_1\overline{\gamma}_2\overline{\gamma}_3\overline{\gamma}_4 = -\,\gamma_5
\qquad\quad
\lbrace \overline{\gamma}_5,\,\overline{\gamma}_\mu \rbrace = 0
\end{equation}
in such a way that the projectors on the upper and lower components of a Euclidean bispinor become
\[
\mathbb{P}_{\uparrow}=\textstyle\frac12 ( 1+\overline{\gamma}_5 )\qquad\quad
\mathbb{P}_{\downarrow}=\textstyle\frac12 ( 1 - \overline{\gamma}_5 )
\]
where the upper and lower spinor components belong to equivalent fundamental representations of SU(2).

\bigskip
The Euclidean Action integral for the bispinor field is
\begin{equation}
S_E[\,\psi_E, \overline{\psi}_E\,]\ =\ 
\int\mathrm{d}^{4} x_E\ \overline{\psi}_E(x_E)\left(\partial\!\!\!/_E+M\right)\,\psi_E(x_E)
\label{diraceuclideanaction}
\end{equation}
In this Action $ \psi_E $ is a column bispinor, $ \overline{\psi}_E $ is a transposed  or row  bispinor, but independent of $\psi_E$. It should be noted that here {the overall sign as well as any overall multiplicative factor is arbitrary and irrelevant}, and can always be absorbed into e.g. $\overline{\psi}_E$ - remember that we are allowed to change $ \overline{\psi}_E$
and not $\psi_E$, or \textit{viceversa}. Conversely, 
the lack of the factor $i$ in front of the derivative term is not at all accidental: it must be there so as to ensure that the Euclidean fermion
propagator, a.k.a. the 2-point {\sl Schwinger's function},
is proportional to $(ip\!\!/_E\,-\,M)/(p_E^2\,+\,M^2)\,;$ if it were not
for this $i\,,$ then we would have tachyon poles after transitioning back
to the Minkowski space. 

It is worth noting that the above spinor Euclidean Action integral can be obtained
from the corresponding one in the Minkowski space, after the 
standard replacements (\ref{E1}) and (\ref{E2})
$$
x_4=i\,x_{0},\qquad\quad 
\overline{\gamma}_k\;\equiv\;-\,i\gamma^k\quad
(\, k=1,2,3\,),\qquad\overline{\gamma}_4\;\equiv\;\gamma_0
$$
$$
\psi(x)\,\mapsto\,\psi_E\left(x_E\right)\qquad\quad 
\overline{\psi}(x)\,\mapsto\,e^{\,i\theta}\,\overline{\psi}_E (x_E)\quad(\,0\le\theta<2\pi\,)
$$
so the we can always write
\begin{eqnarray}
i S[\,\psi,\overline{\psi}\,] &\mapsto& S_E[\,\psi_E,\overline{\psi}_E\,]
= \int \mathrm{d}^4x_E\ \overline{\psi}_E(x_E)\,\left(\,\partial\!\!\!/_E+M\,\right)\,\psi_E(x_E)
\end{eqnarray}

Furthermore, the Euclidean Dirac operator
$\left(\,\partial\!\!\!/_E+M\,\right)$ is exactly the one that gives rise, according
to the definition (\ref{eucldiracprop}), to the inversion 
formula for the 2-point Schwinger function 
$$
\left(\,\partial\!\!\!/_E+M\,\right)_{\alpha\beta}\,S^{\,E}_{\beta\eta}(x_E)\
=\ 
\delta\left(x_E\right)\,\delta_{\alpha\eta}
$$

\medskip
Finally here is a summary of the basic formulas:
\begin{eqnarray}
&&S_E[\,\overline{\psi}\,,\,\psi_E\,]=\int \mathrm{d}^4 x_E\
\overline{\psi}(x_E)\left(\,\partial\!\!\!/_E+M\,\right)\psi_E(x_E)\\
&& S^{\,F}_{\,\alpha\beta}(\, -i x_4,{\bf x})\ \rightarrow\ -\,S_{\,\alpha\beta}^E(x_E)\0\\
&& S_{\,\alpha\beta}^E(x_E)\ =\ \int \frac{\mathrm{d}^4 p_E}{(2\pi)^4} \ \exp\{ i p_E\cdot  x_E\}\ 
\left\lgroup \frac{i}{-\,p\!\!/_E+ iM} \right\rgroup_{\,\alpha\beta}
\\
&& \left(\,\partial\!\!\!/_E+M\,\right)_{\,\alpha\beta}\,S_{\,\beta\eta}^E(x_E)\ =\ 
\delta(x_E)\,\delta_{\,\alpha\eta}\0
\end{eqnarray}
It is important to highlight that  the Euclidean Action integral for the bispinor field is by no means a real quantity.

Summarizing, we have shown above how one can possibly solve the problem of  deriving a Euclidean field theory from a Minkowski one via Wick rotation. According to the rules prescribed here, in order to define such a Euclidean theory one must double the fermion fields:  more precisely the $\bar \psi$ in the Action has to be replaced with a bispinor independent of $\psi$.  Anyhow the so obtained action is not real. But, what is more, following the above prescriptions there is no room for a Euclidean field theory of Weyl fermions. Other more refined prescriptions have been put forward \cite{nicolai78,Mehta1990,Kupsch1,Kupsch2,vanNieuwen1996,wetterich2010,PCT}. For instance one can supplement the Wick rotation with a rotation of the spinor fields. With this enrichment one can impose hermiticity on the Action for a Euclidean Dirac spinor. One can also define a Euclidean action for Weyl spinors, but for them a hermitean action is out of reach.  So that, among other drawbacks, the corresponding path integral is complex.

The gist of this Section can be summarized with the following warning: 
when a field theory contains Weyl fermions one may be tempted to turn it into a classical Euclidean field theory and quantize this theory, 
instead of the original Minkowski one, with the aim of obtaining automatically Wick-rotated Feynman diagrams, already prepared for dimensional regularization. 
Nonetheless, this procedure does not make sense, because we have seen above that the Euclidean field theory one obtains in this way does represent a completely different physics, if any. 
This fact shares some close resemblance with the problem, illustrated above, of PV regularization: one cannot formulate a local field theory that automatically produces 
PV regularized  Feynman diagrams, for the trivial reason that the inverse of the sum of kinetic operators is not the sum of their inverses. {As we have already pointed out, Wick rotation is a tool that can be applied to Feynman diagrams and, in general, to amplitudes formulated in a Minkowski spacetime, in order to rewrite them in a form fit for regularization. Once the regulated integral is performed one returns to the Minkowski metric by applying the inverse Wick rotation. Throughout this full operation the formulation of a classical Euclidean theory is fortunately unnecessary.} 

{
\section{The family's index theorem and Weyl fermion propagators}

{The existence or non-existence of a Weyl fermion propagator has been one of the main issues discussed in the previous sections. Here we would like to link it with another basic concept, that of anomaly in relation to quantization. More precisely we wish to connect the existence or non-existence of a fermion propagator in a theory of Weyl fermions with the presence of dangerous anomalies in a theory. First let us recall a well known fact: any gauge or gravity theories may have anomalies in the conservation of a current or in the divergence or trace of the energy-momentum tensor. Such anomalies are of two types: they may be dangerous for the quantum theory or innocuous. In the first case they may put in jeopardy the the renormalization and unitarity of the theory, in the second case they are honest quantum effects. In this section we would like to argue that the first class of anomalies, which we call type O (O for obstructive) , are linked to the non-existence of a Weyl propagator. This point of view on anomalies is not new, it goes back to the 80's of the last century, see \cite{atiyahsinger1984}, but is not widespread in the literature. It has been put forward again recently in the book \cite{LB} and in the review \cite{BG24} (see more references there). 

Let us start from the two type of anomalies (here, for reasons of space, we limit ourselves mainly to gauge anomalies, but the discussion can be easily extended to gravitational anomalies). Obstructive anomalies are easily recognizable because they appear in the divergence of the internal currents of a theory, i.e. currents that couple minimally to the gauge potentials. They are also characterized by their consistency property, for they satisfy the Wess-Zumino consistency conditions. Non-obstructive anomalies appear in currents that do not couple minimally to the gauge potentials (but may couple minimally to auxiliary axial potentials, like the one introduced above); they are also called ABJ or covariant anomalies, as they are characterized by their covariant behavior under gauge transformation, instead of consistency\footnote{For lack of space we skip the discussion of trace anomalies, which can be both chiral (obstructive) and non-chiral (non-obstructive), \cite{LB,BG24}.}.
Now we will show that there exists a mathematical criterion that encompasses all O type anomalies and excludes all the others. It is provided by the family's index theorem, of Atiyah and Singer, \cite{atiyahsinger1984,atiyah1989}. This theorem is based on very advanced mathematics and it is impossible to present it here in a satisfactory way. Therefore we end this paper with a summary we hope will be enough to appreciate the importance of the point and the rule that comes out of it.

In the short presentation below spinors are represented by sections of a bundle which is the tensor product of the spin bundle and the gauge bundle. We shall consider spinor of definite chirality. The Dirac-Weyl kinetic operator is a differential operator that maps sections of one bundle with definite chirality to sections of the bundle with the opposite chirality, \cite{Lawson}. In more detail} in a gauge field theory we deal with spinors that transform according to a definite representation
 of the gauge group $\sfG$. So, we consider the tensor product of a spinor bundle $S^\pm_{\mathbb C}$ with a vector bundle $E$ corresponding to a representation $\rho$ of the structure group $\sfG$ of ${\sfP}(X,\sfG)$: $S^\pm_{\mathbb C}\otimes E\equiv S_{\mathbb C}(E)$. A spinor bundle is the associated bundle to the principal fiber bundle with structure group $Spin(\sf d)$ ( tag $\mathbb C$ indicates that we are considering complex spinors). The relevant connection will be the spin connection plus a gauge connection $V=V_\mu dx^\mu$ valued in  the representation $\rho$ of the Lie algebra of $\sfG$ with antihermiten generators. The corresponding Dirac operator
\be
\slashed {\ED} = \slashed {D} +i\slashed {V}\label{ED+V}
\ee
acts on the space of sections  $\Gamma[S_{\mathbb C}(E)]$, i.e. the space of spinor fields, and maps it to itself.  The latter splits as
\be
\Gamma[S_{\mathbb C}(E)]= \Gamma[S^+_{\mathbb C}(E)]\oplus \Gamma[S^-_{\mathbb C}(E)]\label{SE+SE-}
\ee
where $^+$  and $^-$ denote chiralities, and eq.\eqref{ED+V} splits accordingly as
\be 
 {\slashed {\ED}}= \left(\begin{matrix} 0 & {\slashed {\ED}}^- \\  {\slashed {\ED}}^+ & 0 \end{matrix} \right)\label{DiracEDsplit}
\ee

The {\it Atiyah-Singer index theorem} reads:
\be
Ind ({\slashed{\ED}}^+) \equiv dim(ker\, {\slashed {D}}^+ )- dim (ker\, {\slashed {D}}^-)= \int_X \, ch(E) \,\hat A(X)\label{indexED}
\ee
$ker{\slashed {D}}^\pm$ denotes the space of zero modes of  ${\slashed {D}}^\pm$ in a  specific spacetime manifold $X$.
The symbol $ch(E)$ indicates the Chern character of the $E$ bundle, i.e. the rational characteristic class given in terms of the curvature $F$ of $V$ by
\be
ch(E)= r + \frac i{2\pi} \tr\, F + \frac {i^2}{2(2\pi)^2} \tr\, F^2+\ldots+ \ldots\label{chE}
\ee
where $r$  is the dimension of the representation $\rho$. The symbol $\widehat A(X)$ denotes the {\it $\widehat A$-genus}, which is the (rational) Pontryagin characteristic class of $X$. It can be expressed in terms of the Riemann curvature $R$ as follows
\be
\widehat A(X)&=& 1 +\frac 1{(4\pi)^2} \frac 1{12}\, \tr \, R^2 + \frac 1{(4\pi)^4}\left[ \frac 1{288}\, \tr\, (R^2)^2 +\frac 1{360}\,\tr\, R^4\right]+\ldots \label{hatAX}
\ee
This is however the index theorem for a fixed potential $V$. In gauge field theories we have to do with a family of potentials $\cal A$, forming orbits of gauge transformations, in which the relevant space is the quotient ${\cal Q}=\frac {\cal A}{\cal G}$, the moduli space of the theory. In this case the $ker\, {\slashed {D}}^\pm$ , i.e. the spaces of zero modes of ${\slashed {D}}^\pm$ is not fixed, but varies in dimension from point to point of  ${\cal Q}$, i.e. they are bundles over ${\cal Q}$. Therefore the index in this case is the difference of two bundles. Differences of bundles are described by the K theory. Using the latter Atiyah and Singer could write down the formula for the index of a family of operators in terms of forms.

In the gauge theory case just outlined in which  the Dirac operator acts between two families of sections $\Gamma(S^+_{\mathbb C}\otimes \EV)$ and  $\Gamma(S^-_{\mathbb C}\otimes \EV)$, $\EV$ being the gauge bundle, the index theorem can be expressed in the form
\be
ch \left(ind ({\slashed \ED}^+)\right) =\int_{\sfM}  ch( {\EV})\cdot \hat A(T{\cal Q})\label{chindfam}
\ee
where $ch$ is the Chern character, which for a vector bundle $V$ is defined by
\be
ch(V)=rank(V)+ c_1(V)+\frac 12 \left(c_1(V)^2 -2 c_2(V)\right)+\ldots\label{cherncharacter}
\ee
In \eqref{chindfam},  $\sfM$ is a generic spacetime manifold, ${\cal Q}\equiv \frac{\cal A}{\cal G}$ is the orbit space of connections, $T{\cal Q}$ is its tangent space and $\EV$ is the gauge bundle. The Chern character of the index measures the extent to which $ker\, {\slashed \ED}^+$ differs from $coker \,{\slashed \ED}^+=ker \,{\slashed \ED}^-$. It is intuitive that, as long as, $ind ({\slashed \ED}^+)$ differs from 0, the inverse of ${\slashed \ED}^+$ cannot exists. {This conclusion although plausible may seem somewhat hasty. However there is more than intuition in it at least in the significant case where D. Quillen's analysis can be applied, \cite {Quillen}. Quillen's determinant bundle is the line bundle constructed by taking the antisymmetruc products of the zero eigenmodes of the kernel tensored with the analogous construction relative to the cokernel. It turns out that, roughly speaking, as long as the family's index does not vanish Quillen's line bundle is nontrivial, thus it does not have an invertible global section, which indicates that the inverse of the Dirac operator does not 
exist.}

The cohomology classes relevant to a gauge theory in 4d are of order 4 and 6. The order 6 ones, via the transgression formula, give rise to the consistent gauge anomalies. Therefore we see that consistent gauge anomalies are in one to one correspondence with cohomology classes that obstruct the invertibility of the Dirac-Weyl operator and for this reason are called obstructive (type O). But there are also the classes of order 4: they are proportional to $\tr (R^2)$ and $\tr (F^2)$, where $R$  is the Riemann curvature and $F$, as usual, the gauge curvature.  They are known as Pontryagin and Chern classes, respectively. In field theory they may appear as odd parity trace anomalies. It is clear that, independently of the field theory interpretation, the corresponding obstructions must be canceled in order to permit the inversion of the Dirac-Weyl operator, i.e. the existence of the corresponding propagator.

Some specifications are in order. When the potential $A$ is  a fixed one and $\sfM$ a precise space-time manifold, like in the case of the ordinary index theorem \eqref{indexED}, one can, disregarding the zero modes, still define a product of nonzero eigenvalues and interpret it as a partition function. But in the case considered here the zero modes vary from point to point of $\cal Q$ and $\sfM$ represents all possible manifolds, see \cite{LB,BG24}, such an operation is generically impossible. The integration over $\sfM$ should not mislead the reader. Contrary to the case of the ordinary index theorem where the topology of spacetime plays a major role, the family's index theorem is formulated in a universal form, in terms of classifying space and classifying bundles, meaning that it holds independently of the topology of the spacetime. This matches exactly the sense of perturbative approach to field theory where locality requires to make the calculations in a local patch, so that the topology of spacetime is irrelevant.

Another due specification is about the proof of the family's index theorem: to the best of our knowledge it is available only in the Euclidean version. Following a well established and sensible tradition we considers its statement valid also in Minkowski spacetimes. This does not contradict what has been said in the previous Section: 
in the present Section we consider the problem of inverting the Dirac-Weyl kinetic operator, in the previous section the object was the construction of a canonical Action involving the same operator as kinetic operator. Both issues are studied in a Euclidean setting, but are different.

In the present context we do not insist on gravitational anomalies (the ones that may appear in the divergence of the energy-momentum tensor), although this could be done, see \cite{LB}, because we know that such anomalies vanish for group theoretical reasons, see \cite{BG24}.

Once these clarifications are  set forth, the family's index theorem seems to be able to establish an uncontroversial rule for perturbative calculations in local field theory.  Any method or procedure that concerns the computations of effective Actions, conservation laws and anomalies must be in agreement with this theorem. The family's index theorem provides the backbone for the correctness of any perturbtive calculation. {It links the existence of O type anomalies with non-invertibility of the Dirac-Weyl operator, i.e. with the absence of the Weyl propagator. So, in particular, it explains the origin of the negative role of such anomalies: no correct quantization is possible without the relevant propagators. If we replace the missing propagator with a fake one we are bound to meet trouble.}

{ Finally an issue on which the familiy's index theorem alone does not provide an answer is whether Weyl fermion theories with vanishing index theorem have all the necessary propagators. We have explicit examples where this index vanishes: a theory of Dirac fermions for instance is a simple example with vanishing index where the propagator is well defined; a more significant example is provided by the stabndard model (and its generatlizations). In the SM the Weyl spinor kinetic terms rearrange themselves to form a set of Dirac fermion ones. Of course in such a case the fermion propagators do exist. Everything indicates that this may be true in general: that is, theories of Weyl fermions of vanishing family's index have all the needed propagators because they can be rearranged as theories of Dirac fermions. But a formal proof of this is lacking.}

\section{Conclusion}

The purpose of this paper was to clarify a few controversial or unclear issues concerning fermions, their definition and their use. 
For this reason we have started in Section 2 and 3 with a review of the their basic properties. In particular, Section 3 is an elementary and detailed construction of Weyl fermion solutions of the 4d Dirac equations, by means of which both Dirac and Majorana massless spinors have been constructed. 
The corresponding quantum fields have been introduced, their specific properties under discrete symmetry transformations have been established and their causal Green's functions properly constructed. 
The general properties of Dirac, Weyl and Majorana fields have been summarized in Section 4. 
All this preparatory material has been brought to bear in the subsequent Sections.
Section 5 is about the perturbative calculation of the effective Action of a Weyl fermion coupled to an Abelian gauge potential. 
This example exhibits one of the difficult issues considered in this review: the non-existence of the Weyl fermion propagator. 
This problem is tackled in the most popular way met in the literature: one adds to the original theory a free Weyl fermion with opposite chirality. 

The example carried out in Section 5 with three different regularization (dimensional, PV and cutoff) shows that an accurate treatment lead to the same result for all three, 
as it should,  but especially that {\sf no mass term can arise from quantization} and that the added Weyl fermion of opposite chirality remain free after quantization, 
so that chirality of the original theory is preserved. 
The case considered in Section 5 has been then analyzed in general in Section 6. 
The main result of this analysis has been that the practice of replacing the Weyl propagator, which does not exist, with a Dirac propagator, 
is however flawed by an underlying logical loophole, so that this trick may accidentally lead to correct results but does not have any probatory value. 
It has been shown that there are ways to circumvent this obstacle by enlarging the space of potentials, 
such as for instance adding an axial potential to the original vector one. 

Section 6 has been devoted to dispelling a not infrequent identification of massless Majorana with Weyl fermions. 
It is argued that while these two types of fermions have a few characteristics in common, they are crucially different in other fundamental aspects: 
Weyl fermions are eigenstates of the chirality matrix, while Majorana fermions do not have any definite chirality and no definite helicity. 
The chirality issue is fundamental in the framework of anomalies: all anomalies with opposite sign for  opposite chiralities (the dangerous ones) are absent for Dirac and Majorana fermions. For the very same reason Dirac and Majorana fermions can have a mass term, 
while for the Weyl ones a mass is forbidden. 
The difference between the Dirac and Majorana mass terms has been subsequently clarified. 
{ The main outcome of our analysis is that, in spite of being build-up by the very same two-component Weyl spinor,
the chiral Weyl bispinor and the massless Majorana spinor do exhibit quite different and distinguished mathematical and physical properties and behaviors.}

There are two cases in which locality of the classical Action is put in jeopardy. One is when the PV regularization is inappropriately transferred from the realm of Feynman diagrams to that of local Actions by confusing the sum of inverses of differential operators (the PV ghost propagators) with the inverse of the sum of the corresponding kinetic differential operators to construct a local Action. This point has been insisted on throughout the paper. There is a similar issue related to Wick rotation. One may be tempted to turn to the Euclidean metric already in the classical Action, but it was shown in Section 7 that the theory one ends up with in this way represents a physical system different from the starting one. The Wick rotation is an algorithm concerning the regularization of Feynman diagrams and related amplitudes, which cannot be transferred back mechanically to the original Action integral.

The final Section is devoted to the family's index theorem. Although anomalies are not a central issue in this review, they nevertheless  loom all through it. 
The anomalies called obstructive (the dangerous ones) pop up in chiral theories as a signal of a disease: the lack of chiral fermion propagators. 
The connection between anomalies and (non-existence of) chiral propagators is provided by the Atiyah-Singer index theorem. 
For this reason we have deemed it useful to give a summary review of it, certainly insufficient but hopefully enough to stir the reader's interest.
\vskip 1cm
{\bf Acknowledgements}. After completing this work we became aware of ref.\cite{dreiner22} which partial overlaps with some of our topics.
\end{document}